\documentclass[preprint]{aastex}
\usepackage{graphicx}
\usepackage{natbib}
\usepackage{lineno}
\begin{document}

\title{Connections between Spectra and Structure in Saturn's Main Rings Based on Cassini VIMS Data}
\author{M.M. Hedman$^{a,*}$, P.D. Nicholson$^a$, J.N. Cuzzi$^b$, R.N. Clark$^c$, \\ G. Filacchione$^d$, F. Capaccioni$^d$, M. Ciarniello$^d$  }
\affil{\it $^a$ Department of Astronomy, Cornell University, Ithaca NY 14853 USA \\
$^b$ NASA Ames Research Center, Moffett Field CA 94035 USA \\
$^c$ US Geological Survey, Mail Stop 964, Box 25046, Denver Federal Center, Denver, CO 80225, USA \\
$^d$ INAF-IAPS di Tor Vergata, via del Fosso del Cavailere, 100, 00133 Rome, Italy}

\maketitle

{\bf Abstract:} Saturn's main rings exhibit variations in both their opacity and their spectral properties on a broad range of spatial scales, and the correlations between these parameters can provide insights into the processes that shape the composition and dynamics of the rings. The Visual and Infrared Mapping Spectrometer (VIMS) instrument onboard the Cassini Spacecraft has obtained spectra of the rings between 0.35 and 5.2 microns with sufficient spatial resolution to discern variations on scales below 200 km. These relatively high-resolution spectral data reveal that both the depths of the near-infrared water-ice absorption bands and the visible spectral slopes are often correlated with structural parameters such as the rings'  optical depth. Using a simplified model for the ring-particles' regolith properties, we have begun to disentangle the trends due to changes in the gross composition of the ring particles from those that may be due to shifts in the texture of the ring particles' regolith. Consistent with previous studies, this analysis finds that the C ring and the Cassini Division possess enhanced concentrations of a contaminant that absorbs light over a broad range of wavelengths. On the other hand, a second contaminant that preferentially absorbs at short visible and near-ultraviolet wavelengths is found to be more evenly distributed throughout the rings. The optical activity of this short-wavelength absorber increases  inwards of 100,000 km from Saturn center, which may provide clues to the origin of this contaminant. The spectral variations identified as shifts in the regolith texture are in some places clearly correlated with the ring's optical depth, and in other locations they appear to be associated with the disturbances generated by strong mean-motion resonances with Saturn's various moons. These variations therefore seem to be controlled by the ring particles' dynamical environment, and may even provide a new avenue for constraining the structure and mass density of Saturn's most opaque ring regions. 


\pagebreak

\section{Introduction}

The particles in Saturn's main rings have long been known to be composed primarily of water ice. The ring's near-infrared spectra contain prominent water ice absorption bands \citep{Kuiper70, Lebofsky70, CM80, Poulet03, Nicholson08, Cuzzi09}. Also, the rings' low emissivity and high reflectivity at radio wavelengths  suggests that the rings may be composed of nearly pure water ice (Pollack {\it et al.} 1973; Biggs 1974; Pettengill and Hagfors 1974; Pollack 1975; Cuzzi and Dent 1975 but see Epstein {\it et al.} 1984 for an alternative perspective).  \nocite{Pollack73, Briggs74, PH74, Pollack75, CD75, Epstein84} However,  multiple spectral and photometric observations clearly indicate that the ring material also includes at least two non-icy contaminants \citep{CE98, Poulet03}. One of these contaminants is required to explain the ring's reddish color at visible wavelengths, while another contaminant is needed to explain the generally low albedos of the particles in the C ring and Cassini Division. Furthermore, variations in the rings' spectral properties on a broad range of spatial scales \citep{Estrada03, Nicholson08} require shifts in the ring particles' composition and/or  regolith texture.  Since 2004, the Visual and Infrared Mapping Spectrometer  (VIMS) onboard the Cassini spacecraft has obtained a large quantity of visible and near-infrared spectral data on the rings, enabling their spectral properties  to be studied in finer detail than previously possible. Indeed, there are now multiple ongoing efforts to model the detailed shape of VIMS spectra and thereby determine the optical properties and composition of the rings \citep{Clark2012, F12}. However, important insights into the rings' texture and composition can also be obtained by  studying how the rings' spectral properties vary with location within the rings. Spectral variations  on scales down to a few hundred kilometers can now be compared with similar-scale variations in the rings' opacity obtained from high-quality stellar occultations, allowing us to explore the relationships between the dynamical state of a ring and the spectral properties of its component ring particles.  

This report describes a preliminary investigation of some of the highest-resolution spectral observations of the rings obtained by VIMS in order to catalog and identify interesting correlations between the rings' spectral properties and various structural features. Section 2 describes the relevant Cassini VIMS data sets and how they were reduced to yield radial profiles of opacity and spectral parameters. Section 3 presents profiles of various spectral quantities and discusses some of the interesting correlations and patterns in these data. Section 4 describes how these spectral parameters are transformed into estimates of the optical activity of the two non-icy contaminants and the ``effective scattering length'' in the ring particle regolith.  Section 5 presents the results of this analysis as profiles of the relevant regolith parameters, which are discussed in more detail in the following two sections. Section 6 examines the compositional trends found by this analysis, while Section 7 discusses the observed  variations in the effective scattering lengths in the ring particles' regolith. While the physical processes responsible for the observed trends are still obscure, they could reflect such diverse processes  as inter-particle collisions, pollution from meteoritic debris, and ballistic transport. These spectral variations should therefore provide new insights into the collisional and dynamical properties of the rings. 


\section{Observations and data reduction}

The VIMS instrument is described in detail in \citet{Brown04}. This analysis will use both ring spectra derived from selected imaging observations and measurements of the ring's opacity taken from stellar occultations. In order to facilitate comparisons between the ring's opacity and its spectral properties, both these data sets have been reduced to produce profiles of the ring's opacity and spectral properties  as a function of ring radius (distance from the planet's spin axis). As the data reduction procedures for the spectral and occultation data are quite different, they are described separately below.

\subsection{Spectral data}

In normal operations, VIMS uses  two separate co-aligned channels to obtain spatially-resolved spectra of a given scene. The VIS channel (covering the wavelength range  of 0.35-1.05 microns in 96 channels) uses a long-slit spectrometer and CCD array to acquire spectra for a row of up to 64 pixels simultaneously, and an one-axis scanning mirror to build up an image that can be as large as $64\times64$ pixels. On the other hand, the IR channel (covering the wavelength range from 0.85 to 5.20 microns in 256 channels) uses a InSb linear array detector to obtain spectra of a single pixel and a two-axis scan mirror system to form an image up to $64\times 64$ pixels in size. Nominally, each pixel observed by both channels is 0.5 by 0.5 mrad, but both instruments can also operate in a high-resolution mode. For VIS, the high-resolution pixels are 0.17 by 0.17 mrad, while for IR, high-resolution pixels are 0.25 by 0.5 mrad.  The data from both channels are packed together into ``cubes'' with two spatial dimensions and one spectral dimension. 

A typical VIMS ring observation is a radial mosaic composed of multiple cubes targeted at different locations in the rings. Prior to assembling the data from each of these observations into radial profiles of spectral parameters, the relevant cubes need to be calibrated and geometrically navigated.  Calibration of the relevant cubes uses standard routines that remove backgrounds, apply flat-fields and convert the raw Data Numbers to $I/F$, a standardized measure of reflectivity that is unity for a Lambertian surface viewed at normal incidence (the specific flux calibration being RC17, see Clark {\it et al.} 2012).  The calibrated cubes are geometrically navigated using the appropriate SPICE kernels to predict where various ring features would appear in the images. These predictions can be displaced from the actual observed ring features by amounts comparable to a VIMS pixel, likely because of small uncertainties in the orientation of the spacecraft. The observation geometry was therefore refined by making small adjustments to the assumed pointing. Due to VIMS' low spatial resolution, these adjustments could be made simply by aligning the predicted ring features ``by eye''. For some of the highest-resolution scans (e.g. the Rev 008 RDHRCOMP observation discussed below), additional effort is needed to ensure that features seen in multiple cubes are properly aligned.\footnote{In general, features observed in multiple cubes are aligned to within a pixel.  However, forcing features to align across multiple cubes causes the absolute radial position of the features to deviate from their fiducial locations. This probably reflects small errors in either the pixel scale or distortion matrix, which are currently under investigation. For the present analysis, we can align features in different cubes with each other and with the fiducial radius scale by introducing deliberate pointing offsets that shift ring features in the azimuthal direction by a fraction of the field of view. Since we average the data over all longitudes, these artificial longitudinal adjustments have no practical effect on this study.}

After calibration and geometric navigation, we generate radial profiles of various spectral parameters, including $I/F$ levels, brightness ratios between two wavelength ranges, band depths and continuum slopes (see below). To generate any of these profiles, we
first compute the desired spectral parameter for every pixel in all the appropriate cubes, along with the ring-plane radius of the center of each IR and VIS pixel in the relevant cubes for each observation. A well-known offset in the pointing between the VIS and IR channels \citep{Brown04}  is accounted for in these calculations, but small differences in the pointing among the different VIS wavelength channels are ignored at this step (see below). The relevant spectral data for all the pixels in all the cubes are then sorted by radius and grouped into a series of evenly spaced radial bins. The widths of the bins are set so that the average bin contains around 10 spectra. For each bin, we compute the mean value of the spectral parameter of all pixels in that bin. We estimate the statistical uncertainty on this value as the standard error on the mean of the relevant measurements. Finally we estimate the correlation coefficients between certain spectral parameters from the relevant covariances.

After computing the spectral profiles, some additional processing is needed to remove various sub-pixel residual pointing errors. For example, the VIS slit, diffraction grating and CCD are slightly misaligned, so different wavelength channels map to slightly different spatial locations on the CCD  \citep{F06}. This means that the VIS brightness profiles derived above do not precisely line up with each other or with the IR profile. These misalignments are relatively small and are still not perfectly characterized, so rather than attempt to correct the pointing estimates on individual pixels, we instead adjust the radius scale of the VIS brightness profiles so that they align with the IR profiles. This is accomplished by cross-correlating each VIS brightness profile with an IR brightness profile over the radial range from 80,000 to 90,000 km (which possesses numerous sharp features), and extrapolating the resulting shifts and scaling factors to the entire profile. This extrapolation was verified by comparing the position of the Encke Gap in the outer A ring among the different spectra.  Note that while these profiles are accurate  to within the 100-200 km resolution of the observations, further small ($\sim$50-100 km) adjustments are required to bring them into alignment with the higher-resolution occultation data (described below).

VIMS has observed the rings numerous times in a broad range of viewing geometries and a variety of resolutions. However, this analysis  focuses on a restricted subset of these observations that are most suitable for exploring radial variations on scales down to a few hundred kilometers. Among the VIMS observations obtained between 2004 and 2011, we searched for data sets that  (1) cover the entire main ring system in an imaging mode\footnote{As opposed to modes where the instrument acquired data for a single line or pixel at a time, which are more difficult to geometrically navigate.}, (2) have resolutions better than 200 km/pixel, and (3) observed the rings at longitudes more than 90$^\circ$ from the sub-solar point (in order to minimize any complications associated with Saturn-shine on the rings). We found roughly 20 data sets that matched all these criteria, which could be divided into four groups based on whether the rings were observed at high or low phase angles and whether the spacecraft viewed the lit or unlit side of the rings. Based on visual inspection of the spectral profiles derived from these observations, we selected one of the highest-quality observations from three of these four groups:  The RDHRCOMP observation from Rev 008\footnote{Rev is a designation for Cassini's orbits around Saturn.} yielded the highest quality low-phase, lit-side spectral data.  The COMPLODRK  observation from Rev 033 is the best-quality low-phase, unlit-side observation that covers the entire rings (the data from Saturn Orbit Insertion described in \citet{Nicholson08} have higher-resolution but do not include any data interior to 86,000 km or between 97,600 and 117,000 km), and the RDHRSCHP observation from Rev 067 is the most stable high-phase, lit-face observation to date. Unfortunately, no comparably high-quality unlit-side, high-phase observation was found in our search. Still, these three observations provide a sufficiently large range of viewing conditions for us to evaluate how the various spectral parameters vary with observing geometry. The relevant parameters all three of these observations are provided in Table~\ref{partab}. 

\begin{table}
\caption{Parameters of the Spectral Observations }
\label{partab}
\resizebox{6in}{!}{\begin{tabular}{|c|c|c|c|c|c|c|c|c|c|c|} \hline
Rev & Observation & Start Time & End Time &
Incidence & Emission & Phase & Obs. long. - & IR Resolution & VIS Resolution 
&  Scan  \\ 
& &  & & Angle & Angle & Angle & Sub-solar long. & (km/pixel) & (km/pixel) & Sampling \\ \hline
008 & RDHRCOMP001   & 2005-140T23:45 & 2005-141T01:54 & 
111.6$^\circ$ & 99.2$^\circ$-106.4$^\circ$ & 12.7$^\circ$-41.1$^\circ$ &
101.9$^\circ$-131.8$^\circ$ & 131-154  & 44-51  & 20 km \\
033 & COMPLODRK001 & 2006-325T04:07 & 2006-325T06:49 & 
104.9$^\circ$ & 58.1$^\circ$-73.3$^\circ$ & 35.4$^\circ$-63.4$^\circ$  &
105.4$^\circ$-122.9$^\circ$ & 148-186  & 146-186 & 50 km \\ 
067 & RDHRSCHP001   & 2008-131T03:21 & 2008-131T05:55 & 
97.1$^\circ$ & 133.1$^\circ$-156.0$^\circ$ & 106.1$^\circ$-120.8$^\circ$ &
 234.9$^\circ$-247.9$^\circ$ & 100-118  & 100-118  & 50 km \\ \hline
\end{tabular}}
\end{table}

\subsection{Occultation data}

In addition to taking spatially-resolved spectra of a scene, VIMS can also operate in an ``occultation mode'' where the VIS channel is turned off, and the IR channel stares at a single pixel targeted at a star, obtaining a series of rapidly-sampled near-infrared stellar spectra. A precise time stamp is appended to each spectrum, facilitating the geometry reconstruction. As the star moves behind the rings, its apparent brightness varies due to variations in the ring's opacity, and these data can be used to obtain a profile of the ring's optical depth as a function of radius.

Unlike the spectral data, for occultations the raw Data Numbers are not converted to $I/F$ values. Instead a constant instrumental background is removed from each spectral channel. Note that the response of the detector is highly linear, and so these DN values are proportional to the stellar flux at each of 31 wavelengths (occultation data are usually spectrally summed over 8 channels to save on data volume). In order to avoid contamination from sunlight scattered by the rings, we focus exclusively on the spectral channel covering the range 2.87-3.00 microns. The rings are sufficiently dark at these wavelengths that ringshine is negligible, so the transmission through the rings 
$T$ is simply the ratio of the observed stellar signal to the signal measured outside the rings. This transmission estimate can be converted into an estimate of the normal optical depth through the rings using the standard formula:
\begin{equation}
\tau_n=-|\sin B_*| \ln T
\end{equation}
where $B_*$ is the elevation angle of the star above the ring-plane.

Using the available SPICE kernels, we can compute the radius where the starlight  pierces the ringplane for each sample in the given occultation. These parameters depend only on the positions of the spacecraft and the star relative to Saturn, which are known to a higher accuracy than the spacecraft orientation. Indeed, the reconstructed geometry is accurate to within a kilometer, which can be verified using sharp edges of gaps and ringlets. 

VIMS has observed many occultations from a variety of geometries over the course of the Cassini mission \citep{Hedman07, NH10}. However, for this analysis we will use only the data from an occultation by the rings of the  bright star $\gamma$ Crucis observed during Rev 089. This occultation not only covers the entire main rings, but was also observed from a very high ring opening angle ($B_*=62^\circ$),  which permits us to obtain detectable signals through relatively high optical regions in the rings. Indeed, the maximum detectable normal optical depths for individual optical depth measurements (which average over less than a kilometer in ring radius) is of order 5, and averaging measurements over regions comparable to the radial sampling of the spectral profiles (20-50 km) yields maximum
detectable optical depths above 6 (However, the optical depth profiles plotted in this paper are usually truncated at a normal optical depth of 5 for the sake of clarity). Also, the high elevation angle ensures that self-gravity wake structures in the rings have little effect on the derived normal optical depths.

\section{Radial profiles of spectral parameters}

\begin{figure}
\resizebox{6in}{!}{\includegraphics{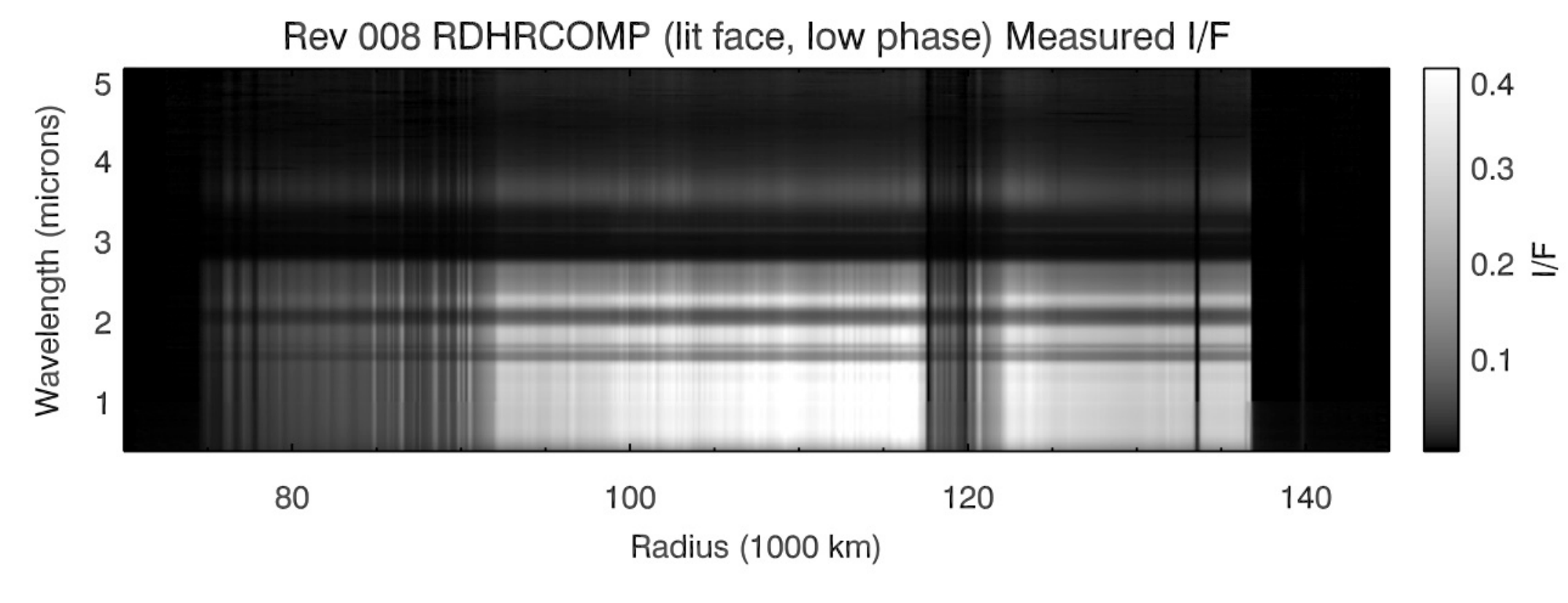}}
\resizebox{6.1in}{!}{\includegraphics{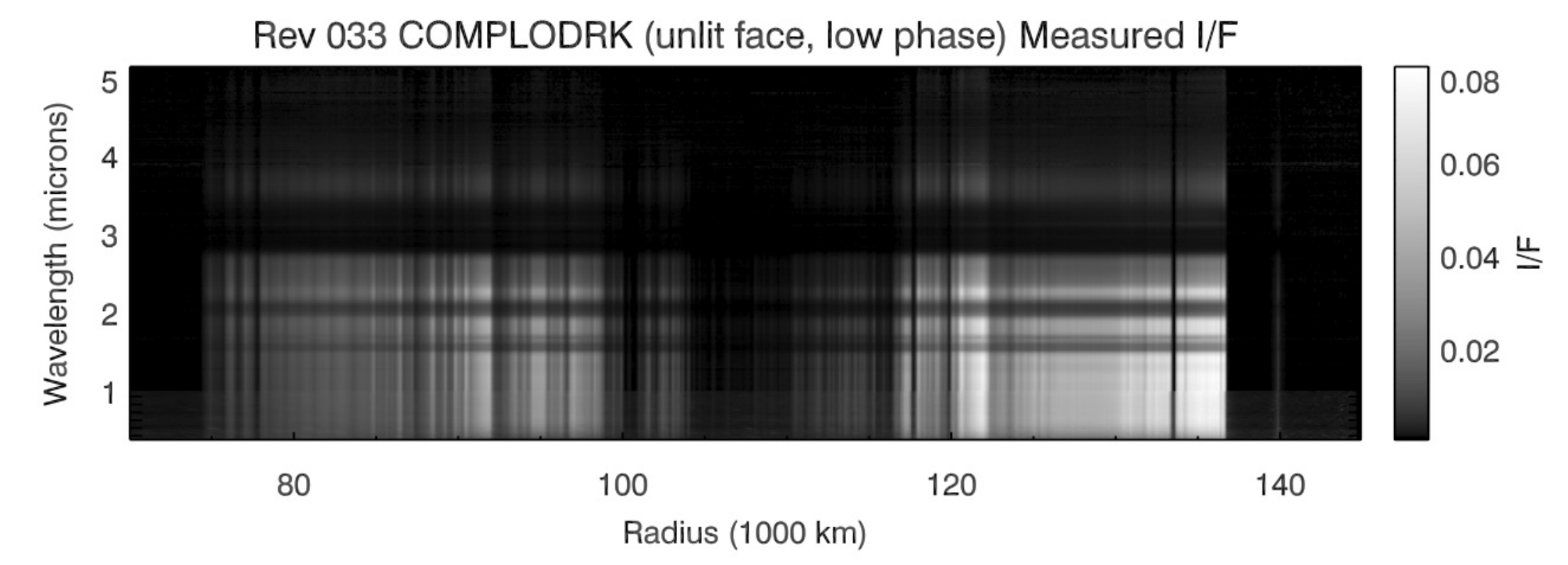}}
\resizebox{6.17in}{!}{\includegraphics{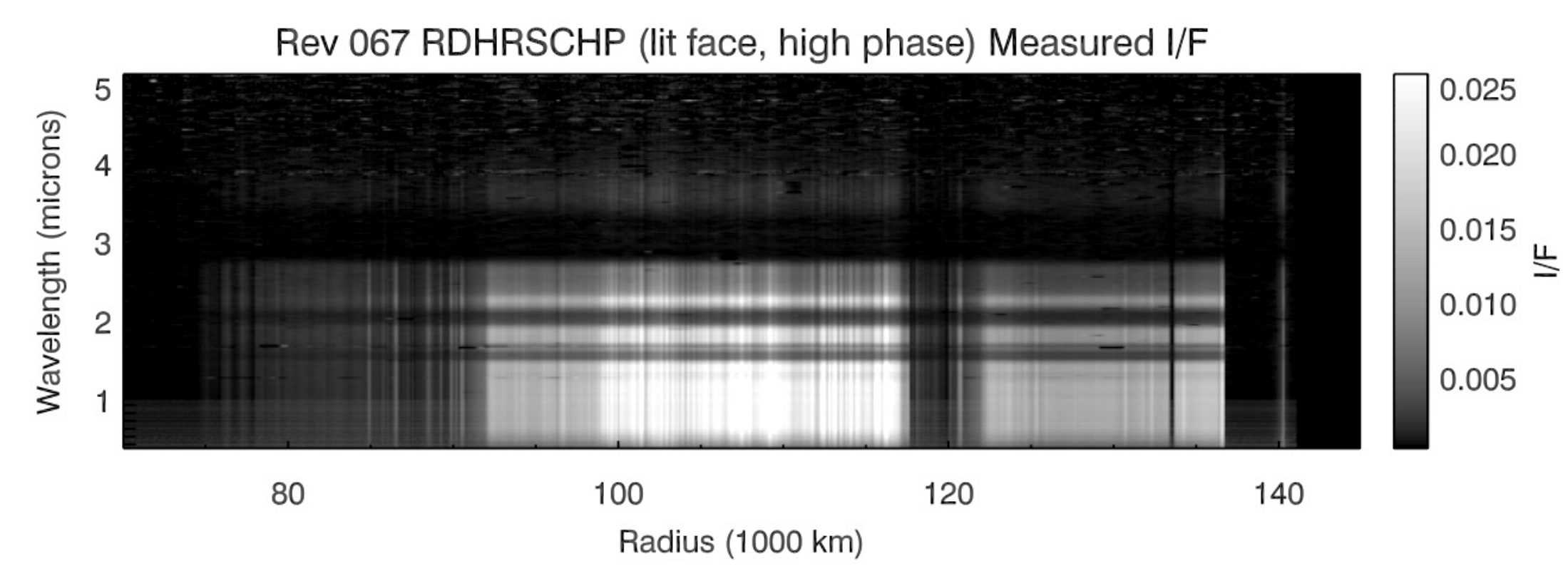}}
\caption{Spectrograms derived from the three observations considered in this analysis.
 Each panel shows the brightness of the rings as a function of radius and wavelength. Note that a different gamma-corrected brightness scale has been used for each observation. See Nicholson {\it et al.} (2008) for more details about interpreting these sorts of images.}
\label{specspaceim}
\end{figure}

\begin{figure}
\resizebox{6in}{!}{\includegraphics{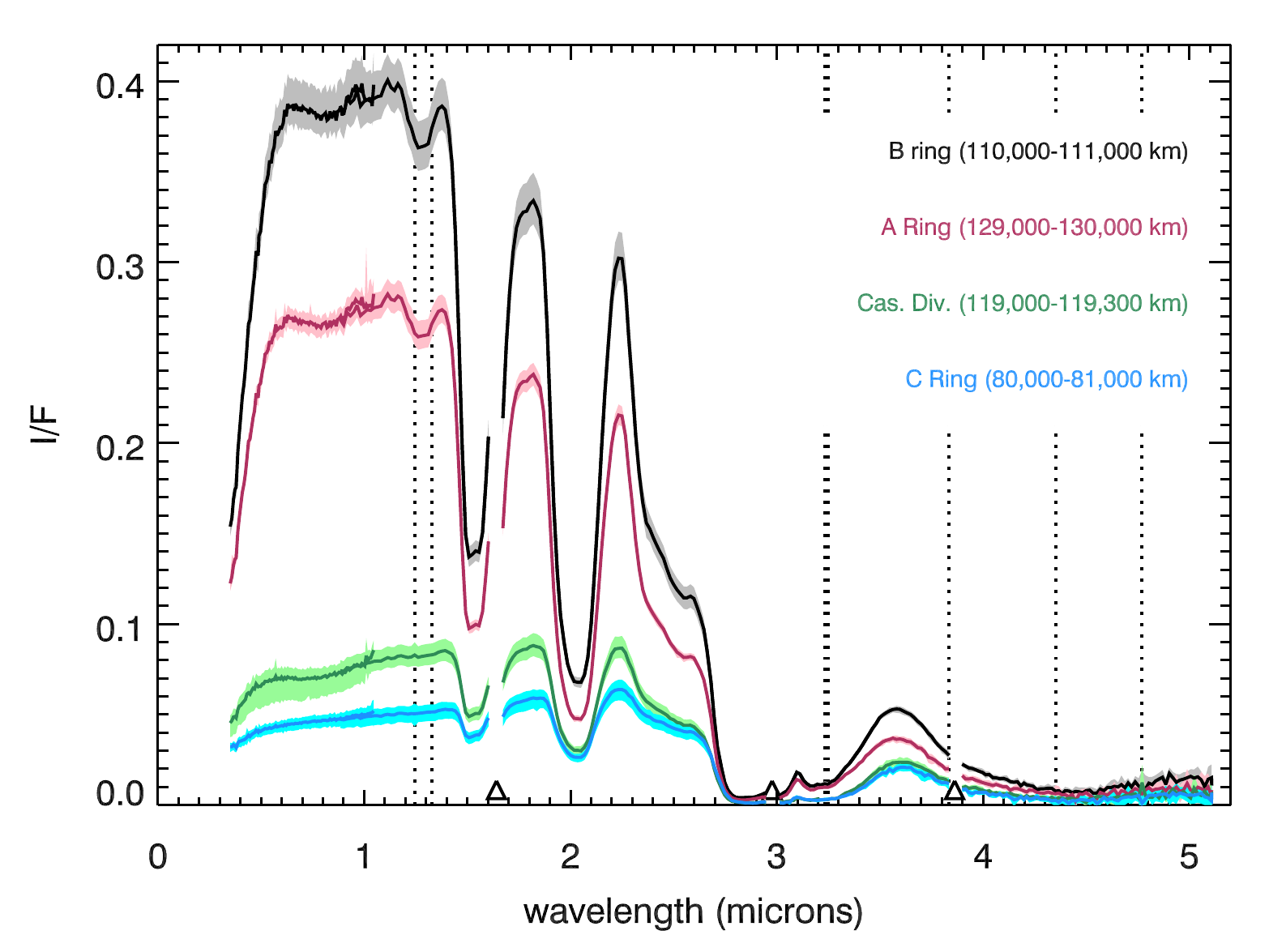}}
\caption{Average spectra of selected regions in the main rings, derived from the low-phase, lit-face RDHRCOMP data illustrated in Figure~\ref{specspaceim}. The vertical dotted lines mark the locations of ``hot pixels'' in the infrared detector with higher than average dark current. The triangles mark the locations of gaps in the order-sorting filter, where the spectral data are less secure and thus not plotted. The solid curves give the mean spectrum for each region, while the shaded bands indicate the range of $I/F$ values observed in each of these locations. These ranges primarily reflect variations in the absolute brightness of the ring, rather than statistical uncertainties in the spectra themselves. The overlapping curves between 0.85 and 1 microns arise because the VIS and IR channels both cover this spectral range. Note the strong water-ice absorption bands at 1.5, 2.0 and 3.0 microns  (weaker ice bands can also be seen in the A and B ring spectra at 1.25 and 1.05 microns) and the steep slope at wavelengths below 0.6 microns. Also note the variations in the spectral slope at continuum wavelengths among the different spectra.}
\label{rdhr8spec}
\end{figure}

The procedures described in Section 2.1 can yield radial profiles for a wide variety of spectral parameters. For example, one can derive separate brightness profiles for each of the 352 wavelength channels to produce an array of spectra called a spectrogram.  Figure~\ref{specspaceim} illustrates spectrograms derived from the three relevant spectral observations as two-dimensional images of brightness versus radius and wavelength. Horizontal slices through one of these spectrograms yield reflectivity profiles at a given wavelength, while vertical cuts provide spectra at a given radial location. Hence these arrays contain nearly all of the relevant spectral information from the original observations. However, in practice we do not need 352 separate brightness profiles to characterize the gross spectral properties of the rings. 

Figure~\ref{rdhr8spec} shows representative examples of main-ring spectra derived from the RDHRCOMP observation, but all three observations show the same basic spectral features. Most obviously, there are water ice absorption bands at  1.25, 1.5,  2.0, 3.0 and 4.5 microns. These spectra also show steep slopes at wavelengths below 0.6 microns, which cannot be attributed to water ice. Instead, these slopes provide evidence for some contaminant material in the ring, which could either be an organic compound or nanometer-sized grains of iron and hematite \citep{Cuzzi09, Clark2012}. The slope at continuum wavelengths longer than 0.6 microns also varies among these spectra, being blue in the A and B rings and red in the C ring and Cassini Division \citep{F12}. These variations in the continuum slope are likely due to another contaminant in the ring that absorbs over a broad range of wavelengths. 

The above spectral features can be quantified using spectral slopes, band depths and/or brightness ratios, and we will consider all these different types of spectral parameters here. First, we compute brightness levels, band depths and visible spectral slopes for all three observations in order to facilitate comparisons with previously published profiles.   
However, it turns out that band depths and spectral slopes are not the most useful parameters to use when deriving information about the ring-particles' composition and regolith texture. Hence we also compute a series of simple ratios of the ring's brightness at different wavelengths (in Section 4 these ratios will be used  to constrain spectral models). Finally, we will summarize some of the salient variations in the rings'  spectral properties. 

\subsection{Brightness levels, band depths and spectral slopes for the three observations}

There are a number of different ways to quantify the strength of an absorption band. Since this analysis focuses upon the spatial variations in gross spectral properties across the rings rather than the detailed shape of the water-ice absorption bands, we will
use a simple, standard expression for band depths \citep{CR84}:
\begin{equation}
D_B=\frac{I_{cont}-I_{band}}{I_{cont}}
\end{equation}
where $I_{band}$ is the brightness in the middle of the chosen band, and $I_{cont}$ is a continuum brightness level inferred from regions outside the band (In practice, the relevant brightness ratios are derived from ratios of the observed reflectivities $I/F$). Here the depth of the 1.5-micron water-ice absorption band $D_{1.5}$ is computed by setting $I_{band}$ equal to the average brightness between 1.50 and 1.57 $\mu$m, and $I_{cont}$ equal to the average brightness in the two wavelength ranges on either side of the band (1.34-1.41 $\mu$m and 1.75-1.84 $\mu$m). Similarly, the depth of the 2-micron ice band $D_{2.0}$ uses the average brightness between 1.98 and 2.09 $\mu$m for $I_{band}$ and the average brightness in the ranges 1.75-1.84 $\mu$m and 2.22-2.26 $\mu$m  for $I_{cont}$. 

On the other hand, spectral slopes are measured by fitting the shape of the spectrum between the wavelengths $\lambda_i$ and $\lambda_j$ to a straight line and deriving the following quantity from the resulting fit parameters \citep{Cuzzi09, F12}:
\begin{equation}
Sl_{X}=\frac{I(\lambda_i)-I(\lambda_j)}{\Delta \lambda_{ij}\cdot I(\lambda_{i})}
\end{equation}
where $I(\lambda_i)$ and $I(\lambda_j)$ are the estimated signals at the endpoints of the range and $\Delta \lambda_{ij}=\lambda_i-\lambda_j$. For Saturn's rings, the two commonly used slopes are those measured over the ranges 0.35-0.55 $\mu$m and 0.55-0.95 $\mu$m \citep{F12}. These two slopes will be designated the ``blue slope'' ($Sl_B$) and the ``red slope'' ($Sl_R$), respectively.  Note that these names indicate the wavelength range where the slope is measured, they do not indicate the direction of the slope (in both these regions the ring increases in brightness with increasing wavelength). 

In principle, both band depths and spectral slopes can be computed from the  spectrograms illustrated in Figure~\ref{specspaceim}. However, such calculations do not provide a reliable estimate of the errors on these parameters. Most of the variations in the spectra (illustrated by shaded bands in Figure~\ref{rdhr8spec}) are due to variations in the overall brightness of the rings rather than uncertainties in the shape of the spectra at any given location. Hence, one can obtain more reliable estimates of the statistical uncertainties on these parameters by computing them for each pixel in the original cubes and then constructing profiles using the procedures described in Section 2.1 above. Note that these calculations only provide estimates of the statistical error bars on the spectral parameters, and do not account for systematic uncertainties in the instrument's spectral calibration. Fortunately, such calibration errors should affect all spectra equally, and so the computed statistical errors should provide a useful estimate of the uncertainties on any spatial variations observed in the data.

\begin{figure}
\centerline{\resizebox{5in}{!}{\includegraphics{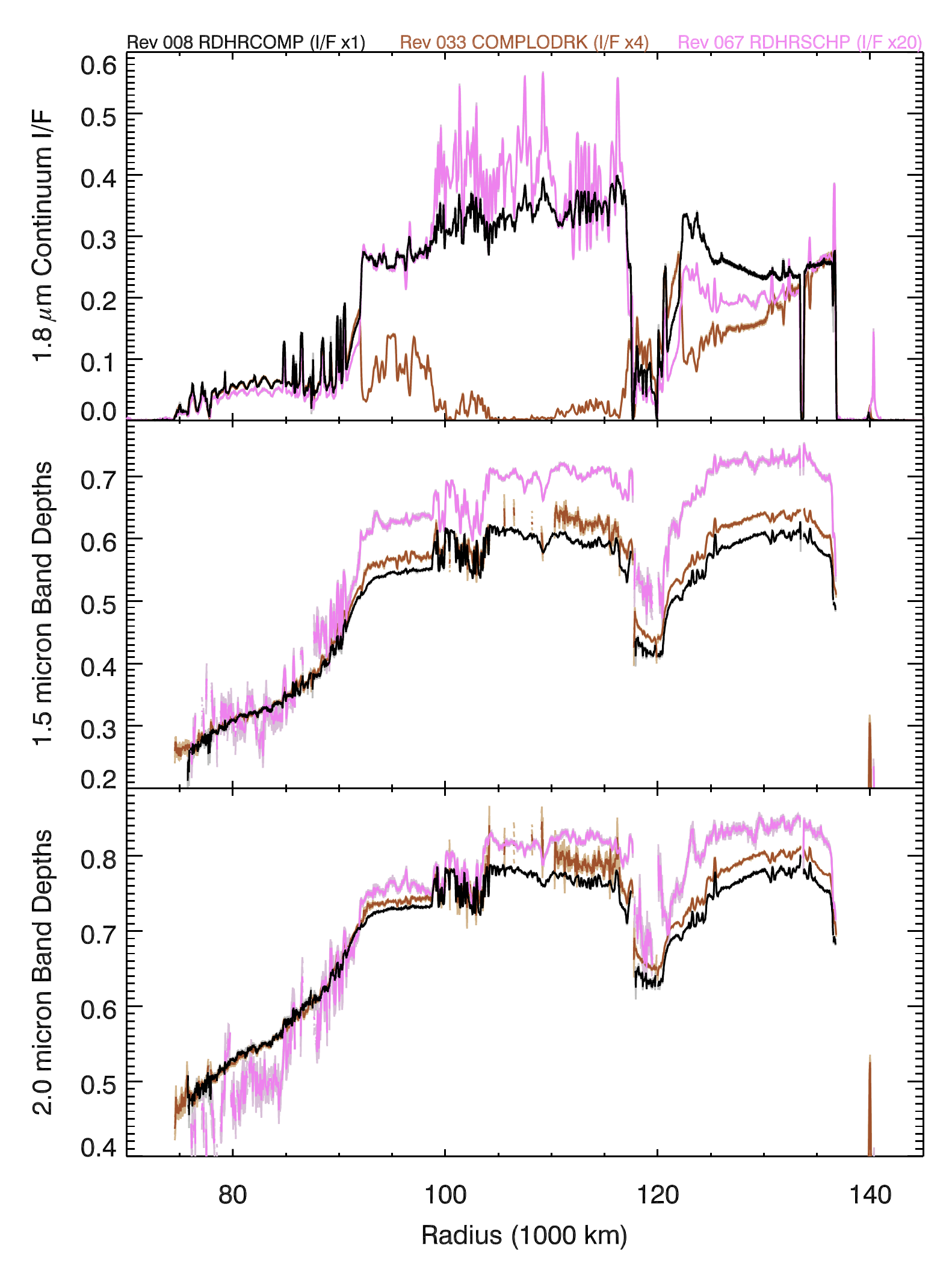}}}
\caption{Infrared spectral parameters derived from the Rev 008 RDHRCOMP, Rev 033 COMPLODRK and Rev 067 RDHRSCHP observations. The top panel shows the average continuum brightness level between 1.75 and 1.84 $\mu$m for all three observations (scaled by the factors given at the top of the plot), while the lower panels show the unscaled 1.5-micron and 2.0-micron band depths. Where visible, light shaded bands indicate the 1-$\sigma$ statistical error bars on these parameters at the relevant sampling scale. Wherever the shaded error bands are not visible, the error bars are less than the line thickness. Gaps in the COMPLODRK and RDHRSCHP profiles correspond to regions where the signal-to-noise was too low at the sampling scale to obtain sensible estimates of the relevant spectral parameters. Outside these gaps, the three observations show very similar trends in the band depths.}
\label{prof3band}
\end{figure}

\begin{figure}
\centerline{\resizebox{5in}{!}{\includegraphics{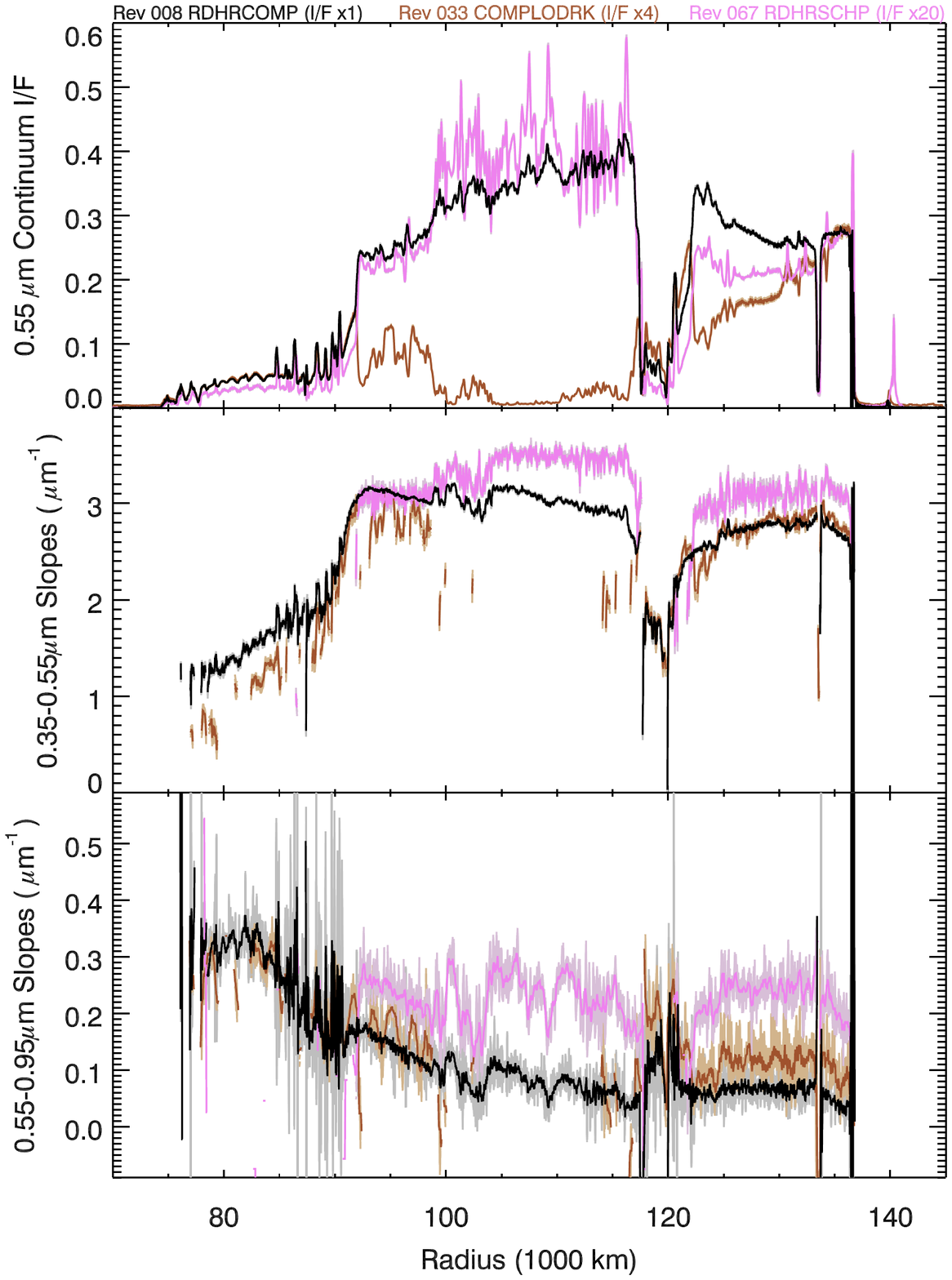}}}
\caption{Visible spectral parameters derived from the Rev 008 RDHRCOMP, Rev 033 COMPLODRK and Rev 067 RDHRSCHP observations. The top panel shows the continuum brightness level between 0.5 and 0.6 $\mu$m for all three observations (scaled by the factors given at the top of the plot), while the lower panels show the unscaled spectral slopes. Where visible, light shaded bands indicate the 1-$\sigma$ statistical error bars on these parameters at the sampling scale.  Wherever the shaded bands are not visible, the error bars are less than the line thickness. Gaps in the COMPLODRK and RDHRSCHP profiles correspond to regions where the signal-to-noise was too low at the sampling scale to obtain sensible estimates of the relevant spectral parameters. Outside these gaps, the three observations show very similar trends in the slopes.}
\label{prof3vis}
\end{figure}

Figures~\ref{prof3band} and~\ref{prof3vis} display the infrared band depths and visible spectral slopes derived from the three observations, along with the  average $I/F$ levels in representative regions outside of strong absorption bands (1.75-1.84 $\mu$m and 0.5-0.6 $\mu$m, respectively).
In comparing the data from these different observations, let us first consider the continuum brightness levels. The overall brightness of the rings varies dramatically among these three observations, but the shapes of the brightness profiles are not so different. All three brightness profiles have essentially the same shapes in the Cassini Division and the C ring, while the COMPLODRK data show reversed contrast for the B and inner A rings, with the rings becoming extremely dark in the B-ring core. This is consistent with expectations based on the analytical expressions for single-scattering by a vertically extended ring given by \citet{Chandra60}, which predict that brightness should decrease with optical depth on the unlit side of the rings when the normal optical depth exceeds a critical value:
\begin{equation}
\tau_{max}=\frac{\mu_0\mu}{\mu-\mu_0}\ln\left(\frac{\mu}{\mu_0}\right),
\end{equation}
where $\mu_0$ and $\mu$ are the cosines of the incidence and emission angles, respectively.
For the COMPLODRK data, $\tau_{max} \simeq 0.3$, so it is sensible that the C-ring and Cassini Division appear in normal contrast, while the B ring and A ring appear in reversed contrast (cf. Nicholson {\it et al.} 2008). 

In addition to the contrast reversal between the lit-side and unlit-side data, there are at least two other interesting differences among these profiles. In the B ring, the high-phase RDHRSCHP profile shows larger fractional fine-scale brightness variations than the low-phase RDHRCOMP data. This suggests that the phase function of the ring particles varies slightly across this ring. Meanwhile, in the A ring there are interesting differences in the expression of the ``core-halo'' complexes associated with the strong density waves \citep{Nicholson08}. These will be are discussed in more detail in Section 6.2 below. 

Turning to the infrared ice-band depths and visible spectral slopes, we should first note the differences in the quality of the three data sets. The RDHRCOMP data exhibits extremely small statistical error bars (of order 0.01 in the band depths and $0.03 \mu$m$^{-1}$ in the spectral slopes for each 20-km-wide radial bin) that are often less than the thickness of the lines in these plots. These small errors are consistent with the high signal levels obtained with this lit-side, low-phase viewing geometry. The noise in individual VIMS spectral channels is typically of order 1 data number in the uncalibrated data, and the typical signal in these observations is several hundreds of data numbers per spectral channel in each pixel. The statistical error bars per radial bin for the other two observations are generally larger, as is to be expected given the lower signal levels in these geometries (which is only partially compensated for by the use of longer exposure times).  Furthermore, there are several regions in the rings where the COMPLODRK or the RDHRSCHP observations do not have sufficient signal-to-noise to provide useful measurements of the spectral parameters at the scale of the individual 50-km-wide radial bins. For example, we cannot obtain sensible band depth estimates for much of the central B ring in the COMPLODRK data because the rings are too dark in this region when viewed from the unlit side. Similarly, parts of the C ring and Cassini Division are sufficiently faint that both the COMPLODRK and the RDHRSCHP data do not provide useful estimates of the band depths or spectral slopes at the level of individual radial bins (the error bars exceed 0.1 in band depth or 0.1$\mu$m$^{-1}$ in spectral slopes). The signal-to-noise on these band depths could be improved by averaging over larger radial regions, but since the focus of this analysis will be radial variations on both large and small scales, we have elected to simply not plot these data here.

If we restrict our attention to the data with acceptable signal-to-noise, then these three observations yield band-depth profiles and spectral slopes with remarkably similar shapes. Most obviously, the C ring and Cassini Division exhibit reduced infrared band depths and blue slopes in all three data sets. However, even fine-scale variations in the band depths repeat from observation to observation. For example, the complex variations in the band depth between 100,000 and 105,000 km in the B ring are similar  in both the low-phase RDHRCOMP and the high-phase RDHRSCHP data (most of the differences between the two data sets can be attributed to differences in the effective resolution of the observations), while the fine-scale variations within the A ring are seen in both the lit-side RDHRCOMP data and the unlit-side COMPLODRK data. The shapes of these curves are also very similar to recently published profiles of band depths and spectral slopes derived from other VIMS observations and Voyager color data \citep{Cuzzi09, F11, F12}.  

The only obvious  observation-geometry-dependent trend in these data is that the high-phase RDHRSCHP observation yields systematically deeper band-depths and steeper slopes than the lower-phase observations in the A and B rings. Similar trends with phase angle have also been observed by \citet{F11, F12}, and probably arise from differences in the fraction of the observed light that was scattered multiple times from different ring particles\footnote{Different proportions of single and multiple scattering among the regolith grains of the ring particles may also contribute to this effect}. The ring particles are believed to be highly back-scattering, so at low phase angles on the lit side of the rings most of the observed light should be singly scattered from individual ring particles. By contrast, the light observed at high phase angles or from the ring's unlit side is expected to include a larger fraction of light scattered multiple times between multiple ring particles. Such multiply-scattered light would exhibit much deeper ice bands than singly-scattered light  and thus would contribute to the observed differences in the band-depth profiles. However, the differences between the three observations are rather modest, which implies that only a small fraction of the light undergoes multiple scattering events before escaping the rings,  which is consistent with previous spectral and photometric analyses \citep{Dones93, Cuzzi09}. This lack of multiply-scattered light  is also  consistent with the ring particles being confined to a very flat layer, which greatly reduces the efficiency of such inter-particle scattering  

As a practical matter, the similarities in the shapes of the  band-depth profiles with viewing geometry means that the observed radial variations in the spectral parameters do not primarily represent variations in the rings' phase function, but instead mostly indicate changes in the ring-particles' wavelength-dependent albedo. Hence we can use the low-phase, lit-side data to infer ring properties like regolith structure and composition and not be too concerned that we are missing features visible in other viewing conditions. Thus for the rest of this analysis, we will focus largely on the extremely high signal-to-noise RDHRCOMP data.

\subsection{Quantifying spectral trends with brightness ratios }

\begin{figure}
\centerline{\resizebox{5in}{!}{\includegraphics{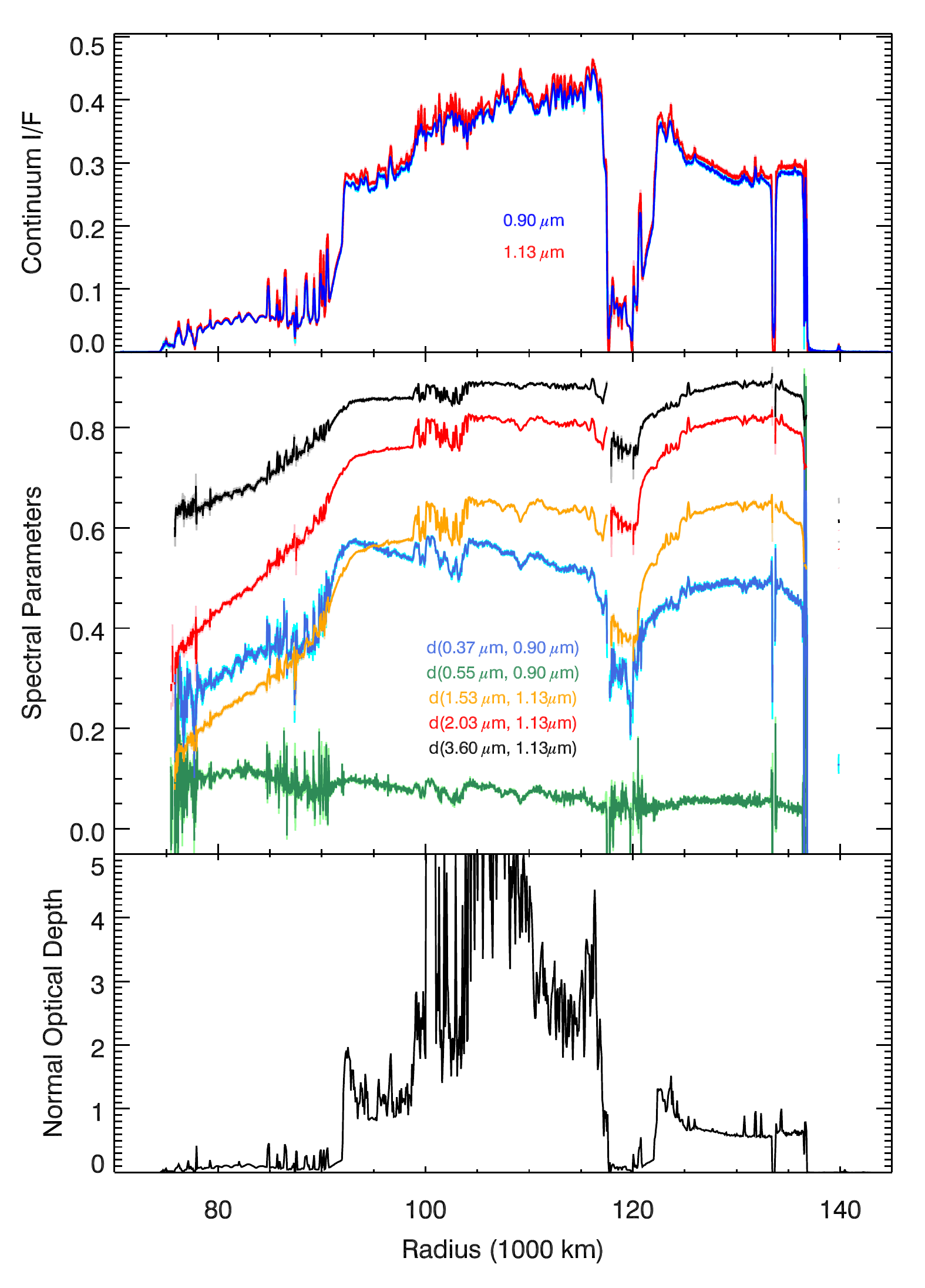}}}
\caption{Spectral parameters derived from the Rev 008 RDHRCOMP
observations, compared with the Rev 089 $\gamma$ Crucis occultation data.
The top panel shows the continuum brightness level in the VIS (blue) and IR (red) channels, and the middle panel shows five spectral parameters derived from spectral ratios. Where shown, light shaded regions indicate the 1-$\sigma$ statistical error bars on these parameters per 20-km-wide radial bin (elsewhere these error bars are less than the line thickness). The bottom panel shows, for comparison, the optical depth profile derived from the occultation data, smoothed to approximately the resolution of the spectral data. Close-up views of various ring regions can be found in Appendix B (Figures~\ref{rdhr8plotc}-~\ref{rdhr8plota}).}
\label{rdhr8plot}
\end{figure}

For our detailed analysis of the RDHRCOMP data, we will not use the band depths or spectral slopes computed above. Instead, we will employ proxies for these quantities consisting of simple ratios of the observed brightness at two different wavelengths:
\begin{equation}
r(\lambda_i, \lambda_j)=\frac{I(\lambda_i)}{I(\lambda_j)}
\end{equation}
where $I(\lambda_i)$ and $I(\lambda_j)$ are the measured brightness levels (i.e. $I/F$) at the two wavelengths  $\lambda_i$ and $\lambda_j$. These sorts of brightness ratios have a few advantages over band depths and spectral slopes. For one, brightness ratios can be more efficiently translated into information about the ring-particles' composition and texture (see Section 4). More immediately, however, brightness ratios are also more generic and therefore facilitate comparisons between multiple spectral parameters. Indeed, for a given brightness ratio, we can easily construct a quantity:
\begin{equation}
d(\lambda_i, \lambda_j)=1-r(\lambda_i, \lambda_j)
\end{equation}
which is analogous to either a band depth (if $\lambda_i$ is at a band center and $\lambda_j$ is at a continuum wavelength) or a spectral slope (if $\lambda_i$  and $\lambda_j$ are at opposite ends of the slope). Table~\ref{rpartab} lists all the brightness ratios used in this paper. Note that the ratios $r(1.53 \mu{\rm m}, 1.13\mu{\rm m})$ and $r(2.03 \mu{\rm m}, 1.13\mu{\rm m})$ quantify the strengths of the 1.5 and 2.0 micron ice bands, respectively. Hence the parameters  $d(1.53 \mu{\rm m}, 1.13\mu{\rm m})$ and $d(2.03 \mu{\rm m}, 1.13\mu{\rm m})$ can serve as proxies for the band depths $D_{1.5}$ and $D_{2.0}$.  Similarly, $r(0.37 \mu{\rm m}, 0.95\mu{\rm m})$ and $r(0.55 \mu{\rm m}, 0.95\mu{\rm m})$ provide measures of steepness of the visible spectral slopes, so $d(0.37 \mu{\rm m}, 0.95\mu{\rm m})$ and $d(0.55 \mu{\rm m}, 0.95\mu{\rm m})$ can stand in for $Sl_B$ and $Sl_R$, respectively\footnote{Technically, $Sl_B$ would be most closely related to the parameter  $d(0.37 \mu{\rm m}, 0.55\mu{\rm m})$. However, in practice there is little difference between $d(0.37 \mu{\rm m}, 0.95\mu{\rm m})$ and $d(0.37 \mu{\rm m}, 0.55\mu{\rm m})$, and the former parameter provides a more useful basis for quantifying the concentration of the ultraviolet absorber (see Section 4). Thus, for simplicity's sake, we will only consider  $d(0.37 \mu{\rm m}, 0.95\mu{\rm m})$ here.}. Indeed, the radial profiles of these four $d$-parameters, shown in Figure~\ref{rdhr8plot} have very similar shapes to those of their analogs in Figures~\ref{prof3band} and~\ref{prof3vis}. However, since all these $d$-parameters have the same units, we can compare them directly to one another and display them together in a single plot. Furthermore, spectral ratios and $d$-parameters can quantify spectral parameters that cannot be  easily represented as band depths or spectral slopes. For example,  the parameter, $d(3.60 \mu{\rm m}, 1.13\mu{\rm m})$,   quantifies the depths of the 3.0 and 4.5 micron ice bands. The similar shapes of the $d(1.53 \mu{\rm m}, 1.13\mu{\rm m})$, $d(2.03 \mu{\rm m}, 1.13\mu{\rm m})$ and $d(3.60 \mu{\rm m}, 1.13\mu{\rm m})$ profiles in Figure~\ref{rdhr8plot} indicate  that the strengths of the 1.5, 2.0, 3.0 and 4.5 micron water-ice absorption bands exhibit similar trends (The 1.25 micron band also shows these trends, but we do not plot this parameter simply to avoid cluttering the graph).

\begin{table}
\caption{Relationships between spectral parameters used in this study}
\label{rpartab}
\begin{tabular}{|c|c|c|c|}\hline
Brightness ratio & Spectral Slope/Band Depth & Illustrated in & Used to derive \\
&  most closely related to $r$ & Figure~\ref{rdhr8plot}? & model parameters$^a$ \\ \hline
$r(0.37 \mu{\rm m}, 0.95\mu{\rm m})$ & $Sl_B$ & * & $r_v \rightarrow f_{UV}\kappa_{UV}$ \\
$r(0.55 \mu{\rm m}, 0.95\mu{\rm m})$ & $Sl_R$ & * &  \\ \hline
$r(1.53 \mu{\rm m}, 1.13\mu{\rm m})$ & $D_{1.5}$ & * & $r_1 \rightarrow f_{C}\kappa_{C}, S_{\rm eff}$ \\
$r(1.79 \mu{\rm m}, 1.13\mu{\rm m})$ & IR continuum slope  &  & $r_2 \rightarrow f_{C}\kappa_{C}, S_{\rm eff}$ \\
$r(2.03 \mu{\rm m}, 1.13\mu{\rm m})$ & $D_{2.0}$ & * & $r_3 \rightarrow f_{C}\kappa_{C}, S_{\rm eff}$ \\$r(2.24\mu{\rm m}, 1.13\mu{\rm m})$ & IR continuum slope & & $r_4 \rightarrow f_{C}\kappa_{C}, S_{\rm eff}$ \\$r(3.60 \mu{\rm m}, 1.13\mu{\rm m})$ & $D_{3.0}, D_{4.5}$ & * &  \\ \hline
\end{tabular}

$^a$ See Section 4.4 

\end{table}

As with the band-depths and the spectral slopes, brightness ratios could be computed from spectrograms, but we instead compute these parameters from the original pixel data in order to obtain statistical error bars and correlation coefficients (again, any systematic uncertainties in the spectral calibration should have little affect on the spatial variations in these parameters). For the RDHRCOMP data, the statistical error bar on a given brightness ratio in a typical 20-km-wide radial bin is less than 1\%, and the average correlation coefficient between any two ratios with a common reference wavelength (e.g. $r(1.53 \mu{\rm m}, 1.13\mu{\rm m})$  and $r(2.03 \mu{\rm m}, 1.13\mu{\rm m})$) is roughly 0.50, as expected (see Table~\ref{rcorrmat} and Section 4.3 below). 

\subsection{Summary of notable spectral variations}
 
Before attempting to translate the observed spectral parameters into constraints on the ring-particles' texture and composition, we first wish to highlight some of the notable variations seen in the rings' spectral properties. Here we focus on the trends seen in the brightness ratios shown in Figure~\ref{rdhr8plot} (or the higher-resolution Figures~\ref{rdhr8plotc}-\ref{rdhr8plota} given in Appendix B), although many of the same trends can also be seen in the band depths and spectral slopes plotted in Figures~\ref{prof3band} and~\ref{prof3vis}.

All the spectral parameters that quantify the depth  of the water ice absorption bands ($d(1.53 \mu{\rm m}, 1.13\mu{\rm m})$,  $d(2.03 \mu{\rm m}, 1.13\mu{\rm m})$, $d(3.60 \mu{\rm m}, 1.13\mu{\rm m})$) indicate that the C ring and Cassini Division have weaker ice bands than the A and B rings. Within both the A and B rings, there is abundant fine-scale variation in the depth of all three bands. In the A ring, there are localized features in these profiles surrounding the  strong Lindblad resonances at 125,400 km, 130,800 km, 132,300 km and 134,300 km. Each of these features is composed of a ``core'' of enhanced band strengths, surrounded by a wider depression in the band depths \citep{Nicholson08}. Meanwhile, in the inner B ring between 98,000 km and 105,000 km, the water-ice band depths shift back in forth in sync with variations in the rings' optical depth. 

Interestingly, the small-scale variations in the ``blue slope'' ($d(0.37 \mu{\rm m}, 0.95\mu{\rm m})$) within the A and B rings are very tightly correlated with those in the ice-band depths, consistent with previous VIMS measurements  \citep{Nicholson08}. This supports the idea that the coloring agent responsible for the blue slope is well mixed with the ice. These data are also consistent with previously published profiles derived from low-phase Voyager data, specifically the Voyager $G/UV$ color ratio curve in \citet{EC96} and ~\citet{Estrada03} (differences between the two curves most likely arise because the Voyager color ratio is basically $(1-d(0.37 \mu{\rm m}, 0.55\mu{\rm m}))^{-1}$ so there is a somewhat non-linear relationship between the  two parameters).   However, closer comparisons of the relevant profiles reveal some interesting differences between the blue slope and the water-ice band-depths. For example, the C-ring plateaux are prominent in the blue slopes, consistent with the Voyager color ratios \citep{EC96, Estrada03}, but the variations are far more subtle in the IR band depths (see Figure~\ref{rdhr8plotc}), consistent with earlier VIMS observations \citep{Nicholson08}. On a broader scale, the blue slope seems to show larger swings than the 1.5 micron band strength across the B ring, indicating that the concentration of this coloring agent varies to some extent across this ring.

The ``red slope''  or $d(0.55 \mu{\rm m}, 0.95\mu{\rm m})$ exhibits very different trends from  all of the other curves in Figure~\ref{rdhr8plot} (cf. Filacchione {\it et al.} 2012). In particular, this curve shows a broad hump in the C ring and a weaker rise in the Cassini Division (which is easier to see in Figure~\ref{prof3vis}, and is also clearly visible in the higher-resolution SOI data presented in Nicholson {\it et al.} 2008). These results are consistent with prior interpretations of this spectral feature by \citet{CE98} and \citet{Nicholson08}, who argued that this slope was due to an extrinsic darkening agent like meteoritic or cometary debris that has polluted the lower-optical depth C ring and Cassini Division more than the denser A and B rings. However, properly interpreting this parameter is difficult because the observed slopes are weak, making them very sensitive to small brightness offsets or calibration errors. The interpretation of this slope is also complicated by other spectral features like the brightness peak around 0.6 microns in the A and B rings (see Figure~\ref{rdhr8spec}), which may reflect diffraction by sub-micron grains in the regolith \citep{Clark2012}. Due to these potential complications, we will not attempt to quantify the distribution of the contaminant responsible for the red slope using visible colors, but instead use brightness ratios between continuum wavelengths  in the infrared (see Section 4.4 below).

\section{Interpretation of the spectral parameters}

The rings' spectral properties can be influenced by both the  composition and the texture of the ring particles' regolith. Interpreting the spectral variations discussed in the previous section is therefore not straightforward. Separate constraints on regolith texture and composition can be derived by fitting the observed VIMS spectra to appropriate light-scattering models. For example, \citet{Clark2012}, \citet{F12} and \citet{C12} place constraints on the composition and texture of icy surfaces in the Saturn system by modeling the detailed shapes of selected high signal-to-noise spectra with Hapke's (1993) light-scattering theory. Here we will take a different, but complementary approach. Assuming a highly simplified model of the ring-particles' regolith, we invert a set of algebraic expressions involving a small number of brightness ratios and solve for the average scattering length in the ring-particles' regolith and the optical activity of the two main non-icy contaminants.  Such calculations can be performed much more rapidly than iterative least-squares fits, which facilitates the analysis of the thousands of spectra in the RDHRCOMP data set and the generation of high-resolution profiles of regolith properties.

Section 4.1 describes how we calculate model brightness ratios using formulae based on the \citet{Shkuratov99} light-scattering model, while Section 4.2 describes our simplified model for the light-scattering properties of the ring-particles' regolith. Section 4.3 discusses how such simplified models permit a small number of spectral parameters to be translated into estimates of regolith properties. This section also  illustrates some of the limitations of our particular models which will prevent us from determining regolith properties for parts of the inner C ring. Section 4.4  (and Appendix A) describes in detail the algebraic calculations and numerical procedures used to derive our estimates of regolith parameters. Finally, Section 4.5 discusses  several checks we have performed to validate these findings. Readers who are not particularly interested in knowing all the details of our calculations may wish to skip Sections 4.4 and 4.5 and proceed directly to Section 5, where the results of these calculations are presented.

\subsection{Modeling brightness ratios}

The measured brightness of the rings at a given wavelength depends on the albedo of the ring particles, the spatial distribution of these particles in the ring, and on the observation geometry (i.e. phase, incidence and emission angles). A complete spectrophotometric analysis would therefore require careful comparisons of spectral parameters and optical depths obtained from a variety of viewing geometries. However, as mentioned above, documenting how the ring's brightness varies with viewing geometry on scales smaller than 1000 km is difficult because there are relatively few  suitably high-quality high-resolution VIMS observations. Fortunately, by considering brightness ratios rather than absolute brightness measurements, we can obtain useful information about the ring particles' regolith texture and composition from a single high-resolution observation like RDHRCOMP.

The key feature of brightness ratios for this analysis is that they are insensitive to viewing geometry. In addition to comparing the three profiles shown in Figures~\ref{prof3band} and~\ref{prof3vis}, we have also conducted preliminary investigations of numerous lower-resolution VIMS observations. These studies indicate that the brightness ratios listed in Table~\ref{rpartab} are essentially independent of the observed incidence and emission angles, and vary by less than 10\% between phase angles of 20$^\circ$ and 100$^\circ$ (i.e. outside the opposition surge and the forward-scattering lobe of small regolith grains, see also Filacchione {\it et al.} 2011). \nocite{F11} This implies that most of the observed light at low phase angles was scattered by a single ring particle, and that the phase functions of individual ring particles are not strong functions of wavelength (see above). In this limit, the ratio of brightnesses at two different wavelengths will not depend directly on the rings' optical depth or on the particles' phase function. Thus we may make the simplifying assumption that for the low-phase RDHRCOMP data, {\it  the ratio of the rings'  brightness at two wavelengths $r$ is approximately equal to the ratio of the ring-particles' average albedo at those two wavelengths.} 
\begin{equation}
r(\lambda_i, \lambda_j)=\frac{I(\lambda_i)}{I(\lambda_j)}
	\simeq \frac{A(\lambda_i)}{A(\lambda_j)}.
\label{rtoa}
\end{equation}
Note that since band depths and spectral slopes are also defined in terms of brightness ratios, the above assumption allows these parameters to be expressed in terms of albedo ratios as well. 

The insensitivity of the measured brightness ratios to observation geometry also means
that {\it the parameters $A$ in Equation~\ref{rtoa}  do not have to be the total amount of light scattered by an individual ring particle, but could instead represent other photometric quantities, including the total amount of light scattered by a small patch of a ring-particle's surface, or the total amount of light scattered by the rings as a whole.} All of these parameters are proportional to each other, with constants of proportionality that depend on the scattering properties of the relevant surfaces and the spatial distribution of the ring particles (see Table 1 of Cuzzi 1985 for a nice summary of these relationships).  \nocite{Cuzzi85} Provided the relevant scattering functions are not wavelength dependent (as seems to be the case here), these proportionality constants should cancel out. Hence the ratio $A(\lambda_i)/A(\lambda_j)$ can be equally well regarded as a ratio of planar regolith surface albedos, ring-particles' single-scattering albedos, or ring system albedos, and we could use expressions for any of these quantities to compute what these ratios should be under various circumstances.

Standard light-scattering models like those described in \citet{Hapke81, Hapke93}, \citet{CE98} and \citet{Shkuratov99}  provide analytical expressions for these various wavelength-dependent  albedos. Such albedos are typically expressed as functions of the product $\alpha S$, where  $S$ is the mean scattering length for the photons within the regolith (also known as the regolith's ``grain size''), and $\alpha$ is the absorption coefficient of the ring material, which is given by the expression $\alpha=4\pi\kappa/\lambda$, where $\kappa$ is the imaginary part of the material's  refractive index.  Hence, with the above approximation we can use these models in order to generate expressions for $r$ as a function of  $\alpha S$ at the two relevant wavelengths.  Of course, since the values of $r$ do vary somewhat with viewing geometry, the values of $\alpha S$ derived from these expressions will vary somewhat depending on the observation geometry. However,  the most obvious changes in the spectral parameter profiles obtained from different viewing geometries amount to offsets which are nearly constant with radius (see Figures~\ref{prof3band},~\ref{prof3vis} and \citet{F11}). Hence, even if the above assumption leads to biases in the estimated average values of $\alpha S$, any trends in these parameters across the rings should be much more robust. 

Cuzzi and Estrada (1998, using a Hapke-based formalism) and \citet{Shkuratov99} provide very different formulas for the albedo as a function of  $\alpha S$, and it is well known that the Hapke and Shkuratov scattering theories can yield different estimates of the composition and effective scattering lengths required to match a given spectrum \citep{Poulet02}.  In part, this is because \citet{CE98} compute the albedo of an individual ring particle, while \citet{Shkuratov99} compute the albedo for a one-dimensional model of a regolith surface (see below). For the particular simplified model of $\alpha S$ that will be used here (see Section 4.2), it turns out that the \citet{Shkuratov99} light-scattering model is more compatible with the observed spectra than the Hapke-model described in \citet{CE98}. Hence for this analysis the expected value of $r$ for a given pair of $\alpha S$ values  is computed using Equations 8-12 from \citet{Shkuratov99}. For the sake of simplicity we assume here that the real part of the grains' refractive index $n = 1.3$ (appropriate for ice-rich material) and the volume filling factor $q=1$ (the subsequent analysis is relatively insensitive to the assumed values of these parameters). 
In this case, the relevant formula can be simplified to:
\begin{equation} 
A_1=\frac{1+r_b^2-r_f^2}{2r_b}-\sqrt{\left(\frac{1+r_b^2-r_f^2}{2r_b}\right)^2-1},
\label{a1eq}
 \end{equation}
where  the parameters $r_b$ and $r_f$ are given by the expressions:
 \begin{equation}
 r_b=R_b+\frac{1}{2}(1-R_e)(1-R_i)R_ie^{-2\alpha S}/(1-R_ie^{-\alpha S})
 \end{equation}
 \begin{equation}
 r_f=(R_e-R_b)+(1-R_e)(1-R_i)e^{-\alpha S}+\frac{1}{2}(1-R_e)(1-R_i)R_ie^{-2\alpha S}/(1-R_ie^{-\alpha S})
 \end{equation}
and the coefficients $R_i$, $R_e$ and $R_b$ are set by our choice of $n$:
\begin{equation}
R_i \simeq 1.04-1/n^2\simeq 0.45
\end{equation}
\begin{equation}
R_e \simeq \frac{(n-1)^2}{(n+1)^2}+0.05 \simeq 0.067
\end{equation}
\begin{equation}
R_b \simeq (0.28n-0.20)R_e \simeq 0.011
\end{equation} 
It is important to note that the albedo parameter derived by \citet{Shkuratov99}  (here denoted $A_1$) is actually the albedo of a one-dimensional model system that corresponds to the ``brightness coefficient'' of a regolith surface viewed at low phase angles \citep{Shkuratov99}. While this is compatible with our assumption that $r(\lambda_i, \lambda_j)=A_1(\lambda_i)/A_1(\lambda_j)$ (at least for the RDHRCOMP data), it also means that the value of $A_1$ at any given wavelength should not be mistaken for  the Bond or single-scattering albedo of an individual ring particle. 

\subsection{Modeling regolith properties}

If we assume $r(\lambda_i, \lambda_j)=A_1(\lambda_i)/A_1(\lambda_j)$, then $r$ can easily be computed from the values of $\alpha S$ at the two relevant wavelengths. However, translating an observed set of $r$ into separate constraints on $\alpha$ and $S$ is not so trivial, and requires an explicit model describing how these parameters can vary with wavelength. 
 
In principle, the mean scattering length $S$ can vary with wavelength because small grains or defects in the regolith can more efficiently scatter photons with smaller wavelengths. However, for the purposes of this analysis we will assume that {\sl the structure of the ring-particles' regoliths can be modeled using a single, wavelength-independent ``effective scattering length'' $S_{\rm eff}$.}  This greatly simplifies the analysis because any variation in $\alpha S$ with wavelength must be due to  $\alpha$. However, this assumption is only likely to be a valid approximation over relatively restricted wavelength ranges. This limitation informs several aspects of the following analysis, and should be kept in mind when interpreting the results of these calculations. Nevertheless, $S_{\rm eff}$ provides a convenient way to parametrize the physical state of the ring-particles' regolith. 

As for $\alpha$, {\sl the effective optical constants of the ring material are computed using the linear mixing approximation and assuming that the dominant component of the ring material is water ice.}  More specifically, the real refractive index of the ring material is assumed to be 1.3 at all wavelengths between 0.35 and 2.3 microns, and the effective imaginary refractive index $\kappa_{\rm eff}(\lambda)$ is assumed to be  the volume-weighted average of three components:
\begin{equation}
\kappa_{\rm eff}(\lambda)=(1-f_C-f_{UV})\kappa_I(\lambda)+f_C\kappa_C(\lambda)
	+f_{UV}\kappa_{UV}(\lambda)
\end{equation}
where $\kappa_I$ is the refractive index of water ice, while $f_C$ and $f_{UV}$ are the concentrations of two different contaminants in the ring, one that absorbs at a broad range of wavelengths (with refractive index $\kappa_C$), and one that only absorbs at short visible wavelengths (with refractive index $\kappa_{UV}$). Assuming the rings are composed primarily of water ice is equivalent to assuming $f_C<<1$ and $f_{UV}<<1$, so that the prefactor in front of $\kappa_I$ can be approximated as unity. With these assumptions, the absorption coefficient can be written as the sum of three terms:
\begin{equation}
\alpha(\lambda)=\alpha_I(\lambda)+\alpha_C(\lambda)+\alpha_{UV}(\lambda)
\end{equation}
where $\alpha_I=4\pi\kappa_I/\lambda$, $\alpha_C=4\pi f_C\kappa_C/\lambda$ and $\alpha_{UV}=4\pi f_{UV}\kappa_{UV}/\lambda$. The value of $\alpha_I$ at any given wavelength can be computed directly from the values of imaginary refractive index tabulated in \citet{Mastrapa09}. However, the identities of the two contaminants are not yet certain, so we cannot say {\it a priori} how $\alpha_C$ or $\alpha_{UV}$ should vary with wavelength. 

In principle, data at many wavelengths could be used to obtain detailed information about the optical properties of the two contaminants. However, for the purposes of this preliminary analysis of the {\sl spatial} distribution of these contaminants, we will instead use a small number of wavelengths and assume that $\alpha_C$ and $\alpha_{UV}$ are simple functions of wavelength. Based of the relatively weak continuum slopes in the near infrared, the absorption length of the broad-band absorber appears to be a relatively smooth function of wavelength. After some experimentation with various functional forms, we found that the relevant near-infrared spectra for most parts of the ring were consistent with a quadratic model for $\alpha_C$:
\begin{equation}
\alpha_C=\alpha_0+\alpha_1\left(\frac{\lambda-\lambda_0}{\lambda_0}\right)+\alpha_2\left(\frac{\lambda-\lambda_0}{\lambda_0}\right)^2.
\label{acmod}
\end{equation}
where $\alpha_0, \alpha_1$ and $\alpha_2$ are wavelength-independent constants, and $\lambda_0$ is some reference wavelength (here assumed to be 1.13 $\mu$m, see below). No explicit form is proposed for $\alpha_{UV}$, instead we will assume that $\alpha_{UV}=0$ at wavelengths longer than about 0.6 microns. The concentration of UV absorber in the ring can thus be parameterized as simply the value of $\alpha_{UV}$ (or, equivalently $f_{UV}\kappa_{UV}$) at a suitably short wavelength (in practice, we use 0.37 $\mu$m below).

 \subsection{Methods for translating spectral ratios into estimates of regolith parameters}
 
 With the above simplifying assumptions, there are just five parameters that need to be estimated from each spectrum: the effective scattering length $S_{\rm eff}$, the parameters $\alpha_0$, $\alpha_1$ and $\alpha_2$ that determine the concentration and spectral properties of the broad-band absorber, and the value of $\alpha_{UV}$ at a selected wavelength. In principle, one could derive estimates of these parameters at each point in the ring using an iterative least-squares approach to fit the full spectrum at the relevant location. However, the finite number of iterations required to converge on the best-fit solution becomes problematic when there are of order a thousand spectra to analyze, as is the case for the RDHRCOMP profiles. Hence we have elected to instead take five suitably chosen brightness ratios $r$ from each spectrum, and solve for the above parameters by numerically inverting the relevant expressions for the $r$ as functions of $\alpha S$. This method is much less computationally expensive than a least-squares approach and enables us  to analyze a large number of spectra more efficiently. The calculations  used to derive all five parameters are presented in the following subsection. However, before discussing such computational details, it is useful to first consider some simpler situations in order to clarify  the utility and the limitations of this approach. 
 
\begin{figure}
\centerline{\resizebox{4in}{!}{\includegraphics{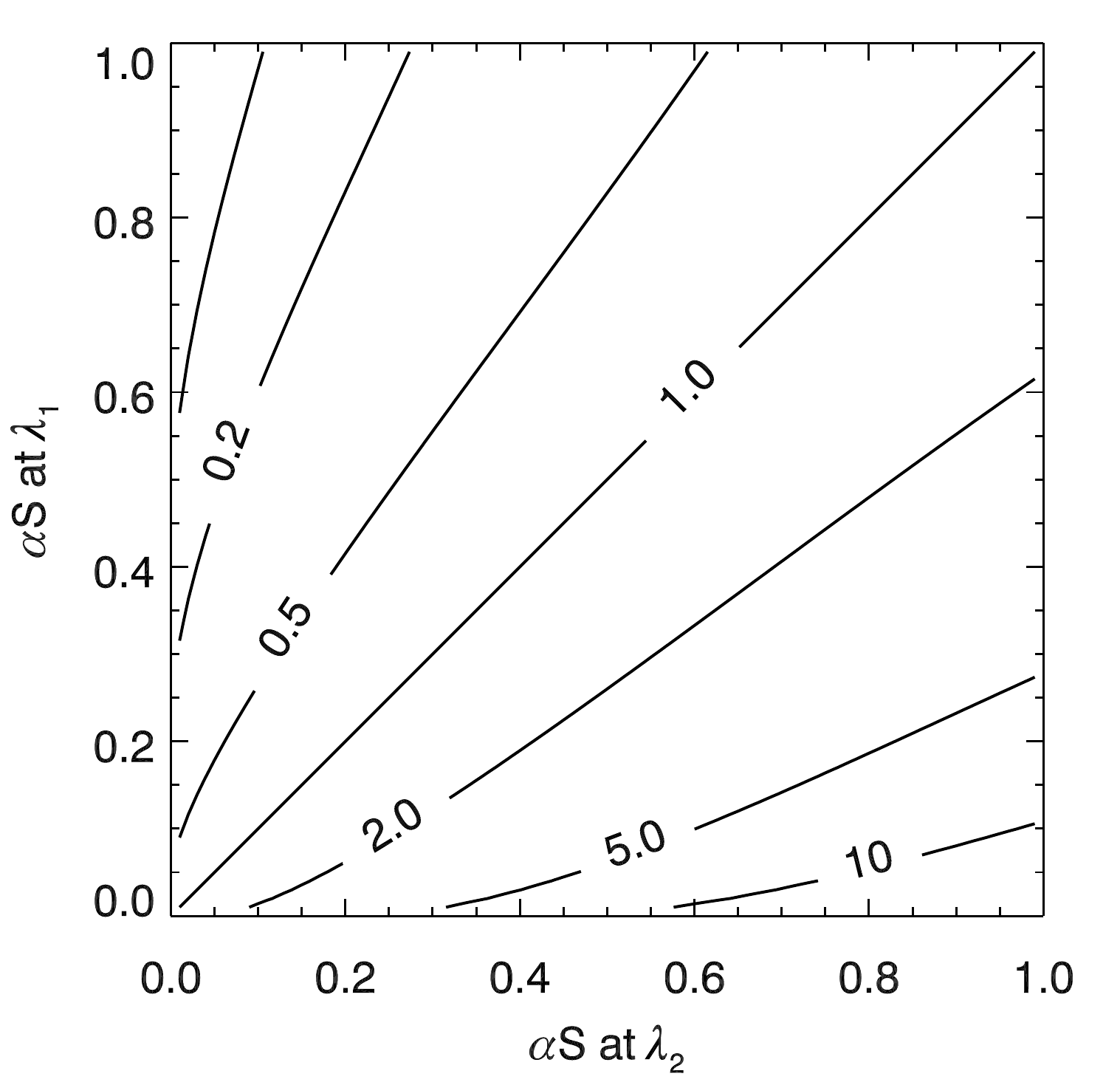}}}
\caption{ Contour plot of the brightness ratio $r(\lambda_1,\lambda_2)$ as a function of the product $\alpha S$ at the two wavelengths computed using the \citet{Shkuratov99} equations, assuming $q=1$ and $n=1.3$. A measured value of $r$ constrains the values of $\alpha S$ at the two wavelengths to fall along one of the curves in this plot.  }
\label{shkrplot}
\end{figure}

Let us begin by considering a very simple situation, where we know a given brightness ratio $r$ and we wish to use this information to constrain  the values of $\alpha S$ at the two relevant wavelengths. Unfortunately, while the \citet{Shkuratov99} expression for $A_1$ is invertible, the corresponding expression for $r$ cannot be easily rearranged to yield an analytical expression for the $\alpha S$ at one wavelength as a function of $r$ and $\alpha S$ at the other wavelength. Fortunately, calculating $r$ for a range of $\alpha S$ is not computationally expensive, so for a given value of $r$, it is straightforward to compute  the required value of $\alpha S$ at $\lambda_2$  as a function of  $\alpha S$ at $\lambda_1$ (see Figure~\ref{shkrplot} for examples of such curves).

By themselves, a set of $r$ values just constrains the values of the product $\alpha S$ at multiple wavelengths, and does not provide any information on the values of $\alpha$ or $S$ at any given wavelength. However, if we assume an explicit model for how $\alpha$ and $S$ should vary with wavelength (like the one described in the previous section), then it becomes possible to simultaneously derive estimates of the unknown parameters, even ones that might at first appear to be largely degenerate. For example, consider the parameters $\alpha_C$ and $S_{\rm eff}$. Since both these parameters affect the depth of the water-ice absorption bands in the near-infrared, it is not immediately obvious whether these parameters can be separately constrained by the spectral data. However, it turns that the degeneracies between these two model parameters do not pose a serious problem if we employ data from both of the moderately strong absorption bands at 1.5 and 2.0 microns. 

\begin{figure}[tbp]
\centerline{\resizebox{3.6in}{!}{\includegraphics{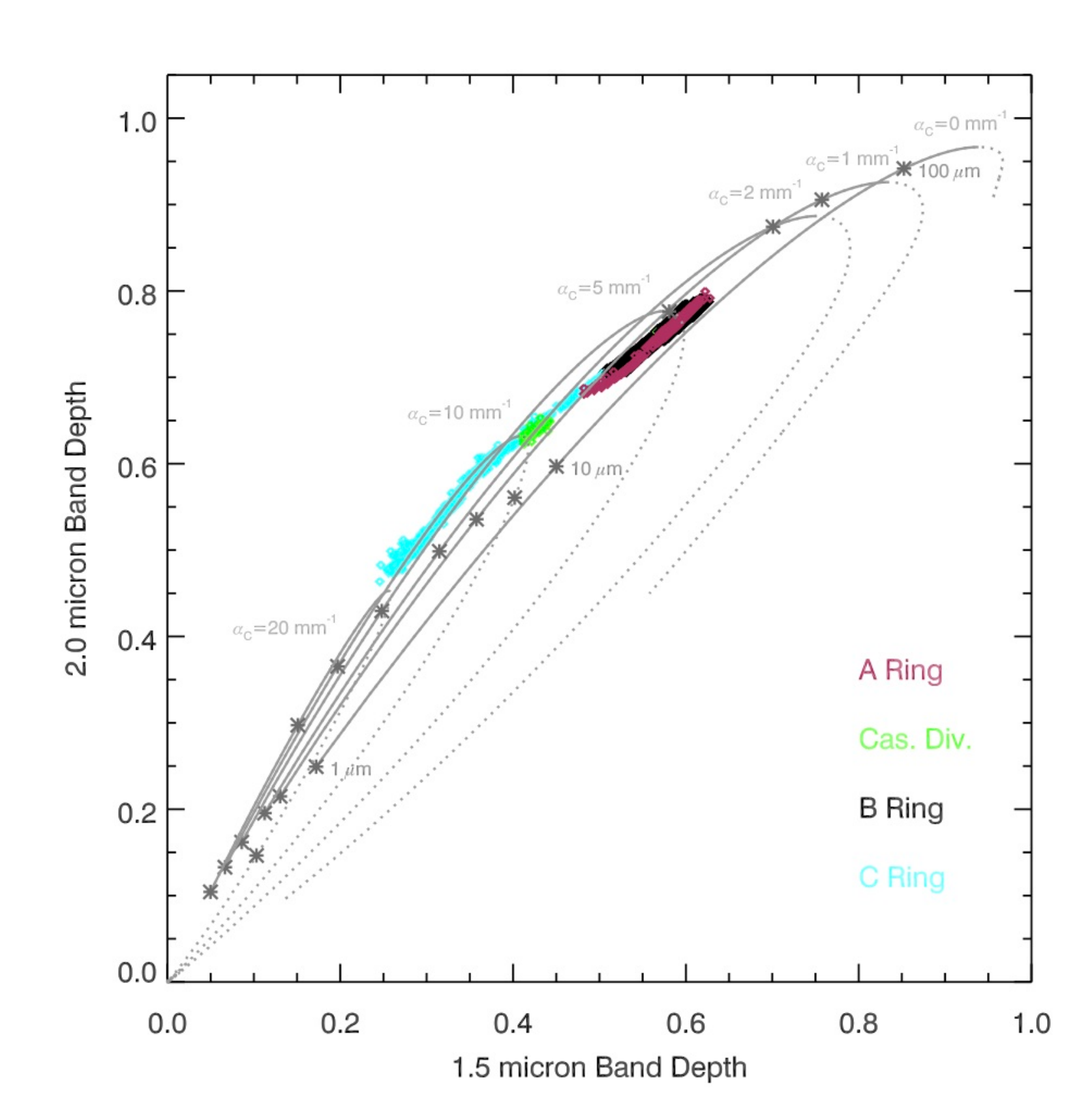}}}
\centerline{\resizebox{3.6in}{!}{\includegraphics{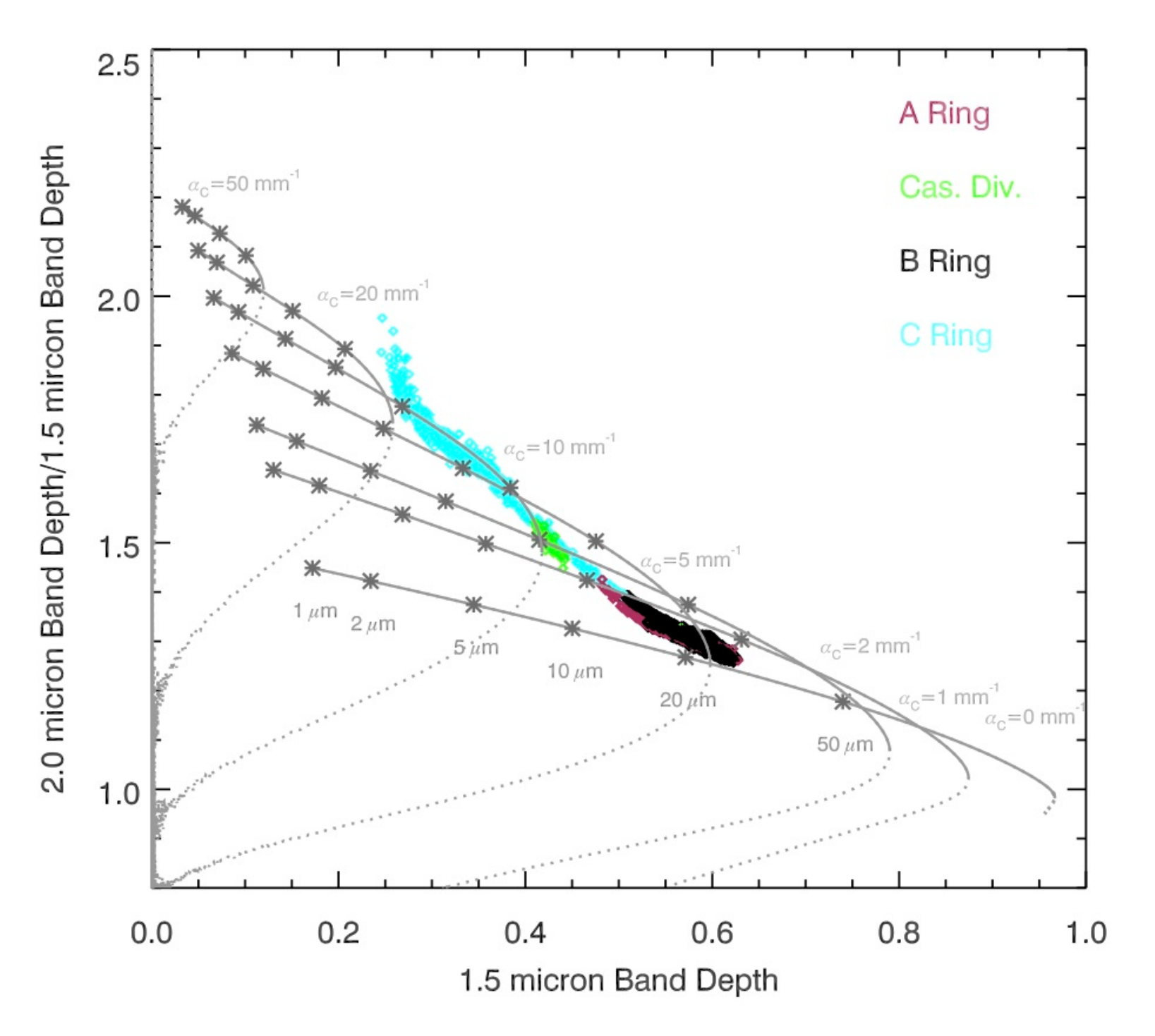}}}
\caption{Plots illustrating how the 1.5 and 2.0 micron band depths vary with $\alpha_c$ and $S_{\rm eff}.$ The upper plot shows  the 2-micron band-depth as a function of the 1.5-micron band depth, while the lower plot shows the ratio of the two band depths as a function of the 1.5 micron band depth in order to make the differences between the various model curves easier to see. In both plots, colored points mark the measured values in the rings derived from the Rev 008 RDHRCOMP observation. The grey  curves are predictions based on a Shkuratov model assuming $\alpha=\alpha_I(\lambda)+\alpha_C$, where $\alpha_C$ is assumed to be a constant. The dark grey stars correspond to particular values of $S_{\rm eff}$, while the lighter grey curves correspond to fixed values of $\alpha_C$. For purposes of clarity, each curve is divided into a solid portion (where the band depths increase with increasing  $S_{\rm eff}$) and a dotted portion (where the band depths decrease with increasing  $S_{\rm eff}$).}
\label{modbandcomp}
\end{figure}

Figure~\ref{modbandcomp} shows how the 1.5 and 2.0 micron band depths predicted by the \citet{Shkuratov99} light-scattering model vary with $\alpha_C$ and $S_{\rm eff}$. $\alpha_C$ is assumed to be independent of wavelength here for the sake of simplicity, and band depths are used instead of brightness ratios because they yield a clearer plot. If  $\alpha_C$ is held fixed while $S_{\rm eff}$ varies, one obtains a curve in band-depth space in which the band depths first become deeper with increasing scattering length, but then eventually fall back to zero as the core (and eventually the wings) of the bands saturate \citep{CR84, Clark2012, F12}. Different values of $\alpha_C$ yield slightly different curves in this parameter space, with higher values of $\alpha_C$ turning around at smaller band depths and yielding higher $D_{2.0}/D_{1.5}$ ratios. 

Let us first consider the solid parts of these theoretical curves,  where the band-depths increase with with increasing $S_{\rm eff}$. These curves cover a small but non-zero part of the possible space of band-depths. Throughout most of this region, there is a one-to-one mapping between the band depths and the parameters $\alpha_C$ and $S_{\rm eff}$. For example, $D_{1.5}\simeq0.35$ and $ D_{2.0}\simeq0.50$ (i.e.  $D_{2.0}/D_{1.5}\simeq1.5$) corresponds to a model with $\alpha_C \sim 1 {\rm mm} ^{-1}$ and $S_{\rm eff} \sim 10 \mu$m. However, this solution is only unique if we only consider the solid parts of the curves. If we also include the dotted parts of the curves (where the bands are saturated and the band-depths decrease with increasing $S_{\rm eff}$), a second solution with $\alpha_C \sim 15 {\rm mm}^{-1}$ and $S_{\rm eff} \sim 50 \mu$m can equally well reproduce the observed band depths. Fortunately, these two solutions predict very different values for  $A_1$ at continuum wavelengths (0.7 and 0.05, respectively). Hence we can distinguish between these two solutions based on the known albedos of the ring material (which correlates with the brightness of the ring at low phase angles $A_1$). The ring particles are known to have relatively high albedos, ranging from 0.2 in the C ring to near unity in the A and B rings  \citep{Doyle89, Cooke91, Dones93, CE98, Porco05, Deau07, Morishima10},  so for this analysis we can safely reject the higher-$\alpha_C$ (i.e. lower albedo) solution and therefore obtain a unique estimate of $\alpha_C$ and $S_{\rm eff}$ whenever the observed band depths fall in the region covered by the model curves. 

Figure~\ref{modbandcomp} also includes colored dots indicating the observed band depths in various parts of the rings. The band-depths derived from the A ring, B ring and the Cassini Division all fall entirely within the parameter space occupied by the model spectra, so for all these regions there is a unique model that matches the observed band depths.  The values of  $\alpha_C$ and $S_{\rm eff}$ for each of these locations can therefore be derived directly from the band depths. Furthermore, the  statistical uncertainties on these parameters can be estimated by mapping the errors on the measured band-depths into $\alpha_C$-$S_{\rm eff}$ space. 

\begin{figure}
\centerline{\resizebox{6in}{!}{\includegraphics{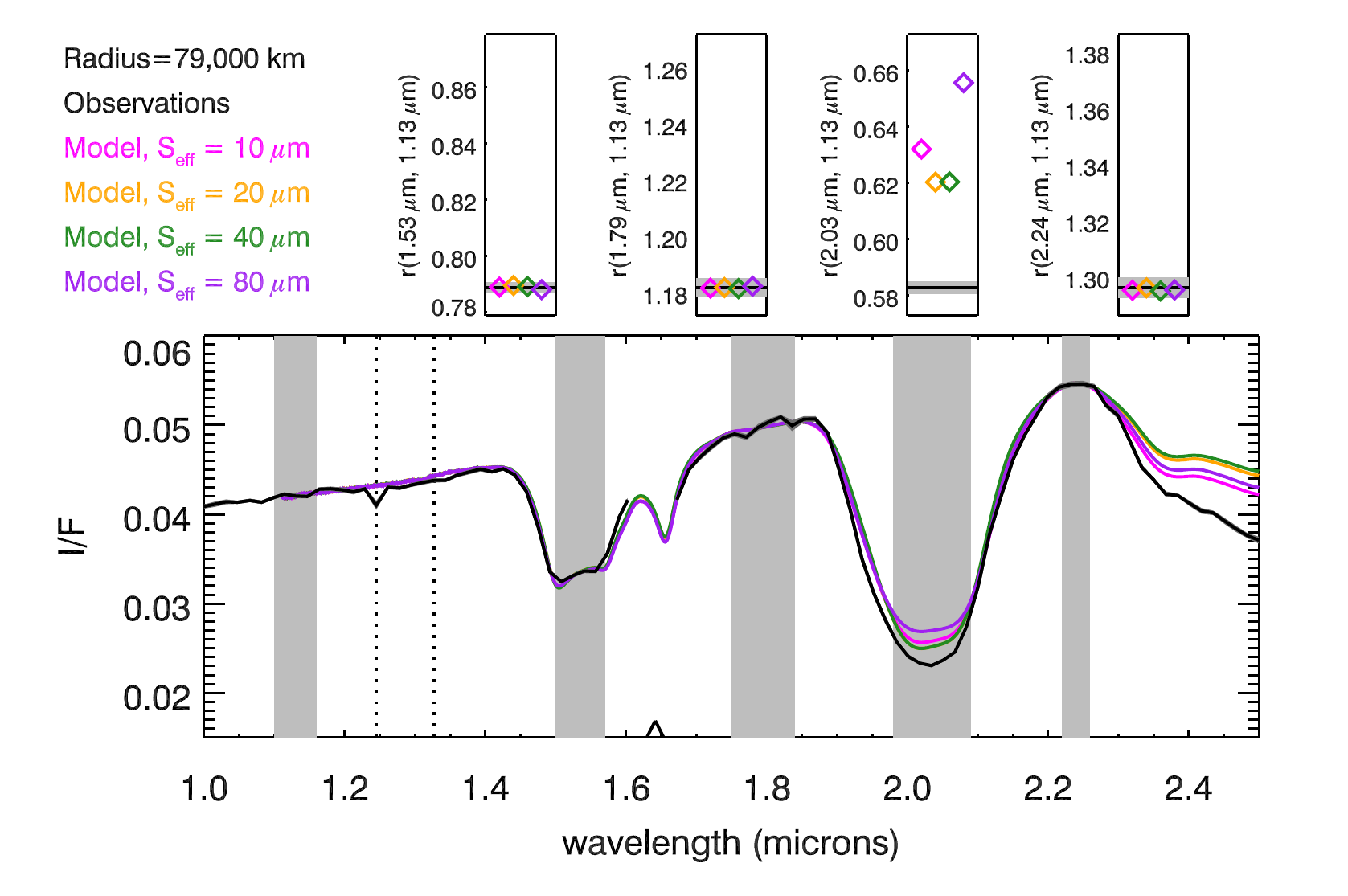}}}
\caption{Illustration of the issues involved in fitting inner C-ring spectra with the simple model used in this analysis. The large panel shows a representative spectrum from the inner C ring as the black curve (the width of the band indicates the 1-$\sigma$ statistical error bars on the data), along with four model spectra which assume different values of $S_{\rm eff}$.  Dotted vertical lines indicate hot pixels and triangles mark the location of filter gaps on the VIMS detector array. The vertical gray shaded bands indicate the wavelength ranges used to  compute four spectral ratios, which are illustrated above in four separate panels. In each of the above panels, the horizontal band is the observed value and 1-$\sigma$ uncertainty on the ratio, while the colored points are the values derived from the different models. Each of the model spectra uses a quadratic model for $\alpha_C$ to match the continuum level and the depth of the 1.5-micron band, but all of these models underestimate the depth of the 2-micron band. There is consequently no combination of $\alpha_C$ and $S_{\rm eff}$ that can adequately reproduce the observed spectrum.}
\label{testcspec}
\end{figure}

The data from the C ring are more problematic. Many of the C-ring band depth measurements
(corresponding to data from the inner half of that ring) skirt the edge of the region spanned by the model spectra. Since lines of constant $S_{\rm eft}$ and $\alpha_C$ run nearly parallel to each other in this part of parameter space, the derived error bars on these regolith parameters will have relatively large error bars (see below). Worse, a number of data points from the innermost C ring  fall entirely outside the region covered by the model spectra. This means that  none of our model spectra perfectly matches the estimated depths of both the 1.5-micron and 2.0-micron ice bands.  Figure~\ref{testcspec} illustrates the problem by comparing an observed inner C-ring spectrum with four model spectra. Each model spectrum assumes a different value of $S_{\rm eff}$, has been scaled to match the ring's average brightness around 1.13 microns, and uses a quadratic model for $\alpha_C$ that has been tuned to match the spectral slopes outside the strong ice bands and the depth of the 1.5 micron water-ice band. All these models clearly under-estimate the depth of the 2.0 micron band  (for comparison, see Figure~\ref{test4spec} below for four cases where a model spectrum can reproduce the depths of both ice bands). This figure also compares the observed and modeled values for four brightness ratios. Again, the models can reproduce three of these ratios, but significantly over-predict the ratio $r(2.03 \mu{\rm m}, 1.13 \mu{\rm m})$, which quantifies the depth of the 2-micron ice band.  The inability of any of the model spectra to reproduce the depth of both ice bands implies that the simple model for $\alpha_C$ and $S_{\rm eff}$ described in Section 4.2 above is  inadequate for describing spectra from the innermost C ring. 

In principle, a more complex model for $\alpha S$ would be able to adequately reproduce  the observed band depths for these spectra.  For example, recent work by \citet{Clark2012} indicates that the presence of small sub-micron grains in the regolith could alter the shapes and relative depths of the 1.5 and 2.0-micron ice bands and yield spectra that better match the C-ring observations.  However, including additional free parameters like the fraction of small grains in the regolith would complicate the analysis. Furthermore, initial examinations of more complex models failed to substantially change the trends observed in parameters like $S_{\rm eff}$ found elsewhere in the rings. Hence we will continue to use the simple model for $\alpha S$ for this preliminary investigation of spatial variations in the rings, keeping in mind that there are aspects of the rings' spectra that we are not yet able to model.  Since this model cannot reproduce the observed spectra in the inner C ring,  we cannot solve for $\alpha_C$ and $S_{\rm eff}$ at these locations. In principle, we could determine a combination of $\alpha_C$ and $S_{\rm eff}$ that best fits the observed data, but Figure~\ref{testcspec} indicates that this solution will still be a poor fit to the observed data. Furthermore, the  parameters become highly uncertain and the uncertainties are almost completely degenerate in these situations, so $\alpha_C$ and $S_{\rm eff}$ are very poorly constrained.  Hence for this analysis, whenever we encounter a spectrum which exhibits a combination of brightness ratios that cannot be reproduced by any spectral model, we do not present any estimate of the regolith parameters at all, and thus leave a gap in the relevant profiles. These gaps, which are fortunately limited to the inner C ring, provide a clear signal of where the assumptions behind this analysis are breaking down, rather than indicating an absence of data.


\subsection{Details of the calculations used to determine regolith parameters}

The previous section provided a simple example of how two regolith parameters (a constant $\alpha_C$ and $S_{\rm eff}$) can be derived from two spectral parameters ($D_{1.5}$ and $D_{2.0}$). Determining all five parameters in the regolith model is a more complex task, but it still amounts to solving a series of equations to find a unique combination of regolith parameters that reproduces a particular set of 5 brightness ratios. 
In practice, we first derive estimates for the values and statistical uncertainties of $\alpha_0$, $\alpha_1,$ $\alpha_2$ and $S_{\rm eff}$ using four brightness ratios in the near-infrared between 1.0 and 2.5 microns. Then we use another brightness ratio at visible wavelengths (together with the above estimates of $S_{\rm eff}$ and $\alpha_C$), to estimate $\alpha_{UV}$. We also compute statistical error bars on these quantities by mapping the errors on the brightness ratios into regolith-parameter space, taking into account the appropriate covariances. Readers not interested in the details of these computations may wish to skip this subsection.

In order to estimate all four parameters $\alpha_0, \alpha_1, \alpha_2$ and the effective grain size $S_{\rm eff}$, we need four brightness ratios $r_i$, where $i=1,2,3,4$. Each of these ratios is defined to be the average observed ring brightness over a wavelength range centered on $\lambda_i$ divided by the average brightness in a reference wavelength range centered on $\lambda_0$, so $r_i=r(\lambda_i, \lambda_0)$. In principle, we could use many different combinations of five wavelength bands $\lambda_0, \lambda_i$ for this sort of analysis. In practice, we chose $\lambda_0 =1.10-1.16 \mu$m, $\lambda_1=1.50-1.57\mu$m, $\lambda_2=1.75-1.84 \mu$m, $\lambda_3=1.98-2.09\mu$m and $\lambda_4=2.22-2.26 \mu$m,  (in each case, the brightness is the average $I/F$ over the indicated range). Thus $\lambda_0$, $\lambda_2$ and $\lambda_4$ are continuum wavelengths, while $\lambda_1$ and $\lambda_3$ are in the middle of the moderately strong water ice absorption bands at 1.5 and 2.0 microns. Indeed, $r_1$ and $r_3$ are the ratios used to quantify the strength of the water-ice bands in Figure~\ref{rdhr8plot} (see also Table~\ref{rpartab}). According to \citet{Mastrapa09}, the average imaginary refractive indices of water ice at these five wavelengths are: $\kappa_I(\lambda_0)<0.000001$, $\kappa_I(\lambda_1)=0.00056$, $\kappa_I(\lambda_2)=0.000041$, $\kappa_I(\lambda_3)=0.0017$, and $\kappa_I(\lambda_4)=0.000086$.  The range of wavelengths was kept relatively small because of the above-mentioned concerns that the effective scattering length could vary with wavelength. We also did not want to use measurements beyond 2.5 microns in order to avoid regions around the very strong 3-micron absorption band where the assumptions that the real refractive index of water ice is around 1.3 and the imaginary refractive index is small would break down.  The above choice of wavelengths also includes a broad range of ice absorption coefficients, which helps ensure this analysis returns unique solutions for the relevant parameters. In particular, including data from two moderately strong ice bands allows us to evade the worst of the degeneracies between $S_{\rm eff}$ and $\alpha_0$ (see Section 4.3 above).

Using Equation~\ref{a1eq}, each of the ratios $r_i$ can be converted into a curve specifying the value of $\alpha S$ at  $\lambda_i$ as a function of $\alpha S$ at $\lambda_0$ (cf. Figure~\ref{shkrplot}). We can then seek a value of $\alpha S$ at $\lambda_0$ where all the $\alpha S$ values are consistent with the regolith model described in Section 4.2.  If this condition can be satisfied,  then we can solve the appropriate equations to obtain estimates for the relevant parameters. The detailed algebra behind these calculations is presented in Appendix A. For most of the Rev 008 RDHRCOMP spectra, these calculations yield two possible solutions for the regolith parameters.  These two solutions are equivalent to the two  different possible solutions for $\alpha_C$ and $S_{\rm eff}$ discussed in the previous subsection. As in that case,  we consistently select the solution with the lower $\alpha S$ as the correct solution, because the other solution would correspond to a spectrum with large concentrations of contaminants and nearly saturated ice bands in the A and B rings, which is inconsistent with the high albedo of the ring material. Interior to about 79,500 km in the C ring, there is no solution that yields a spectrum consistent with our assumed model for $\alpha S$. These spectra are like the one shown in Figure~\ref{testcspec}, and no further attempt is made to determine the regolith parameters at these locations.

\begin{table}
\caption{Average correlation coefficients between the spectral ratios used to determine $\alpha_0, \alpha_1, \alpha_2$ and $S_{\rm eff}$ (average values for the entire ring system).}
\label{rcorrmat}
\resizebox{6in}{!}{\begin{tabular}{|c|cccc|} \hline
 & $r(1.53 \mu m,1.13 \mu m)=r_1$ &  $r(1.79 \mu m,1.13 \mu m)=r_2$ & $r(2.03 \mu m,1.13 \mu m)=r_3$ &  $r(2.24 \mu m,1.13 \mu m)=r_4$\\ \hline 
$r(1.53 \mu m,1.13 \mu m)=r_1$  & 1.00 & 0.62 & 0.57 & 0.52 \\
$r(1.79 \mu m,1.13 \mu m)=r_2$  & 0.62 & 1.00 & 0.52 & 0.71  \\
$r(2.03 \mu m,1.13 \mu m)=r_3$  & 0.57 & 0.52 & 1.00 & 0.51 \\
$r(2.24 \mu m,1.13 \mu m)=r_4$  & 0.52 & 0.71 & 0.51 & 1.00  \\ \hline
\end{tabular}}
\end{table}

Statistical error bars on the parameters $\alpha_0, \alpha_1, \alpha_2$ and $S_{\rm eff}$ are derived from the statistical error bars on the four observed brightness ratios $r_i$. Recall that each value of $r_i$ is an average value derived from $\sim$10 pixels in the RDHRCOMP observation data set. The standard error on the mean for those pixels furnishes an error bar on each of the ratios, which are generally below 1\%. However, since all four ratios use the same reference wavelength $\lambda_0$, the errors on these parameters are correlated. We therefore also compute a covariance matrix for all four ratios at each pixel. Unfortunately, with only a handful of pixels in each bin, individual estimates of the correlation coefficients are very noisy, so we do not use the raw covariance matrix for each pixel.  Instead, we compute the average correlation coefficients for each pair of brightness ratios, which are given in Table~\ref{rcorrmat}. These average coefficients are mostly around 0.5, which is consistent with what one should expect for ratios with a common denominator. The covariances of any two brightness ratios for a given bin can therefore be estimated as the products of the appropriate two standard errors and the matching average correlation coefficient. The resulting covariance matrix of the ratios is designated $\mathcal{C}_r$.
The covariance matrix for the regolith parameters $\mathcal{C}_p$ is then derived from $\mathcal{C}_r$ using the standard linear error propagation:
\begin{equation}
\mathcal{C}_p={\mathcal J}{\mathcal C}_r{\mathcal J}^T
\label{cmat}
\end{equation}
where $\mathcal{J}^{-1}$ is the jacobian matrix whose elements are the partial derivatives of the ratios $r_i$ as functions of the parameters $\alpha_0, \alpha_1, \alpha_2$ or $S_{\rm eff}$, evaluated at the observed value of $r_i$. Equation~\ref{cmat}  yields accurate estimates of the statistical uncertainties in the regolith parameters provided the errors on the spectral ratios are sufficiently small, as is the case for the RDHRCOMP data (see below). Still, it is important to remember that ${\mathcal C}_p$ can only provide statistical errors on the regolith parameters assuming the instrument's calibration and our model for $\alpha S$ is correct.

After using the above methods to determine both $\alpha_C$ and $S_{\rm eff}$ over most of the rings, we can finally turn our attention to estimating $\alpha_{UV}$. Constraining this parameter  requires one additional brightness ratio $r_v=r(\lambda_b,\lambda_r)$, where  $\lambda_b=0.35-0.40 \mu$m and  $\lambda_r=0.85-0.95 \mu$m (i.e. the ratio that was used to quantify the blue slope in Figure~\ref{rdhr8plot}). Note that both brightness measurements come from  the VIS channel to avoid any issues involved in the differing spatial resolutions between the VIS and IR channels.  Since $\alpha_I$ is negligible at visible wavelengths and $\alpha_{UV}$ is assumed to be zero beyond 0.6 microns, the value of $\alpha S$ at $\lambda_r$ can be computed  from the previously-derived estimates  for $\alpha_C(\lambda_r)$ and $S_{\rm eff}$\footnote{Note that both the $r_v$ data and the $\alpha_CS_{\rm eff}$ estimates are smoothed to a common resolution prior to doing this computation}. We can therefore use the \citet{Shkuratov99} equations to determine what the value of $\alpha S$ at $\lambda_b$  must be to produce the observed brightness ratio $r_v$. With this value of $\alpha S$, and assuming the same model for $\alpha_C$ and $S_{\rm eff}$, we can then calculate $\alpha_{UV}(\lambda_b)$. Furthermore, we can use standard error-propagation to translate the statistical errors on $r_v$ into statistical error bars on $\alpha_{UV}$.   Of course, this requires extrapolating our quadratic model for $\alpha_C$ to shorter wavelengths, which introduces some additional uncertainty in the estimate. We therefore solved for $\alpha_{UV}(\lambda_b)$ using both the full quadratic model and assuming that $\alpha_C=\alpha_0$ for all visible wavelengths in order to demonstrate that the trends we observed were not very sensitive to the model for $\alpha_C$. This calculation  requires assuming the regolith has the same effective scattering length at 1.0-2.5 microns and 0.4 microns, which is also questionable. However, the lack of obvious correlations between the derived  $S_{\rm eff}$ and $\alpha_{UV}$ profiles (see below) suggests that changes in the effective scattering length are not producing spurious variations in $\alpha_{UV}$.

\subsection{Testing the regolith parameter calculations}

\begin{table}
\caption{Regolith parameters and uncertainties for four locations in the rings.}
\label{ptab}
\centerline{\begin{tabular}{| c | r | r | rrrr |}\hline
 Parameter & Value & Error & \multicolumn{4}{c|}{Correlation Coefficients} \\
 & & & $S_{\rm eff}$ & $\alpha_0$ & $\alpha_1$ & $\alpha_2$ \\ \hline
\multicolumn{7}{|c|}{Radius=130,000 km (A ring)} \\ \hline
$S_{\rm eff}$ & 30.66 $\mu$m & 0.33 $\mu$m & 1.00 & 0.96 & 0.62 & -0.36 \\
$\alpha_0$ & +0.233 mm$^{-1}$ & 0.018 mm$^{-1}$& 0.96 & 1.00 & 0.62 & -0.38 \\ 
$\alpha_1$ & +0.123 mm$^{-1}$& 0.031 mm$^{-1}$& 0.62 & 0.62 & 1.00 & -0.91 \\
$\alpha_2$ & +0.004 mm$^{-1}$& 0.029 mm$^{-1}$& -0.36 & -0.38 & -0.91 & 1.00 \\ 
$\alpha_{UV}$ & 2.366 mm$^{-1}$& 0.014 mm$^{-1}$& &&& \\ \hline
\multicolumn{7}{|c|}{Radius=119,200 km (Cas. Div.)} \\ \hline
$S_{\rm eff}$ & 18.61 $\mu$m & 1.02$\mu$m & 1.00 & 0.99 & 0.02 & -0.09 \\
$\alpha_0$ & +2.249 mm$^{-1}$& 0.205 mm$^{-1}$& 0.99 & 1.00 & 0.05 & -0.13 \\
$\alpha_1$ & -1.524 mm$^{-1}$& 0.048 mm$^{-1}$& 0.02 & 0.05 & 1.00 & -0.96 \\
$\alpha_2$ & +0.627 mm$^{-1}$& 0.050 mm$^{-1}$& -0.09 & -0.13 & -0.96 & 1.00 \\ 
$\alpha_{UV}$ & 3.312 mm$^{-1}$& 0.124 mm$^{-1}$& &&& \\ \hline
 \multicolumn{7}{|c|}{Radius=102,000 km (B ring)} \\ \hline
$S_{\rm eff}$ & 26.65 $\mu$m & 0.39 $\mu$m & 1.00 & 0.95 & 0.53 & -0.30 \\
$\alpha_0$ & +0.304 mm$^{-1}$& 0.025 mm$^{-1}$& 0.95 & 1.00 & 0.53 & -0.33 \\ 
$\alpha_1$ & +0.005 mm$^{-1}$& 0.038 mm$^{-1}$& 0.53 & 0.53 & 1.00 & -0.92 \\
$\alpha_2$ & +0.023 mm$^{-1}$& 0.039 mm$^{-1}$& -0.30 & -0.33 & -0.92 & 1.00 \\ 
$\alpha_{UV}$ & 3.488 mm$^{-1}$& 0.035 mm$^{-1}$& &&& \\\hline
\multicolumn{7}{|c|}{Radius=89,000 km (Outer C ring)} \\ \hline
$S_{\rm eff}$ & 21.18 $\mu$m & 0.69 $\mu$m & 1.00 & 0.99 & 0.12 & -0.32 \\
$\alpha_0$ & +3.265 mm$^{-1}$& 0.189 mm$^{-1}$& 0.99 & 1.00 & 0.11 & -0.34 \\ 
$\alpha_1$ & -1.488 mm$^{-1}$& 0.053 mm$^{-1}$& 0.12 & 0.11 & 1.00 & -0.91 \\
$\alpha_2$ & +0.355 mm$^{-1}$& 0.062 mm$^{-1}$& -0.32 & -0.34 & -0.91 & 1.00 \\ 
$\alpha_{UV}$ & 4.083 mm$^{-1}$& 0.242 mm$^{-1}$& &&& \\ \hline
\end{tabular}}
\small
All error bars are $1-\sigma$ and the values of $\alpha_{UV}$ assume $\alpha_C=\alpha_0$ for all visible wavelengths

 \end{table}

\begin{figure}
\centerline{\resizebox{5in}{!}{\includegraphics{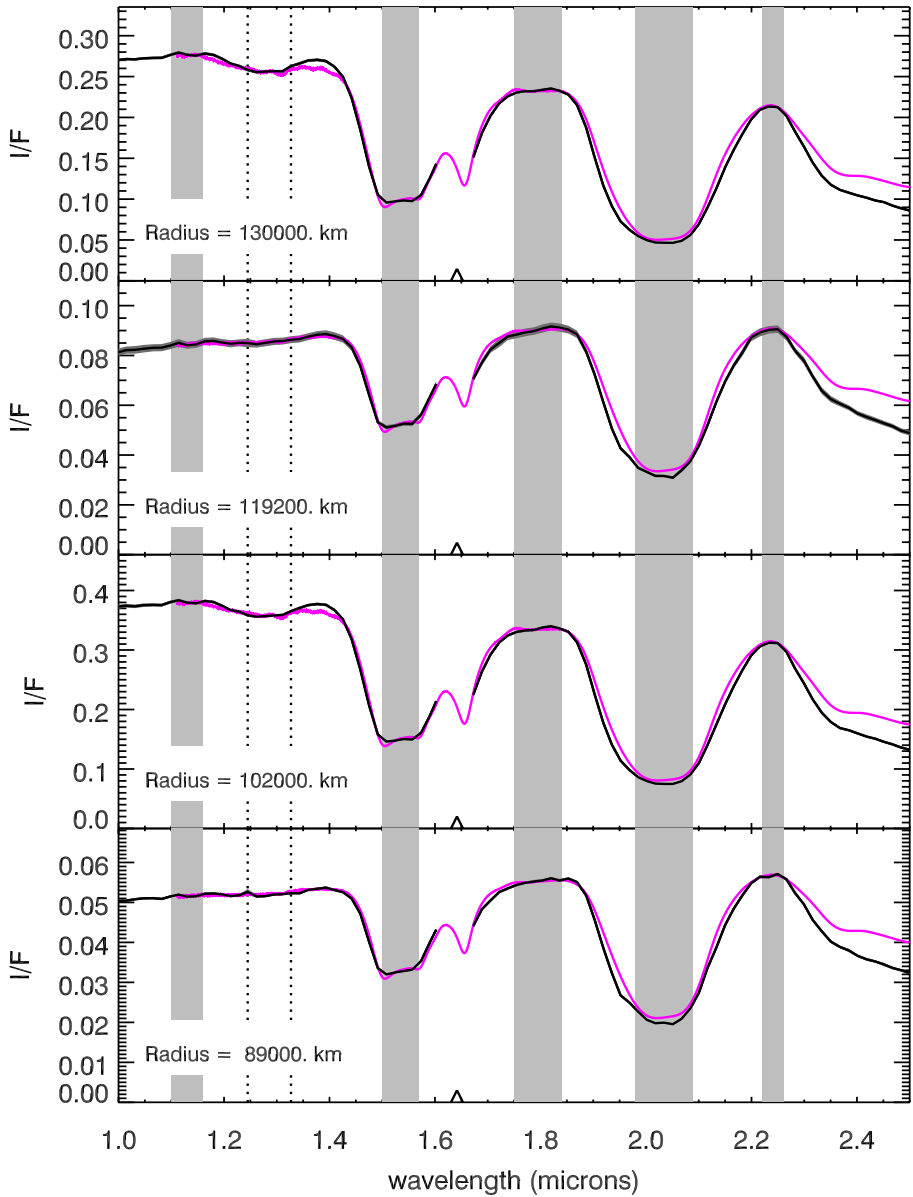}}}
\caption{Comparisons of the selected observed spectra (in black, with a barely-visible dark grey band indicating the $1-\sigma$ statistical error bars) with the Shkuratov model spectra  (in magenta).  Dotted vertical lines indicate hot pixels and triangles mark the location of filter gaps on the VIMS detector array. Each model spectrum is computed assuming the regolith parameters given in Table~\ref{ptab}, and is scaled to match the average brightness of the observed spectrum in the wavelength range indicated by the leftmost vertical grey band. The other vertical grey bands indicate the four wavelength ranges used to derive the spectral ratios employed for this analysis. Note that despite using only the four brightness ratios, the model spectra match the shape of the observed spectra quite well shortward of 2.25 microns. The most obvious differences occur in the shape of the 2 micron band, which is not modeled in detail here.}
\label{test4spec}
\end{figure}

\begin{figure}
\centerline{\resizebox{5in}{!}{\includegraphics{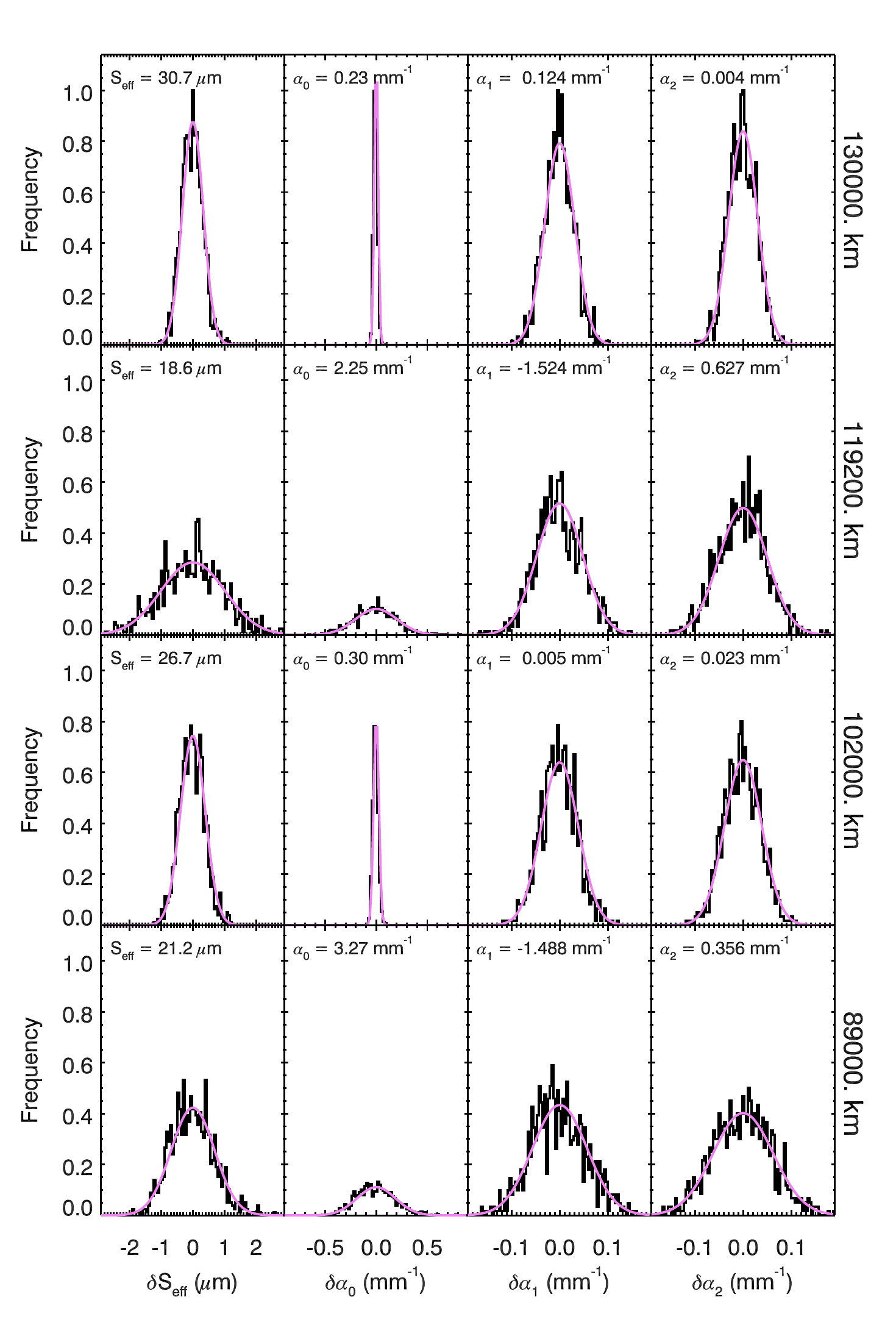}}}
\caption{Estimates of the statistical uncertainties in the regolith parameters. Each row of plots corresponds to a location in the rings (indicated at the right), whose average spectrum is shown in Figure~\ref{test4spec}. The four panels in each row show two different estimates of the probability distribution functions for $S_{\rm eff}$, $\alpha_0$, $\alpha_1$ and $\alpha_2$. In each panel, the black histograms are the result of Monte Carlo simulations, while the magenta curves are the derived using linear error propagation (Equation~\ref{cmat}).  The match between the two sets of distributions is excellent.}
\label{test4err}
\end{figure}

 The above  calculations of $\alpha_C$ and $S_{\rm eff}$ involve some rather opaque algebraic manipulations, so we performed several checks in order to determine whether the estimates and errors on the parameters derived with these methods were computed properly (The derivation of $\alpha_{UV}$ was a much simpler calculation and thus does not require such checks). In particular, we looked at four representative spectra from the four major ring regions, and verified that we have correctly identified the model that most closely matches the observed spectral ratios, and that the statistical error bars on the parameters for this regolith model are appropriate.

Table~\ref{ptab} provides the estimates, statistical uncertainties and correlation coefficients for the regolith parameters derived from representative spectra in the A ring, Cassini Division, B ring and outer C ring  (the C-ring spectrum was chosen to be from a region where the above procedures yielded a viable solution). Figure~\ref{test4spec} illustrates the observed  spectrum at each of these locations, along with the model spectrum computed using the derived parameters. Note that the observed spectra have a range of different continuum slopes and band depths, but in all cases the model spectra match the observed spectra very well between 1 and 2.25 microns. This demonstrates that we have correctly identified a set of regolith parameters that reproduces each of the observed sets of brightness ratios. The model curves deviate from the theoretical curves beyond 2.3 microns, probably because the assumptions behind our model (such as $n=1.3$ and $\kappa<<n-1$) are beginning to break down. The model spectra also fail to reproduce the detailed shape of the 2 micron ice band, but given the relative simplicity of our model, we do not regard this as a major shortcoming of this analysis. 

 Turning to the error bars and correlation coefficients, we may note that the correlation coefficients among the various parameters are not  small, and in particular the correlation coefficient between $S_{\rm eff}$ and $\alpha_0$ can approach 1. These correlations are accounted for in the calculation of the error bars on all the quantities, but one might be concerned that linear error propagation would underestimate the statistical uncertainties on these parameters. To check this possibility, we performed monte-carlo simulations, generating a series of one thousand replicates for each set of brightness ratios  for the four representative spectra. These replicates have a probability distribution determined by the appropriate covariance matrix for the given $r_i$.  Figure~\ref{test4err} shows the distributions of the regolith parameters derived from these replicates, compared with the distributions predicted by the errors in Table~\ref{ptab}. The match between these two distributions is excellent, so our estimates of the statistical uncertainties in the regolith parameters are reliable. However, one should not forget that these statistical certainties assume that our underlying model for $\alpha S$ in Section 4.2 is correct. There are also additional systematic uncertainties that are dependent on our choice of light-scattering and regolith models, which are more difficult to quantify.

\section{Results and general discussion}

\begin{figure}
\centerline{\resizebox{4.5in}{!}{\includegraphics{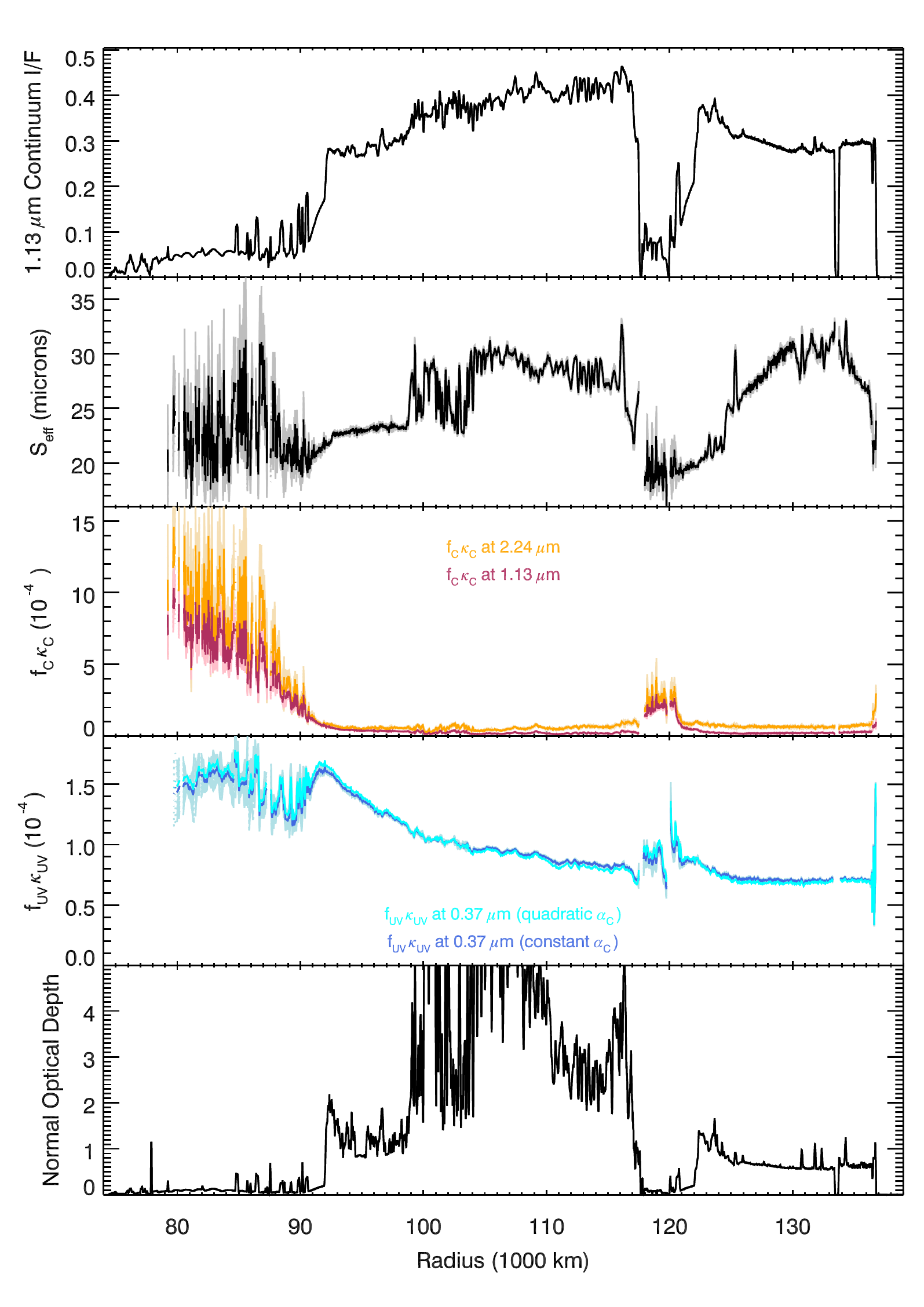}}}
\caption{Profiles of regolith properties across the rings derived from the low-phase, lit-face RDHRCOMP observations. The top panel shows the continuum brightness of the ring, while the second panel gives the estimated effective scattering length. The third panel contains profiles of the imaginary refractive index at two continuum wavelengths in the near-infrared, while the  fourth panel gives profiles of the imaginary part of the refractive index at 0.37$\mu$m that can be attributed to the UV absorber, assuming the parameters given in the panels above. Note that gaps in the derived parameters correspond to locations where our simplified model for the regolith properties cannot reproduce the observed spectral properties of the ring. Where visible, shaded bands indicate the 1$\sigma$ statistical uncertainties in these parameters at the 20-km  sampling scale (elsewhere, these uncertainties are less than the thickness of the lines).  The bottom panel shows an optical depth profile derived from the $\gamma$ Crucis occultation, which has been smoothed to approximately the same resolution of the spectral data. See Figures~\ref{profovc}-\ref{profova} in Appendix B for close up views of various rings.}
\label{profov}
\end{figure}

Figure~\ref{profov} shows profiles of the various ring-particle regolith properties derived using the above methods as a function of ring radius (Figures~\ref{profovc}-\ref{profova} in Appendix B show detailed views of the A, B and C rings). Specifically, we provide profiles of the effective scattering length $S_{\rm eff}$, the product $f_C\kappa_C$ at two different wavelengths in the infrared (derived from $\alpha_C$), and the product $f_{UV}\kappa_{UV}$ at 0.37 microns (derived from $\alpha_{UV}$).\footnote{See Appendix C for a profile of the photometric parameter $A_1$ derived from these quantities} Again, gaps in these profiles indicate locations where no model was able to reproduce the observed band depths. The statistical uncertainties on these parameters are indicated with shaded bands. Recall that these error bars are computed assuming that the current spectral calibration and our model for $\alpha S$ is correct, and thus do not represent all the systematic uncertainties in these parameters. Nevertheless, these bands should indicate whether a given feature in the profiles corresponds to a statistically significant variation in the rings' spectral parameters.

The three parameters show quite different trends. The scattering length shows abundant fine-scale structure in the A and B rings, indicating that this parameter is responsible for most of the band-depth variation within these dense rings. Meanwhile, $f_C\kappa_C$ appears to be strongly elevated in the C ring and Cassini Division, which explains the comparatively weak ice bands and red continuum slopes of these regions. Finally, $f_{UV}\kappa_{UV}$ appears to be relatively uniform across the outer part of the ring system, indicating that this contaminant is well-mixed in the ice \citep{Nicholson08}. These distinctive behaviors (discussed in more detail in the following sections) demonstrate  that our relatively simple treatment of the spectral data can yield useful information about the structure and composition of the ring particles' regolith and how these parameters vary across the rings. The values for these parameters are also reasonably consistent with those derived in previous work. Our estimates of $S_{\rm eff}$ are roughly comparable to those obtained from recent studies of Saturn's rings in the near-infrared \citep{Nicholson08, Cuzzi09, F12}. Also, our estimates of $f_C\kappa_C$ are fairly consistent with previous estimates of the effective $\kappa$ of the ring material as a whole at continuum wavelengths by \citet{CE98}, who estimated that $\kappa \simeq 0.8*10^{-4}$ in the inner B ring (96,500 km) and $\kappa \simeq 4.8*10^{-4}$ in the outer C ring (84,500 km) at long visible wavelengths. This gives us some confidence that these parameters are tracing the desired ring properties.

Despite these encouraging findings, we still must caution the reader against over-interpreting the parameters derived from this analysis. For one, even though the statistical uncertainties in the regolith parameters are relatively small, the absolute values of these parameters can have substantial systematic uncertainties. For example, it is well known that  Hapke and Shkuratov light-scattering theories can yield different estimates of the composition and effective scattering lengths for a given spectrum \citep{Poulet02}. Furthermore, the grains in the ring-particles' regolith probably have a range of sizes, and it is not obvious how the effective  scattering length relates to the moments of the full grain-size distribution. These issues probably explain why this analysis yields effective scattering lengths of 20-40 $\mu$m, which is shorter than other analyses of similar near-infrared spectral data  \citep{Poulet03, Cuzzi09, F12}, and longer than recent studies based on ultraviolet data  \citep{Bradley10}. Thus, for the rest of this study we will not examine the absolute value of the observed scattering lengths, but instead focus on trends and fractional variations across the rings, which should be less sensitive to the details of our simplified regolith model. Similarly we will use appropriate caution in our interpretations of the absolute values of the compositional parameters $f_C\kappa_C$ and $f_{UV}\kappa_{UV}$.

More generally, the results of this analysis depend upon the various simplifying assumptions used above. Thus, it is possible that a more sophisticated spectrophotometric analysis will alter the interpretation of the trends and variations identified here. For example, the variations in the spectra that are here attributed to differences in the mean effective scattering length or composition could instead arise from more complex variations in the regolith structure of the ring particles. Recent studies have indicated that the presence of sub-micron-sized grains can significantly affect the relative strength of the water-ice absorption bands \citep{Clark2012}, so the patterns interpreted here as variations in the mean scattering length may instead reflect changes in the amount of dust adhering to the ring particles. Similarly, some of the compositional trends could be due to radial variations in the distribution of the various contaminants within the particle regolith, which cause  the contaminants to be more or less intimately mixed at different locations. For example, very small grains of iron-rich compounds could potentially act  as either an ultraviolet absorber or a broad-band absorber, depending on its concentration in the regolith \citep{Clark2012}.

Changes in the ring-particles' spatial distribution could also potentially masquerade as variations in regolith parameters if the rings' phase and scattering functions are not completely independent of wavelength. While the insensitivity of the observed spectral ratios to viewing geometry suggests that this should be not a major concern for most regions of the rings, there is some evidence that the rings' scattering function is influencing its spectral properties in the B-ring core (see below). 

Fully exploring these complications will likely require both more sophisticated spectral modeling and careful comparisons with other published spectrophotometric observables like the rings' phase function and albedos \citep{Doyle89, Cooke91, Dones93, CE98, Porco05, Deau07, Morishima10}. Such analyses are well beyond the scope of this initial study, and thus must be the subject of future work. Nevertheless, the trends derived  with the above analysis are sufficiently interesting that their potential implications deserve to be considered in some detail in the next two sections.

\section{Variations in ring particle composition}

Changes in the ring-particle composition across the rings are traced by the parameters $f_C\kappa_C$ and $f_{UV}\kappa_{UV}$. One interpretation of these parameters' very different radial profiles is that they correspond to two different contaminants with different origins and spatial distributions within the rings. In that case, the observed trends could either represent variations in these contaminants'  concentration $f$ or their optical properties $\kappa$. The variations in $f_C\kappa_C$ across the rings seem consistent with variations in the concentration of an extrinsic contaminant. However, the variations in $f_{UV}\kappa_{UV}$ could also be interpreted as changes in the intrinsic properties of the UV-absorbing contaminant.  In principle, one might even be able to explain the trends in both these parameters in terms of variations in the concentration and optical properties of a single contaminant (cf. Clark {\it et al.} 2012), but for the sake of simplicity we will not explore such possibilities in detail here.

\subsection{Distribution of the broad-band absorber}

\begin{figure}
\resizebox{6in}{!}{\includegraphics{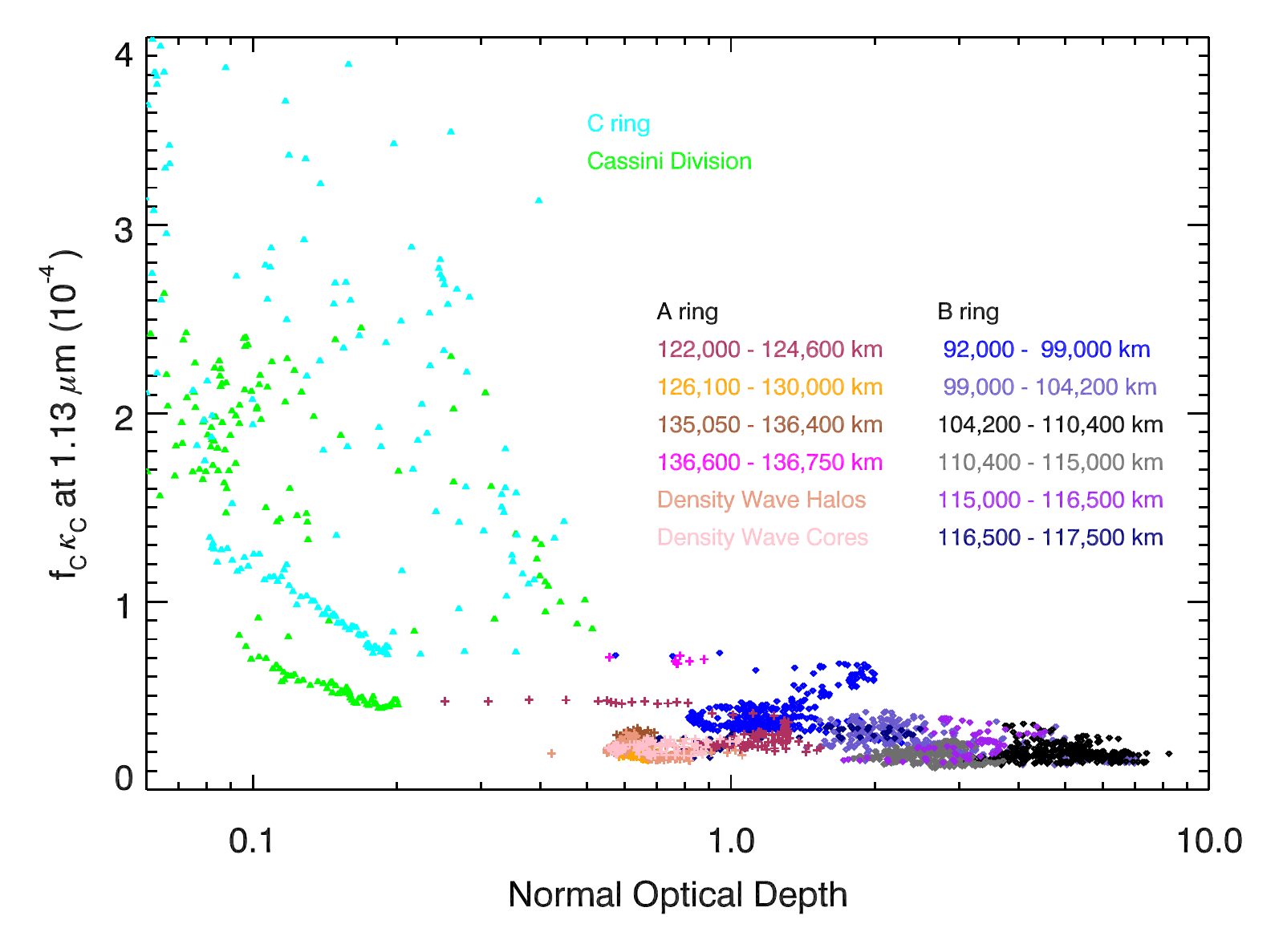}}
\caption{Plot of $f_C\kappa_C$ at 1.13 $\mu$m in the rings as a function of optical depth.}
\label{occbbcont}
\end{figure}

Consistent with previous studies of visible color ratios and  albedo variations across the rings  \citep{Cooke91, CE98}, we find that the concentration of broad-band absorber is much higher in the C ring and Cassini Division than it is in the A and B rings. Hence on large scales $f_C\kappa_C$ is anti-correlated with optical depth (see Figure~\ref{occbbcont}). However, if we examine the $f_C\kappa_C$ profile more closely, we can also identify trends that depend more on radial location than on local optical depth.  For example, the concentration of this contaminant increases smoothly and gradually with decreasing radius across the inner boundaries of both the A and B rings.  For the Cassini Division, the amount of contaminant rises rather quickly to a maximum near the location of the Laplace gap at 120,000 km, and then remains roughly constant throughout the rest of the Cassini Division (see also Figure~\ref{profova}). On the other hand, the concentration of this contaminant seems to steadily increase towards the planet throughout the C ring until we can no longer obtain sensible solutions for this parameter (see also Figure~\ref{profovc}). 

The larger concentration of broad-band absorber in the lower-optical depth parts of the main rings has been interpreted as the result of extrinsic meteoroid bombardment \citep{CE98}. In this scenario, the pure ice rings are progressively darkened by  interplanetary dust grains crashing into the rings from outside the Saturn system. Since the C ring and Cassini Division have a higher surface-area to mass ratio due their lower optical depths, this process darkens these rings more quickly than the denser A and B rings. Ballistic transport of material across the inner edges of the A and B rings then produces the smooth transitions in the concentration of this contaminant across the boundary between the A ring and the Cassini Division or the B ring and the C ring. In principle, the detailed shape of the $f_C\kappa_C$ curves in these regions could be used to refine ballistic transport models and meteoroid exposure ages from these rings \citep{CE98}, but such a detailed analysis is beyond the scope of this report.

We may also consider alternatives to the above scenario, where the differences in the broad-band contaminant are not just due to extrinsic bombardment, but instead represent a lag deposit analogous to those found on comets (thanks to M. A'Hearn for suggesting this idea). The albedo differences between the higher and lower optical depth regions could then be due to thermal ice migration processes similar to those that have been invoked to explain the albedo dichotomy on Iapetus \citep{SD10}. Water molecules are constantly being lost from the ring-particles' surfaces either by thermal sublimation or perhaps by sputtering. These free water molecules are very likely to contact another ring particle and become stuck to the surface again. The relevant sublimation rates are very sensitive to temperature, dropping by a factor of 1000 per degree K at 100 K \citep{Andreas07, SD10}. The darker Cassini Division and C ring can be over 10 K warmer than the A and B rings \citep{Spilker06, Cuzzi09, Morishima10}, so it is possible that water molecules are currently being lost from the lower-optical depth rings and collecting on the higher-optical depth rings. One may therefore conceive of scenarios where a slight albedo difference generated by extrinsic contaminants could be amplified by this thermal migration. Unfortunately, it appears that the relevant sublimation rates are too slow for this mechanism to work.  For the peak temperatures of about 100 K in the C ring \citep{Spilker06, Cuzzi09}, one could in principle obtain sublimation rates as high as 1 $\mu$m per $10^6$ years. This is much less than the rates of impact gardening that would mix sublimated material with the subsurface, which are of order centimeters per $10^6$ years \citep{Durisen92, SD10}, and thus it appears unlikely that variable sublimation rates would affect the rings' albedo at near-infrared wavelengths.

In addition to the broad-scale differences between the major ring regions, there are also so more subtle variations within the A and B rings. For example, the region exterior to the Keeler Gap (at 137,500 km) in the outer A ring appears to have roughly twice as much of this contaminant as the rest of the A ring. Similarly, the low-optical depth regions in the inner B ring appear to have slightly higher concentrations of this contaminant. Since these trends are strongly anti-correlated with the recovered variations in scattering length, they may be artifacts of our regolith model instead of real variations in the composition of the ring. However, the outermost A ring and regions of lower optical depth in the B ring  could also be somewhat more contaminated by infalling debris if the radial mixing is sufficiently slow. Thus any interpretation of these trends will have to be very tentative pending further analysis.

One place where this model almost certainly fails to capture the contaminant's spatial distribution is the B-ring's core between 105,000 km and 110,000 km. Our analysis indicates that the contaminant is concentrated in two regions around 107,500 km and 109,000 km, which seems inconsistent with the observed brightness of these features.  The entire region between 105,000 and 110,000 km is essentially opaque, so peaks in the absorption coefficient should yield dips in the ring's overall continuum brightness, but instead these two regions correspond to brightness maxima. Most likely, these inconsistencies arise because this is one of the few places in the rings where the spatial variations in spectral parameters are strongly phase dependent. Figure~\ref{prof3band} (and Figure~\ref{pp32} below) indicates that the dips in the band depth at 107,500 km and 109,000 km are much more noticeable in the high-phase RDHRSCHP data than they are in the low-phase RDHRCOMP data.  These unusual spectral variations will be discussed in more detail below in the context of features associated with strong mean-motion resonances, but for now we will refrain from ascribing any compositional significance to these peaks.

\subsection{Distribution of the ultraviolet absorber}

\begin{figure}
\resizebox{6in}{!}{\includegraphics{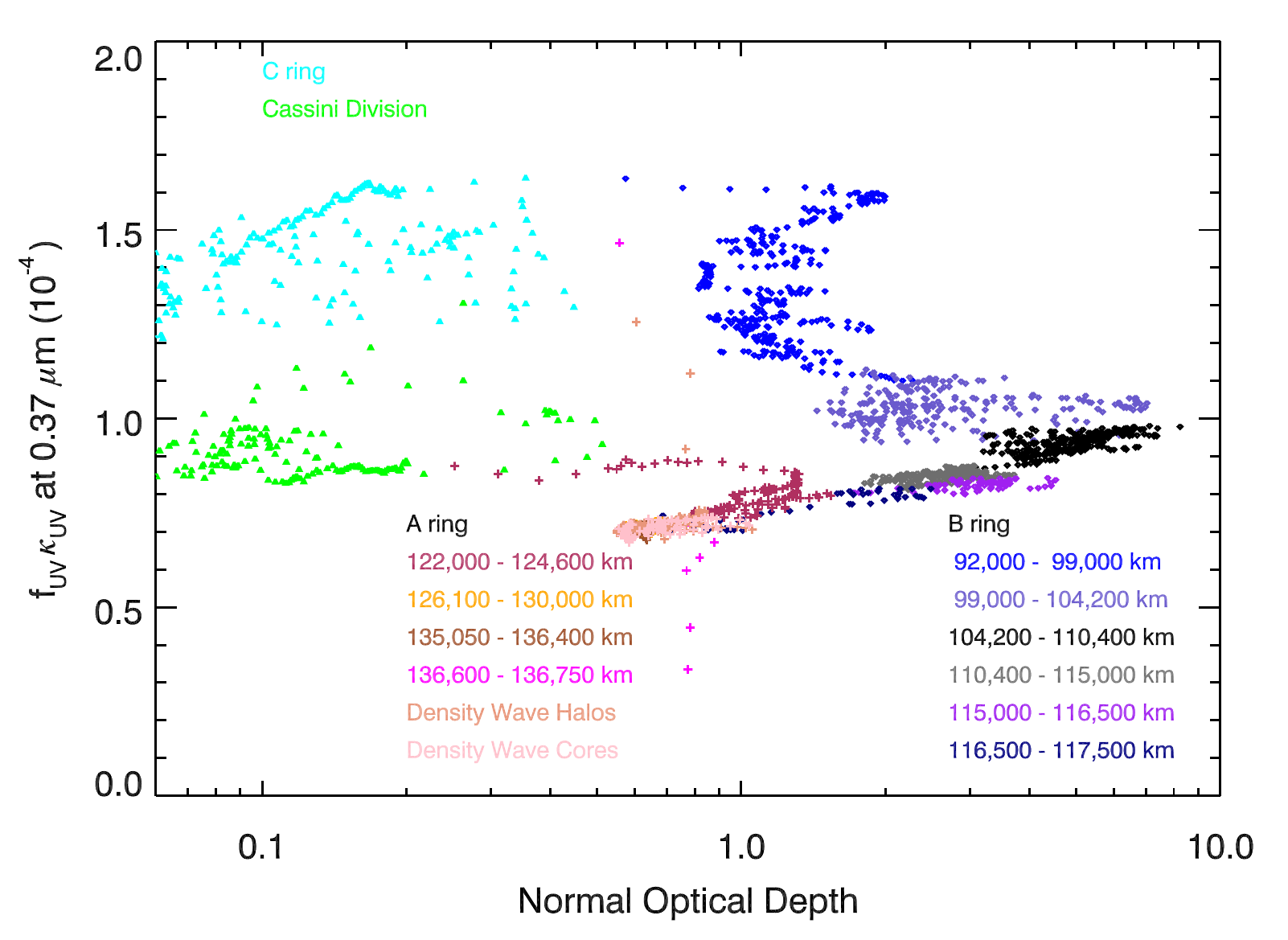}}
\caption{Plot of $f_{UV}\kappa_{UV}$ at 0.37 $\mu$m  as a function of optical depth.}
\label{occuvcont}
\end{figure}

Unlike the broad-band absorber, the ultraviolet absorber does not seem to vary much across the A ring, Cassini Division and outer B ring. This relatively uniform distribution would be more consistent with the ultraviolet absorber being an intrinsic component of the icy ring material than with it being an extrinsic contaminant like the broad-band absorber (cf. Cuzzi and Estrada 1998, Nicholson {\it et al.} 2008). 
Interior to about 100,000 km,  however, $f_{UV}\kappa_{UV}$ smoothly rises towards the planet, reaches a rather sharp peak near the inner edge of the B ring, and remains elevated throughout much of the C ring. Interestingly, there are no small-scale variations in $f_{UV}\kappa_{UV}$ in the inner B ring, where the optical depth varies substantially.  If additional absorber was somehow being added to the rings close to the planet, then one might expect the concentration of the ultraviolet absorber would depend upon the ring's optical depth. However, there is no strong correlation between $f_{UV}\kappa_{UV}$ and optical depth (see also Figure~\ref{occuvcont}). This provides another piece of evidence that the distribution of the ultraviolet absorber is not controlled purely by the infall of material into the rings. 

If we assume that the ultraviolet absorber is intimately mixed in the ring material, then the observed smooth trend in $f_{UV}\kappa_{UV}$  across the B ring could represent a primordial compositional gradient across the rings. In principle, whatever processes that generated the rings could have implanted materials with different chemical compositions at different locations in the rings. Over time, radial diffusion would smooth out these variations (cf. Salmon {\it et al.} 2010), leaving only broad-scale trends like those observed in $f_{UV}\kappa_{UV}$. \nocite{Salmon10} A potential difficulty with such an explanation is that the radial transport of material needs to be sufficiently efficient to remove small-scale compositional variations, but also slow enough so that large-scale trends in the inner B ring are preserved.  

Alternatively, the optical properties of the contaminant could be changing with distance from the planet. Such explanations are motivated by the observation that $f_{UV}\kappa_{UV}$ begins to change noticeably around 100,000 km from Saturn center. \citet{Crary10} has pointed  out that negative ions will be rapidly removed from the rings interior to about 99,000 km due to instabilities in the relevant particle orbits \citep{NH83, Jontof-Hutter12}. In principle, the changing ionization state of the ring material could oxidize or reduce the UV-absorbing molecules to yield a stronger absorption at short wavelengths. Of course,  it is not obvious that such changes in the UV-absorbing material are chemically plausible, especially since there is still much debate about the nature of this contaminant \citep{Cuzzi09, Clark2012}. However, if the spectra of  either candidate material (nano-phase iron particles or organic tholins) underwent the appropriate changes when ionized, this would not only lend some credence to this idea, it would also clarify the nature of the UV absorber.

Finally, we may note that the C ring is the only place where $f_{UV}\kappa_{UV}$ varies on small spatial scales (see Figure~\ref{profovc}). While we were only able to derive sensible regolith parameters for part of the C ring, the broad plateaux between 84,000 and 89,500 km appear to exhibit significant enhancements in the UV absorber  (The situation for the outer two plateaux is less clear). This is consistent with the blue slope being generally enhanced in the plateaux (see Figure~\ref{rdhr8plotc}, also Estrada {\it et al.} 2003).  The implications of these small-scale variations for the compositional evolution of the rings  are still unclear.

\section{Variations in ring particle regolith texture}

According to our analysis, essentially all of the fine-scale structure in the band depths of the A and B rings appears to result from variations in the effective scattering length in the particles' regolith. In some locations, the variations in scattering length are strongly correlated with the ring's optical depth, and in other places there are clear scattering-length variations in the vicinity of  strong mean-motion resonances. While model-dependent, these findings suggest that the ring-particles' surface texture could be strongly influenced by aspects of the local dynamical environment. These spectral data can therefore provide novel probes of dynamical phenomena operating in the rings.

\subsection{Textural changes correlated with optical depth}

\begin{figure}
\centerline{\resizebox{4in}{!}{\includegraphics{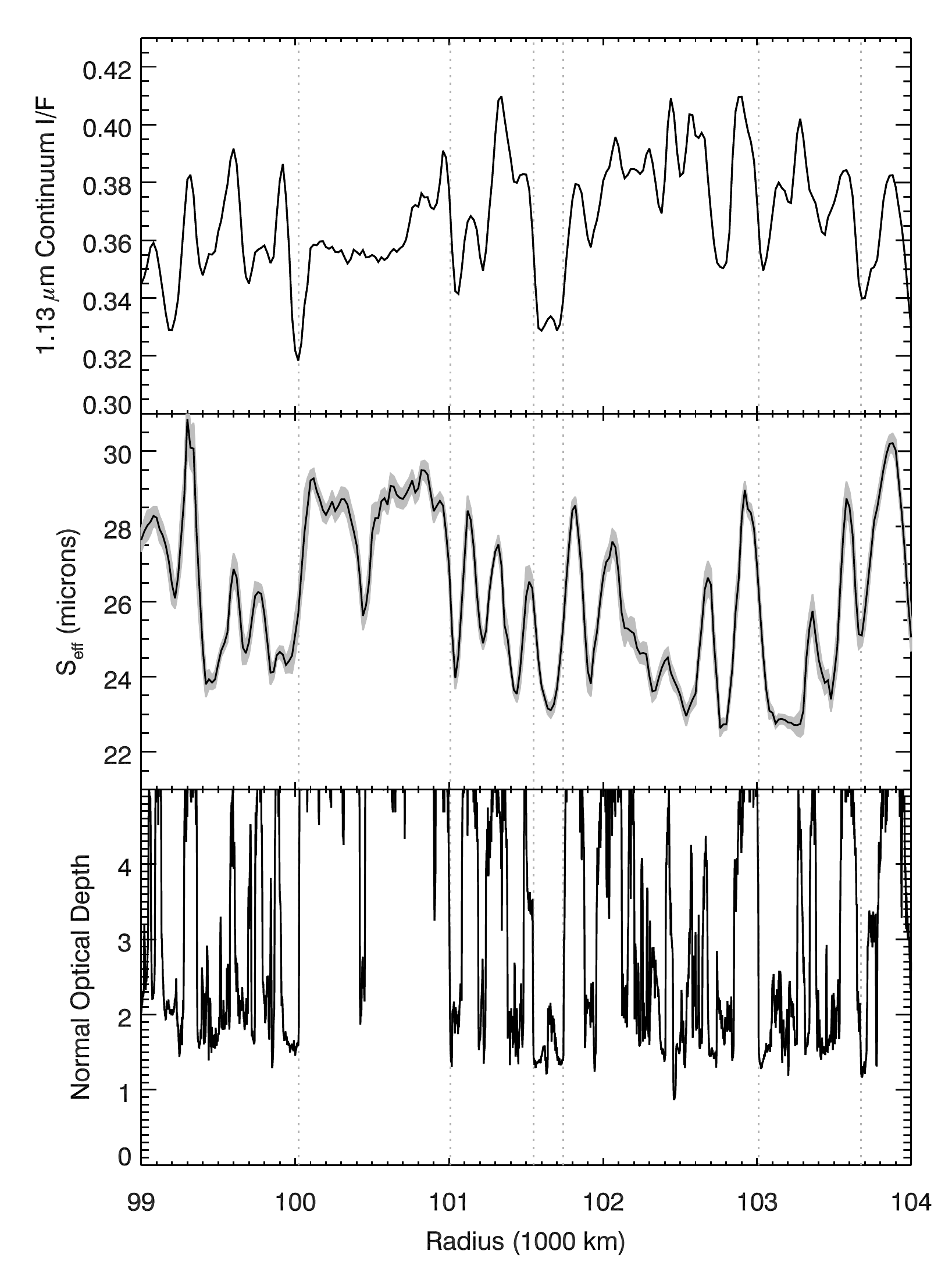}}}
\caption{Close up of the scattering-length variations in the BII region of the inner B ring derived from the RDHRCOMP observatons, along with the optical depth and brightness measurements for reference. The grey band indicates the 1$\sigma$ statistical error bars on the scattering length estimates at the 20-km sampling resolution. Note  the strong correlation between the scattering length and the optical depth throughout this region. By contrast, the brightness of the rings  appears to show variations that are not so clearly correlated with either the optical depth or the scattering length.}
\label{grainib}
\end{figure}

\begin{figure}
\centerline{\resizebox{4in}{!}{\includegraphics{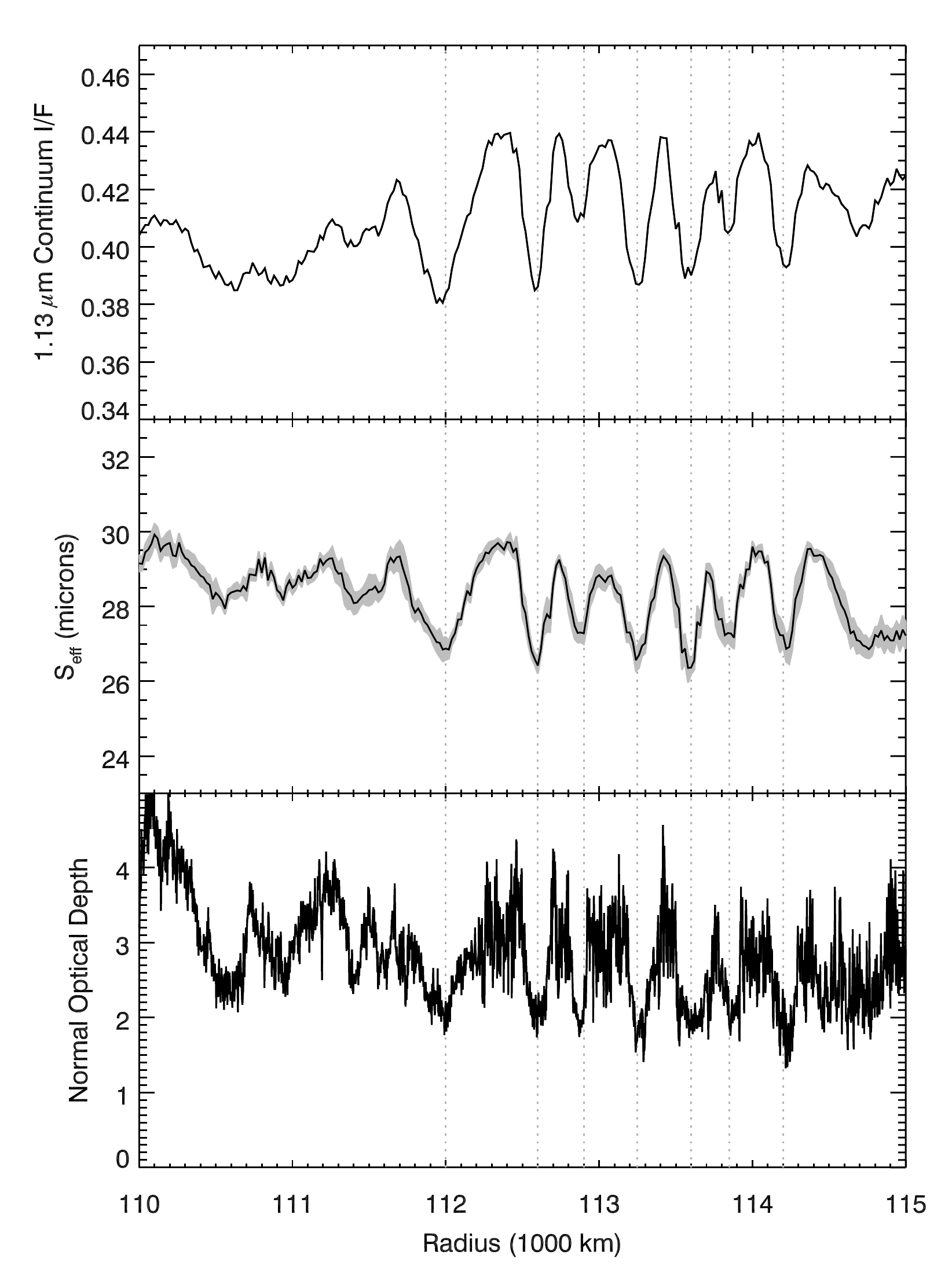}}}
\caption{Close up of the scattering-length variations in the BIV region of the outer B ring derived from the RDHRCOMP observatons, along with the optical depth and brightness measurements for reference. The grey band indicates the 1$\sigma$ statistical error bars on the scattering length estimates at the 20-km sampling resolution. Note the strong positive correlation between the rings' brightness, scattering length and  optical depth in this region.}
\label{grainob}
\end{figure}

\begin{figure}
\centerline{\resizebox{4in}{!}{\includegraphics{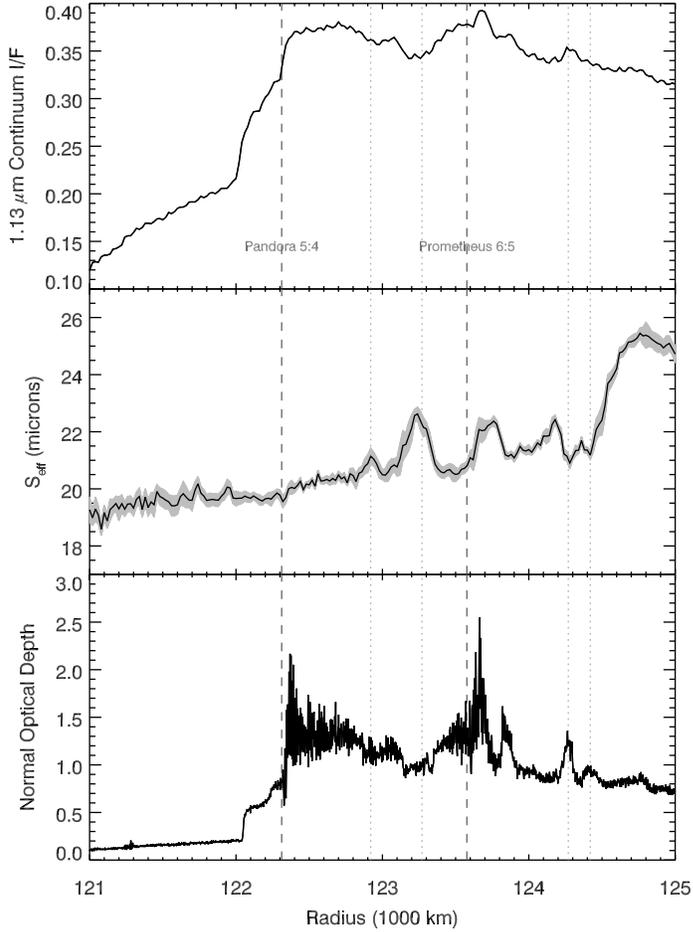}}}
\caption{Close up of the scattering-length variations in the inner A ring derived from the RDHRCOMP observatons, along with the optical depth and brightness measurements for reference. The grey band indicates the 1$\sigma$ statistical error bars on the scattering length estimates at the 20-km sampling resolution. Note that the peaks in scattering length at 122,900 km and 123,200 km correspond to dips in the optical depth and brightness, while the dips in scattering length at 124,250 and 124,400 km  correspond to peaks in optical depth and brightness. The peak in scattering length around 123,700 km may be correlated with the density wave just exterior to the Prometheus 6:5 resonance (visible in the bottom panel).}
\label{grainia}
\end{figure}

\begin{figure}
\resizebox{6in}{!}{\includegraphics{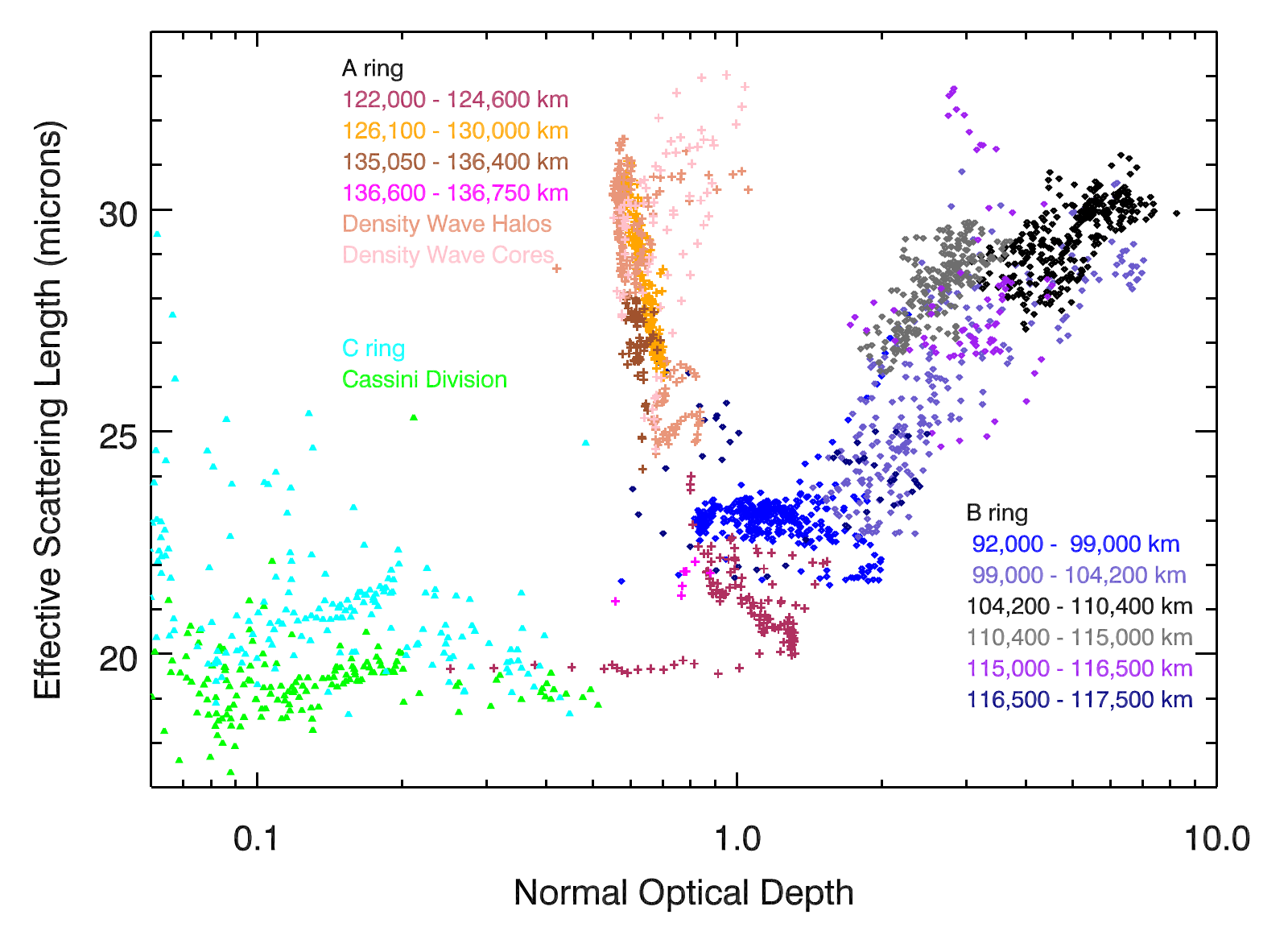}}
\caption{Plot of effective scattering length in the rings as a function of optical depth.}
\label{occgrain}
\end{figure}

Throughout the entire B ring, there is a very strong correlation between the ring's effective scattering length and its optical depth. These correlations are most apparent in the region between 97,000 km and 105,000 km from Saturn center (see Figure~\ref{grainib}). This region, designated BII \citep{Colwell09}, shows a series of abrupt transitions between two optical depth states, one with a typical optical depth between 1 and 2, and the other with an optical depth above 4.  These states are associated with two different scattering lengths in this region. Each transition between the high and low optical depth states matches exactly a transition between the larger and smaller scattering-length states, indicating a close relationship between these two parameters. This is especially remarkable given that the continuum brightness of the rings in this region does not exhibit the same series of transitions as the scattering length or the optical depth.

Correlations between optical depth and scattering length can also be observed in the outer B ring (see Figure~\ref{grainob}). In the BIV region between 110,000 km and 115,000 km, there is a quasi-periodic variation in optical depth with a wavelength of order 300 km. A similar periodic modulation can be seen in the scattering length, and just like in BII, regions of enhanced optical depth have a one-to-one correspondence with regions with increased effective scattering length. Interestingly, here the ring's brightness is also correlated with the optical depth variations. Correlations between optical depth and scattering length can even be found in the BIII region between 105,000 km and 110,000 km, where the ring is nearly entirely opaque. Here there are a series of $\sim$10 narrow regions where the optical depth drops to about 2. Each one of these narrow dips in optical depth corresponds to a narrow dip in scattering length (see also Figure~\ref{pp32} below).

There also appears to be a correlation between the optical depth and scattering length in the inner A ring (see Figure~\ref{grainia}). However, unlike the B ring, where more opaque regions show larger scattering lengths, in the A ring regions of lower optical depth (and lower brightness) seem to have the larger scattering lengths.  For example, there is a sharp increase in the scattering length around 124,500 km where the ring optical depth drops below 0.8. Additional anti-correlated structures in the two profiles can be seen interior to this step. For example, there are peaks in the scattering length around 123,000 km which correspond to weak dips in the optical depth. Also, around 124,420 km, there are dips in the scattering length that seem to correspond to peaks in the optical depth. 

While the correlations between optical depth and effective scattering length seem to be different in the A ring and the B ring, it is interesting that the lower scattering lengths in both rings seem to be primarily associated with regions where the optical depth is between 1 and 2. This is illustrated more clearly in Figure~\ref{occgrain}, which shows the scattering length as a function of optical depth in both the A and B rings. Both rings show reduced scattering lengths when the optical depth is between 1 and 2, with larger effective scattering lengths at higher and lower optical depths. Very low optical depth regions like the  C ring and Cassini Division also tend to exhibit small effective scattering lengths, but these estimates have relatively large statistical  uncertainties. At present, we have no simple explanation for these trends. Perhaps the variations in the effective scattering length reflect differences in the frequency or severity of the impacts between particles within regions of different optical depth.

\subsection{Textural changes associated with strong resonances}

\begin{figure}
\centerline{\resizebox{4in}{!}{\includegraphics{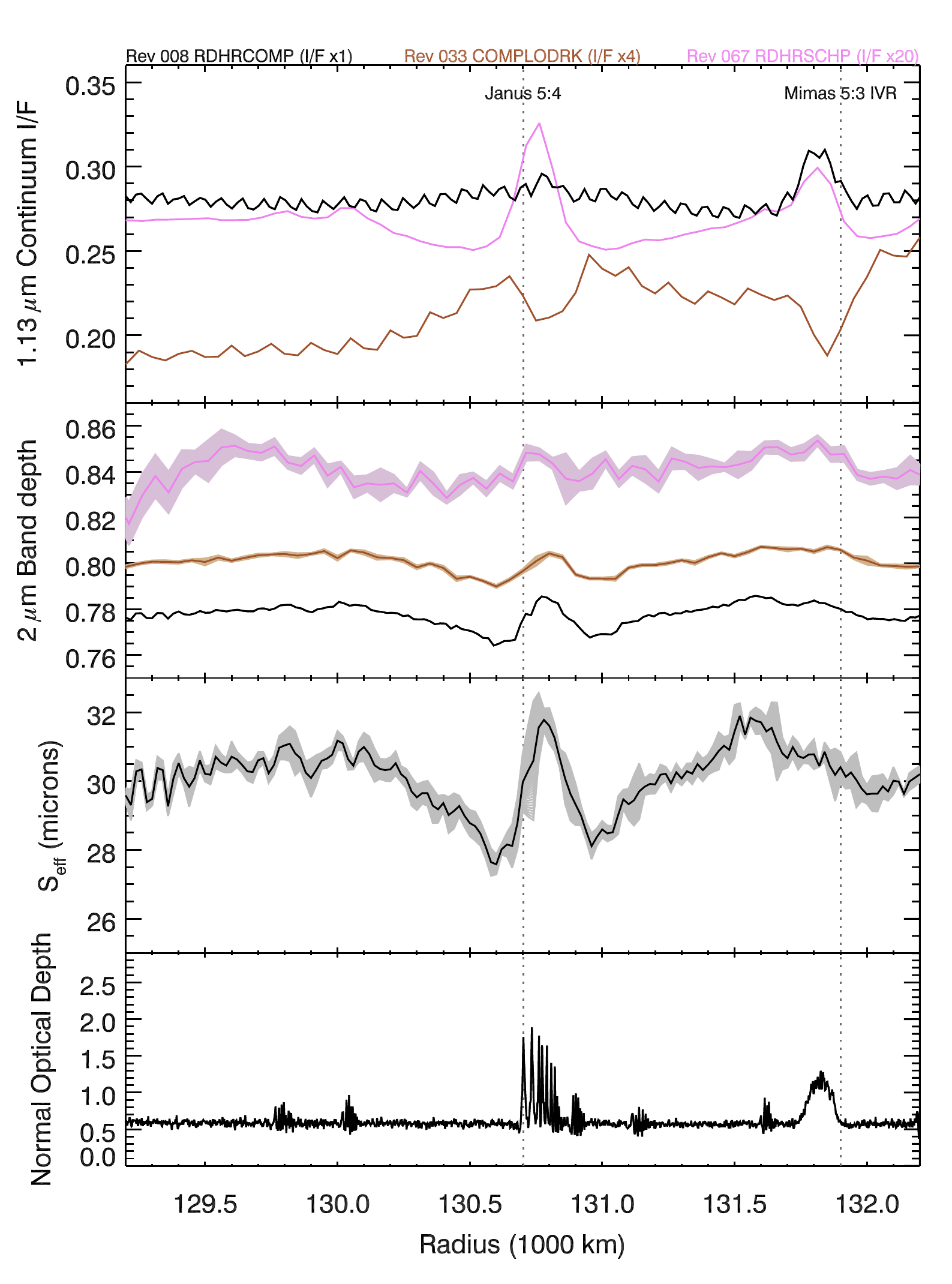}}}
\caption{Close up of the spectral and scattering-length variations in the region around the Janus 5:4 resonance in the middle  A ring. The top panel shows the brightness at 1.8 microns from the 3 observations (scaled by the indicated factors), while the second panel shows the 2-micron band depths (unscaled). The third panel shows the scattering length derived from the spectral analysis of the RDHRCOMP data. Shaded bands indicate the 1$\sigma$ statistical error bars on the parameter estimates at the sampling resolution. Finally, the bottom panel shows the ring optical depth for reference. Note that at the location of the wave itself both the band depths and the scattering lengths are enhanced, and the ring appears brighter at high phase angles and darker on the dark face at low phase angles (however, this feature is not obvious in the lit-side, low-phase profile). Surrounding the wave is a ``halo'' of reduced band depths and scattering lengths, reduced brightness at high phase angles and increased brightness at low phase angles. Note that in the top panel the saw-toothed appearance of the RDHRCOMP and COMPLODRK data is an artifact due to variations in the ring's brightness with longitude.}
\label{ja54}
\end{figure}

The most obvious variations in effective scattering length that are not  correlated with optical depth are found around the locations of the four strongest Inner Lindblad Resonances in the A ring: the Janus 4:3 resonance near 125,400 km, the Janus 5:4 resonance near 130,800 km, the Mimas 5:3 resonance near 132,300 km and the Janus 6:5 resonance near 134,300 km. Previous studies had identified these regions as having anomalous spectral and photometric properties \citep{Dones93, Nicholson08}. Figure~\ref{ja54} illustrates the variations in the ring's brightness, spectral properties and inferred scattering lengths in the vicinity of one of these resonances. These data show that  the ``core" of the complex, which corresponds to a narrow peak in the band depths, appears as a narrow peak in the brightness in the lit-side, high-phase data, a narrow dip in the dark-side low-phase data, and is only weakly expressed in the lit-side, low-phase data.  By  contrast, the broader halos appear to be bright in the dark-side, low-phase data (cf. Fig 10. in Nicholson {\it et al.} 2008), dark in the high-phase, lit-side data and either bright or dark depending on the wave in the low-phase, lit-side data. The present analysis indicates that these characteristic spectral signatures of strong density waves could be  due to alterations in the particles' regolith texture, with a narrow ``core" of increased effective scattering length surrounded by a broader ``halo'' of reduced scattering lengths (see Figure~\ref{ja54}). Note that stellar occultation measurements demonstrate that the characteristics of the rings' self-gravity wakes are also modified in the vicinity of these resonances \citep{Colwell06, Hedman07, NH10}, perhaps indicating changes in the particle size distribution in these regions. 

\begin{figure}
\centerline{\resizebox{4in}{!}{\includegraphics{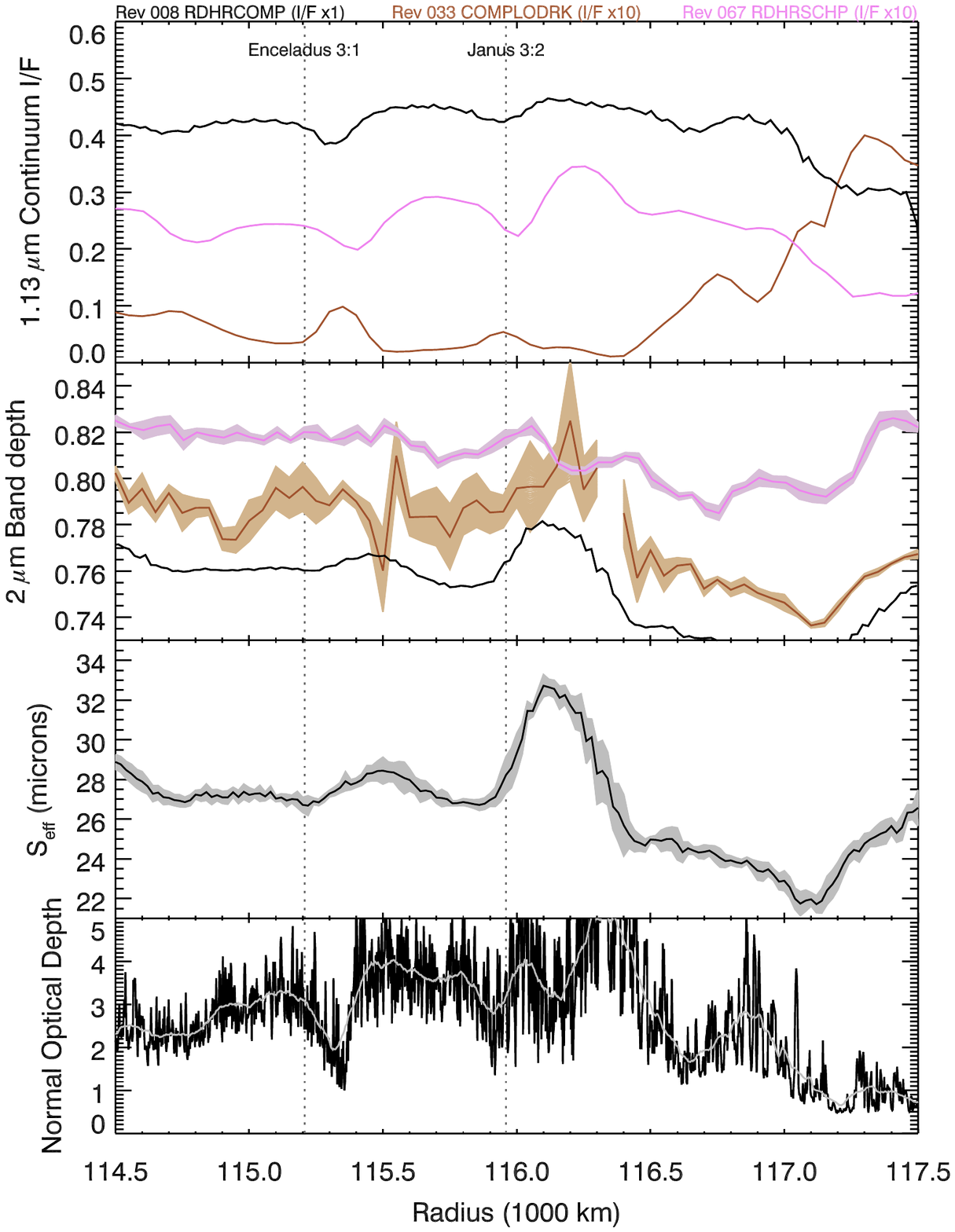}}}
\caption{Close up of the spectral and scattering-length variations in the region around the Janus 3:2 and Enceladus 3:1 resonances in outer B ring. The top panel shows the brightness at 1.8 microns from the 3 observations (scaled by the indicated factors), while the second panel shows the 2-micron band depths (unscaled, gaps in the COMPLODRK data correspond to regions where the signal-to-noise is insufficient to obtain sensible estimates of the band depth). The third panel shows the scattering length derived from the spectral analysis of the RDHRCOMP data. Shaded bands in these panels indicate the 1$\sigma$ statistical error bars on the parameter estimates at the sampling resolution. Finally, the bottom panel shows the ring optical depth for reference (a smoothed version of the data is shown as a gray line). Note the regions of enhanced band depths and scattering lengths just exterior to the two resonance locations. (There is also a hint of a broader halo around the Janus 3:2 signature.) These are very similar to the cores of the spectral variations associated with the strong density waves in the A ring.}
\label{ja32}
\end{figure}

\begin{figure}
\centerline{\resizebox{4in}{!}{\includegraphics{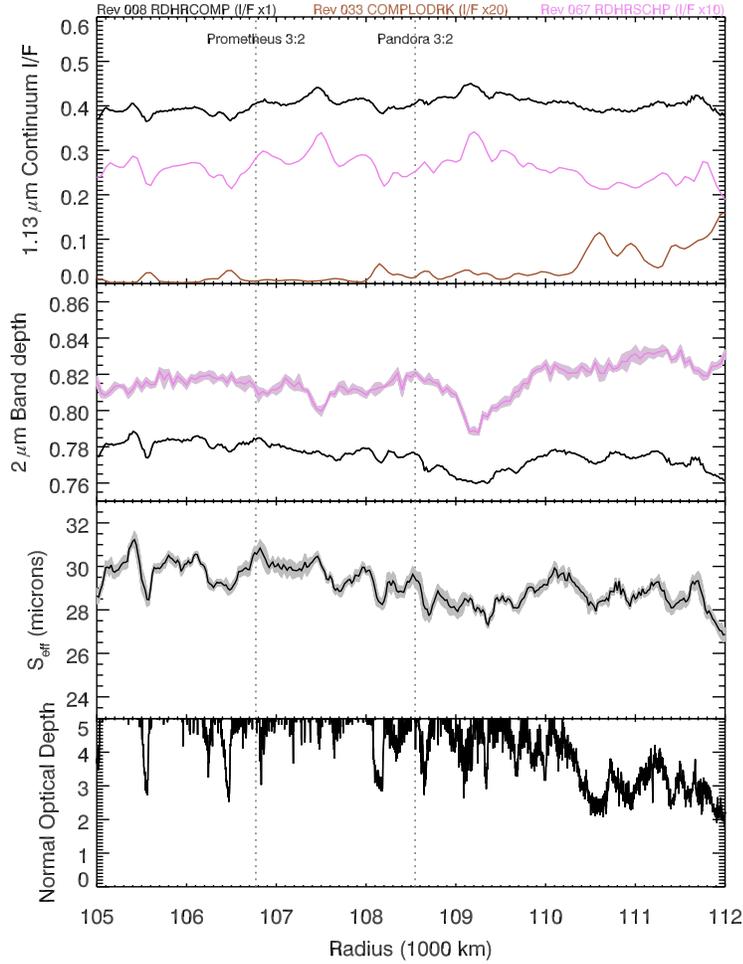}}}
\caption{Close up of the spectral and scattering-length variations in the region around the Prometheus 3:2 and Pandora 3:2 resonances in outer B ring. The top panel shows the brightness at 1.8 microns from the 3 observations (scaled by the indicated factors), while the second panel shows the 2-micron band depths (unscaled,  no COMPLODRK data is shown because signal-to-noise is insufficient in this region  to obtain sensible estimates of the band depth). The third panel shows the scattering length derived from the spectral analysis of the RDHRCOMP data. Shaded bands in these panels indicate the 1$\sigma$ statistical error bars on the parameter estimates at the sampling resolution.  Finally, the bottom panel shows the ring optical depth for reference. Note the regions of reduced band depths (and possibly scattering lengths) about 600 km exterior to the Prometheus 3:2 and Pandora 3:2 resonances.  }
\label{pp32}
\end{figure}

Similar variations in scattering length can be found in the vicinity of some of the strong mean-motion resonances in the B ring. These signatures can be distinguished from other scattering length variations by the lack of correlation between scattering length and optical depth. For example, between 115,000 and 117,000 km there are two sharp peaks in scattering length, each approximately 500 km wide (see Figure~\ref{ja32}). These two  features do not occur at the same locations as any of the peaks in optical depth in this complex region. However, they do occur slightly exterior to the Enceladus 3:1 and Janus 3:2 resonances. We therefore tentatively  interpret these two peaks as analogous to the peaks in scattering length  associated with the cores of strong density waves in the A ring, even though the waves themselves are not visible in the optical depth profile.

Suspicious features in the spectral/regolith parameter profiles can also be found in the densest part of the B ring (see Figure~\ref{pp32}). At both 107,500 and 109,200 km there are two 1000-km wide regions with reduced water-ice band depths. As discussed above, the model interprets these reduced band depths  as regions of increased $f_C\kappa_C$. However,  this result is questionable because such an increase in $f_C\kappa_C$ should lead to a reduction in the ring's brightness at these opaque locations, when in reality these two regions are the brightest parts of the rings. Thus our spectroscopic model is probably not capturing all the optical properties of these densest parts of the rings. Indeed, the difference in the strength of the band-depth dips between the RDHRCOMP and RDHRSCHP data indicates that this is one case where the spectral variations are phase-dependent. These regions are especially interesting because their centers lie roughly 600 km exterior to the strong 3:2 resonances with Prometheus and Pandora. These spectrally distinctive regions may therefore be generated by these resonances. Given that the band depths seem depressed in these regions, they may be analogous to the halos seen surrounding the density-wave signatures in the A ring, even though there is no corresponding peak at the expected wave position.

It is also worth noting that the 2:1 resonance with Janus at 96,500 km in the inner B ring does not appear to produce an obvious feature in the spectral parameters, even though it is very strong and produces an obvious density wave. This is also the only strong wave found in a region of the rings where the effective scattering length is in the low state corresponding to optical depths between 1 and 2\footnote{A possible exception is the peak in scattering length  in Fig.~\ref{grainia} visible around 123,700 km in the A ring, near the location of the relatively strong Prometheus 6:5 resonance.}  (see Section 7.1). Hence this wave's particular environment may be inhibiting the formation of a clear spectral signature.

The outer edges of the A and B rings also show distinctive effective scattering lengths.  The outermost 500 km in the B ring has an enhanced scattering length compared to other nearby regions (see Figures~\ref{profovb} and~\ref{ja32}). This could either be attributed to the strong perturbations in this region associated with  nearby 2:1 resonance with Mimas, or might instead be because the optical depth in this region falls below unity (cf Figure~\ref{occgrain}). Meanwhile, the outermost part of the A ring between the Keeler Gap and the A ring outer edge shows reduced effective scattering lengths (see Figure~\ref{profova}), which are
likely connected with the other spectral and photometric anomalies associated with this region \citep{Dones93, Nicholson08}.

\subsection{Using spectral signatures to constrain the structure and dynamics of the B ring}

As discussed above, our interpretation of the ``effective scattering length'' as a measure of the texture in the ring particle regoliths may be model-dependent, and more sophisticated spectral analysis could reveal that the above variations actually correspond to changes in other aspects of the rings' regolith, such as the nature of the mixing on different scales. Despite these uncertainties, the obvious relationships between $S_{\rm eff}$ and the local dynamical environment documented above are certainly suggestive and worthy of deeper investigation. In particular, the structures in the $S_{\rm eff}$ profile near the strong resonances in the middle and outer B ring can provide important new information regarding the density and dynamics of this almost opaque ring.

In the A ring, the locations and radial extents of the regions with {\it enhanced} scattering lengths closely match those of the visible density variations in the wavetrains (see Figure~\ref{ja54}). The density variations associated with any given density wave first increase with distance exterior to the resonance before they fade away due to inter-particle collisions. The radial extent of such waves is thus quantified in terms of a damping length $x_d$. For all the waves that produce obvious core-halo signatures (Mimas 5:3, Janus 4:3, 5:4 and 6:5), the measured damping lengths are between 140 and 160 km \citep{LE83}, which is comparable to the radial offsets between the resonance and the centers of the core-halo feature, as well as the half-width of the core in these features. Thus, regardless of how these features are generated, these spectral signatures provide an independent way to estimate the damping length of strong density waves. Hence the similar spectral features associated with the strong resonances in the B ring could  provide valuable new information about the resonance-induced disturbances in these highly opaque regions, where the waves themselves have not been detected in images or occultations.

The spectral signatures associated with the 3:2 Janus and 3:1 Enceladus resonances in the outer B ring correspond to peaks in the scattering length similar to the cores of the core-halo features in the A ring (see Figure~\ref{ja32}). The centers of the features are 250-300 km exterior to the resonances, and their half-widths are of comparable size. Thus we may infer that the damping lengths of these features are 250-300 km.  By contrast, the 3:2 Prometheus and 3:2 Pandora features correspond to regions of reduced water-ice band depths. These features thus more closely resemble the halos in the A ring, and since the extent of the A-ring halos are not obviously tied to the extent of the corresponding waves, the widths of these features cannot provide a clear estimate of the disturbance's  damping length. However, in the A ring both the cores and the halos are centered on a point roughly one damping length downstream of the resonances. Thus the center of these features may provide some indication of the damping length. For both these features the weakest band depths are found  600-700 km exterior to the resonance, so we may estimate the damping lengths of these disturbances as 600-700 km. These B-ring disturbances therefore appear to have  longer damping lengths than those in the A ring. This would be consistent with the $\sim 270$ km damping length of the Janus 2:1 waves found in the inner B ring \citep{LE83}. 

The implications of these long damping lengths are still uncertain, since the optical depths of these regions are above 2. The dynamics of such dense rings are still poorly understood, lying beyond the reach of current numerical simulations \citep{Robbins10}. It is therefore not yet clear if the formulae that describe the  propagation of density waves in the A ring can be extrapolated to apply to disturbances propagating through the denser parts of the B ring. Despite these theoretical limitations, it  is still instructive to consider the implications of such long damping lengths using the classical theory of linear density waves, as this can provide insights into  what properties of the rings these spectral signatures might be able to constrain. 

For a linear density wave (where the fractional density variations are much less than unity) with an azimuthal wavenumber $m>1$, the damping length is given by the expression:
\begin{equation}
x_d=\frac{2\pi G\sigma_0a}{(7K\Omega^4(m-1)^2\nu_{\rm eff} a)^{1/3}}
\end{equation}
where $G$ is the gravitational constant, $\sigma_0$ is the background undisturbed surface mass density, $K$ is the local radial epicyclic frequency, $\Omega$ is the local mean motion, $a$ is the semi-major axis, and $\nu_{\rm eff}$ is an effective kinematic viscosity, which depends on both the bulk and shear viscosity of the ring material, as well as these parameters' dependence on $\sigma_0$ \citep{GT78, Shu84, Schmidt11}.  Thus, the damping length scales like $x_d \propto \sigma_0(m-1)^{-2/3} \nu_{\rm eff}^{-1/3} a^{19/6}.$  Furthermore, simulations indicate that for massive rings, the bulk kinematic viscosity is  given by the following expression \citep{Daisaka01}:
\begin{equation}
\nu\simeq 26\left(\frac{a}{122,000 km}\right)^5\frac{G^2\sigma_0^2}{\Omega^3}
\label{nueq}
\end{equation}
so $\nu \propto \sigma_0^2a^{19/2}$. Hence, if  $\nu_{\rm eff} \sim \nu$, then  $x_d \propto \sigma_0^{1/3}(m-1)^{-2/3} $, and the damping length is only a weak function of the surface mass density. 

While $\nu_{\rm eff}$ is likely to be a more complex function of surface mass density than the \citet{Daisaka01} model for the bulk kinematic viscosity, it is not unreasonable that the longer damping lengths in the B ring indicate higher surface mass densities, especially given the high optical depths of the relevant ring regions.  In particular, the extremely long damping lengths implied by the spectral features associated with the Prometheus 3:2 and Pandora 3:2 resonances suggest that these regions could have extremely high densities. Indeed if we use the above naive scaling relations and assume a surface mass density in the A ring of order 40 g/cm$^2$ (see Colwell {\it et al.} 2009 and references therein), we obtain surface mass densities of 500-1000 g/cm$^2$ for these regions. Of course, this estimate is based on a rather questionable model for $\nu_{\rm eff}$, and the true mass density could be  lower if the effective viscosity was  less than the value predicted by Equation~\ref{nueq}. Further theoretical work is clearly needed before these spectral features can be properly interpreted. Nevertheless, these naive calculations already demonstrate that these features could potentially provide important constraints on the structure and dynamics of the densest parts of Saturn's rings.

\section{Summary}

This preliminary analysis of several high-resolution spectrograms obtained by the VIMS instrument provides many new insights into the ring particles' composition and texture 
and how these parameters vary across the rings. 

\begin{itemize}
\item The spatial variations in spectral parameters such as the depths of the water-ice absorption bands do not vary dramatically with viewing geometry, suggesting that most of the observed spectral variations are primarily due to shifts in the ring-particles' albedo rather than changes in the phase function.
\item Simple models for the ring-particles' regolith can be used to separate out spectral variations due to two different contaminants (one that absorbs over a broad range of wavelengths, one that only absorbs at short visible wavelengths) from those that may due to textural changes in the ring particles' regoliths (although other interpretations of this feature may also be viable).
\item Consistent with prior analyses, the concentration of the broad-band absorber is higher in the C ring and the Cassini Division than it is in the A and B rings. These trends support the idea that this absorber is an extrinsic contaminant due to meteoritic infall.
\item  The concentration of the broad-band absorber increases smoothly across the boundary between the C and B rings, as well as the boundary between the A ring and the Cassini Division. This could be attributed to ballistic transport across both these boundaries.
\item The ultraviolet absorber exhibits a rather uniform distribution across the rings, indicating that the contaminant responsible for this spectral feature is well-mixed with the ice, and may be ``primordial".
\item The concentration or the optical activity of the ultraviolet absorber increases smoothly across the inner B ring, and remains  high throughout the C ring (with a possible broad hump around 83,000 km). This could represent a primordial compositional gradient across the inner rings, or a chemical trend induced by changes in the rings' ionosphere.
\item Systematic differences in the ultraviolet absorber's optical activity can be detected between the plateaux and other parts of the C ring.
\item In the B ring, most of the spectral variations attributed to shifts in the texture of the ring particle regolith are positively correlated with the ring's optical depth. In the inner A ring, similar variations may be anti-correlated with the rings' optical depth. In both rings, the shortest effective scattering lengths occur where the ring optical depth is between 1 and 2.
\item Within the A ring, distinct core-halo spectral signatures are associated with the four strongest density waves. Similar spectral signatures can be found in the B ring associated with the Janus 3:2, Enceladus 3:1, Pandora 3:2 and Prometheus 3:2 resonances. These spectral signatures may provide new insights into the dynamics of these disturbances and the structure of  these dense rings.
\end{itemize}

Profiles of spectral and regolith parameters will be available from the author upon request and will be delivered to the Planetary Data System.

\section*{Acknowledgements} 

We acknowledge the support of the VIMS team, the Cassini Project and NASA.  We would like to thank P.Helfenstein for some useful conversations regarding this work. We also thank S. Brooks and an anonymous reviewer for their comments that helped improve this manuscript.


\section*{Appendix A: Algebra used to determine regolith parameters.}

As discussed in Section 4, the \citet{Shkuratov99} formulae allow each of the observed brightness ratios $r_i$ to be converted into a curve specifying the value of $\alpha S$ at  $\lambda_i$ (which we designate as $\beta_i$ below) as a function of $\alpha S$ at $\lambda_0$ (which we call $\beta_0$). We can then seek a value of $\beta_0$ where all four $\beta_i$ are consistent with the model for $\alpha S$ described in Section 4.2.  If this condition can be satisfied,  then we can solve the appropriate equations to obtain estimates for the relevant parameters.

Finding a value of $\beta_0$ consistent with the above model for $\alpha S$ is equivalent to  finding a value of $\beta_0$ where all the $\beta_i$ can be written as:
\begin{equation}
\beta_i=\alpha_I(\lambda_i)S_{\rm eff}+\alpha_0S_{\rm eff}+\left(\frac{\lambda_i-\lambda_0}{\lambda_0}\right)\alpha_1S_{\rm eff}+\left(\frac{\lambda_i-\lambda_0}{\lambda_0}\right)^2\alpha_2S_{\rm eff}
\label{beteq}
\end{equation}
using the same values of $\alpha_0$, $\alpha_1$, $\alpha_2$, and $S_{\rm eff}$.
In practice, this is most easily done by constructing an algebraic combination of $\beta_i$ and $\beta_0$ that depends only on the known wavelengths and ice absorption coefficients. Viable solutions can then be identified as those that yield the correct value for this parameter. Constructing such a parameter is greatly simplified by the above choice of  $\lambda_0$, which  serves as the reference wavelength not only for the brightness ratios, but also for the variations in $\alpha_C$. At $\lambda_0=1.13 \mu$m  ice has a very small imaginary refractive index ($\kappa_I<10^{-6}$), so $\beta_0$ can be approximated as $\alpha_0S_{\rm eff}$. In that case, the following algebraic combination of the observed $\beta_i$:
\begin{equation}
\mathcal{R}=\frac{
({\beta}_3-\beta_0)
-\frac{(\lambda_3-\lambda_0)(\lambda_3-\lambda_1)}{(\lambda_2-\lambda_0)(\lambda_2-\lambda_1)}(\beta_2-\beta_0)
+\frac{(\lambda_3-\lambda_0)(\lambda_3-\lambda_2)}{(\lambda_2-\lambda_1)(\lambda_1-\lambda_0)}(\beta_1-\beta_0)}
{({\beta}_4-\beta_0)
-\frac{(\lambda_4-\lambda_0)(\lambda_4-\lambda_1)}{(\lambda_2-\lambda_0)(\lambda_2-\lambda_1)}(\beta_2-\beta_0)
+\frac{(\lambda_4-\lambda_0)(\lambda_4-\lambda_2)}{(\lambda_2-\lambda_1)(\lambda_1-\lambda_0)}(\beta_1-\beta_0)}.
\end{equation}
can be expressed as the following combination of ice absorption coefficients and wavelengths  (provided all the $\beta_i$ are consistent with Equation~\ref{beteq}):
\begin{equation}
\mathcal{R}=
\frac{\alpha_I(\lambda_3)
-\frac{(\lambda_3-\lambda_0)(\lambda_3-\lambda_1)}{(\lambda_2-\lambda_0)(\lambda_2-\lambda_1)}\alpha_I(\lambda_2)
+\frac{(\lambda_3-\lambda_0)(\lambda_3-\lambda_2)}{(\lambda_2-\lambda_1)(\lambda_1-\lambda_0)}\alpha_I(\lambda_1)}
{\alpha_I(\lambda_4)
-\frac{(\lambda_4-\lambda_0)(\lambda_4-\lambda_1)}{(\lambda_2-\lambda_0)(\lambda_2-\lambda_1)}\alpha_I(\lambda_2)
+\frac{(\lambda_4-\lambda_0)(\lambda_4-\lambda_2)}{(\lambda_2-\lambda_1)(\lambda_1-\lambda_0)}\alpha_I(\lambda_1)}.
\end{equation} 
Hence we can use the known refractive indices of water ice to compute what $\mathcal{R}$ should be for the given set of wavelengths. For the wavlengths listed in Section 4.4,  $\mathcal{R}\sim 0.436$. Any value of $\beta_0$ that yields this value of $\mathcal{R}$ would then correspond to a set of $\beta_i$ consistent with the model for $\alpha S$, and could be used to estimate the parameters $\alpha_0, \alpha_1, \alpha_2$ and $S_{\rm eff}$.  

When these calculations are done for the Rev 008 RDHRCOMP spectra, $\mathcal{R}$ typically  increases with increasing $\beta_0$ up to some maximal level $\mathcal{R}_{max}$, after which it declines back to zero.  So long as $\mathcal{R}_{max}>0.436$, this yields only  two possible solutions for $\beta_0$. These two solutions are equivalent to the two  different possible solutions for $\alpha_C$ and $S_{\rm eff}$ discussed in Section 4.3. As in that case,  we can consistently select the smaller value of $\beta_0$ as the correct solution, because the other solution would correspond to a spectrum with large concentrations of contaminants and nearly saturated ice bands in the A and B rings, which is inconsistent with the high albedo of the ring material. Interior to about 80,000 km in the C ring, many spectra yield  $\mathcal{R}_{max}<0.436$, so there is no value of $\beta_0$ where the spectrum is consistent with our assumed model for $\alpha S$. These spectra are like the one shown in Figure~\ref{testcspec}, and no further attempt is made to determine the regolith parameters at these locations.

\section*{\ Appendix B: Detailed Profiles}

\begin{figure}[hb]

\centerline{\resizebox{5in}{!}
{\includegraphics{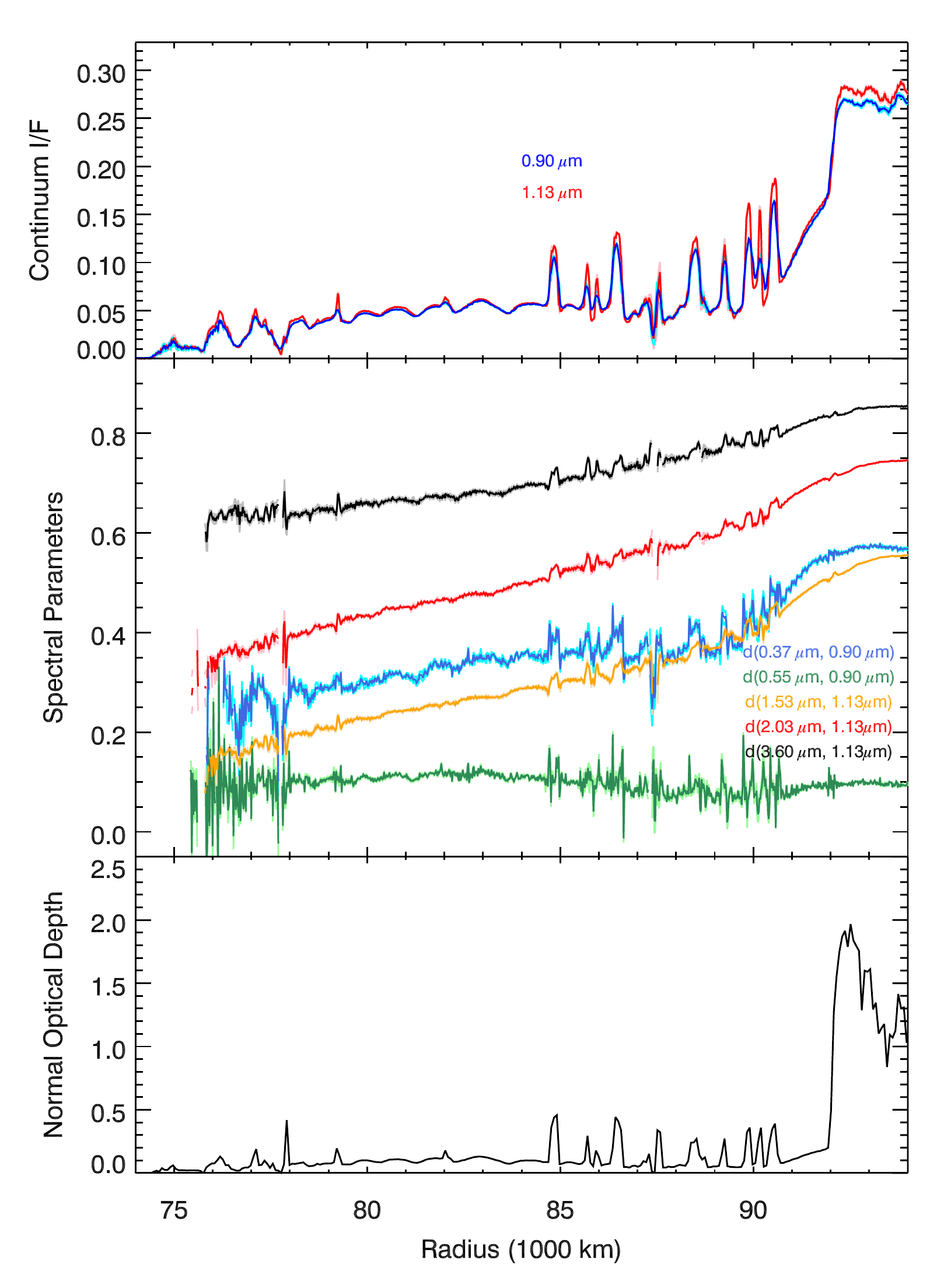}}}
\caption{Close-up view of the  C-ring's spectral parameters derived from the Rev 008 RDHRCOMP
observations, compared with the Rev 089 $\gamma$ Crucis occultation data.
The top panel shows the continuum brightness level in the VIS (blue) and IR (red) channels, and the middle panel shows five spectral parameters derived from spectral ratios. Where shown, light shaded regions indicate the 1-$\sigma$ error bars on these parameters (elsewhere these error bars are less than the line thickness). The bottom panel shows, for comparison, the optical depth profile derived from the occultation data, binned to approximately the same resolution as the spectral data.}
\label{rdhr8plotc}
\end{figure}

\begin{figure}
\centerline{\resizebox{5in}{!}{\includegraphics{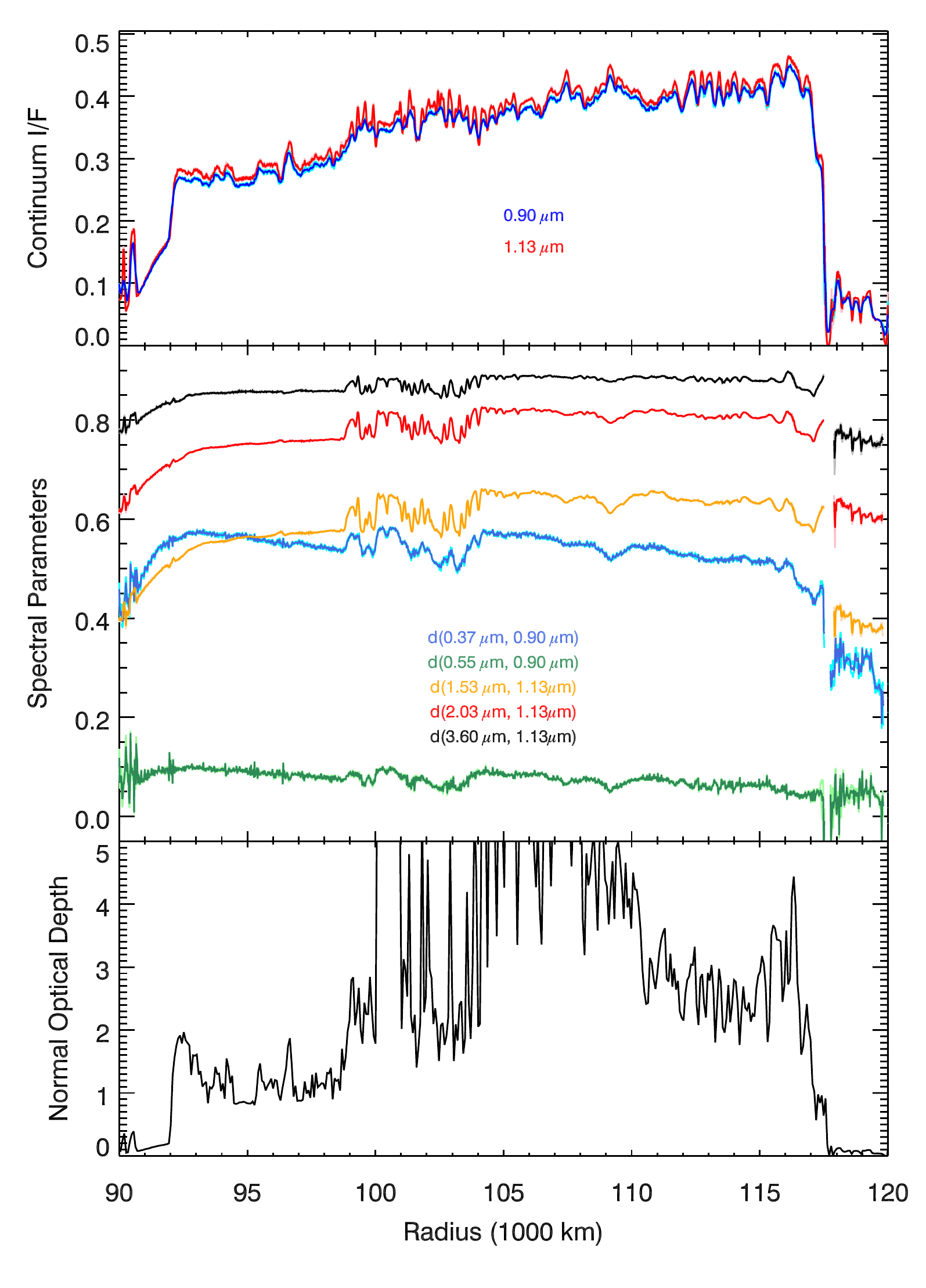}}}
\caption{Close-up view of the B-ring's spectral parameters derived from the Rev 008 RDHRCOMP
observations, compared with the Rev 089 $\gamma$ Crucis occultation data.
The top panel shows the continuum brightness level in the VIS (blue) and IR (red) channels, and the middle panel shows five spectral parameters derived from spectral ratios. Where shown, light shaded regions indicate the 1-$\sigma$ error bars on these parameters (elsewhere these error bars are less than the line thickness). The bottom panel shows, for comparison, the optical depth profile derived from the occultation data, binned to approximately the same resolution as the spectral data.}
\label{rdhr8plotb}
\end{figure}

\begin{figure}
\centerline{\resizebox{5in}{!}{\includegraphics{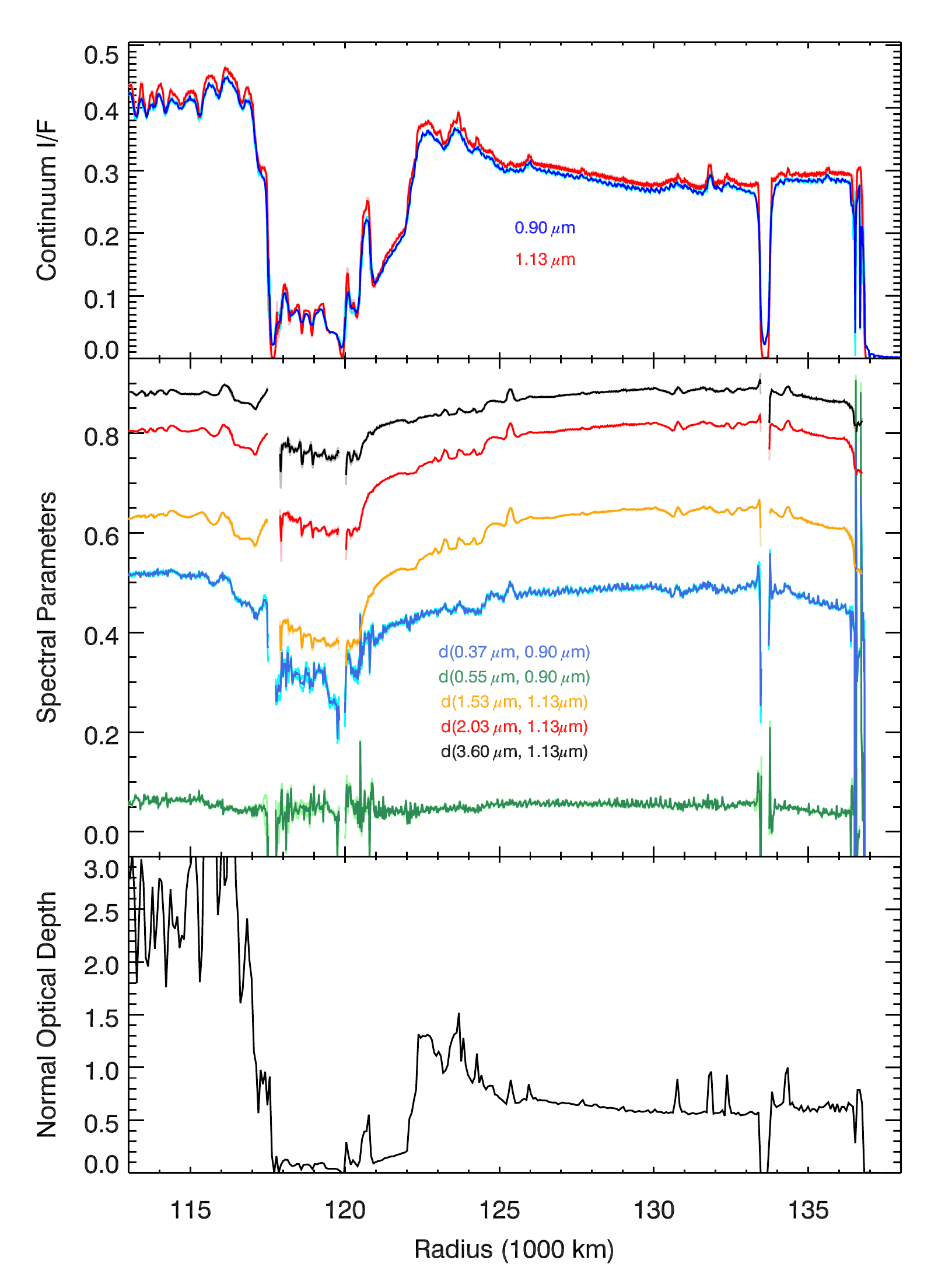}}}
\caption{Close-up view of the spectral parameters in the A ring and Cassini Division derived from the Rev 008 RDHRCOMP
observations, compared with the Rev 089 $\gamma$ Crucis occultation data.
The top panel shows the continuum brightness level in the VIS (blue) and IR (red) channels, and the middle panel shows five spectral parameters derived from spectral ratios. Where shown, light shaded regions indicate the 1-$\sigma$ error bars on these parameters (elsewhere these error bars are less than the line thickness). The bottom panel shows, for comparison, the optical depth profile derived from the occultation data, binned to approximately the same resolution as the spectral data.}
\label{rdhr8plota}
\end{figure}

\begin{figure}
\resizebox{6in}{!}{\includegraphics{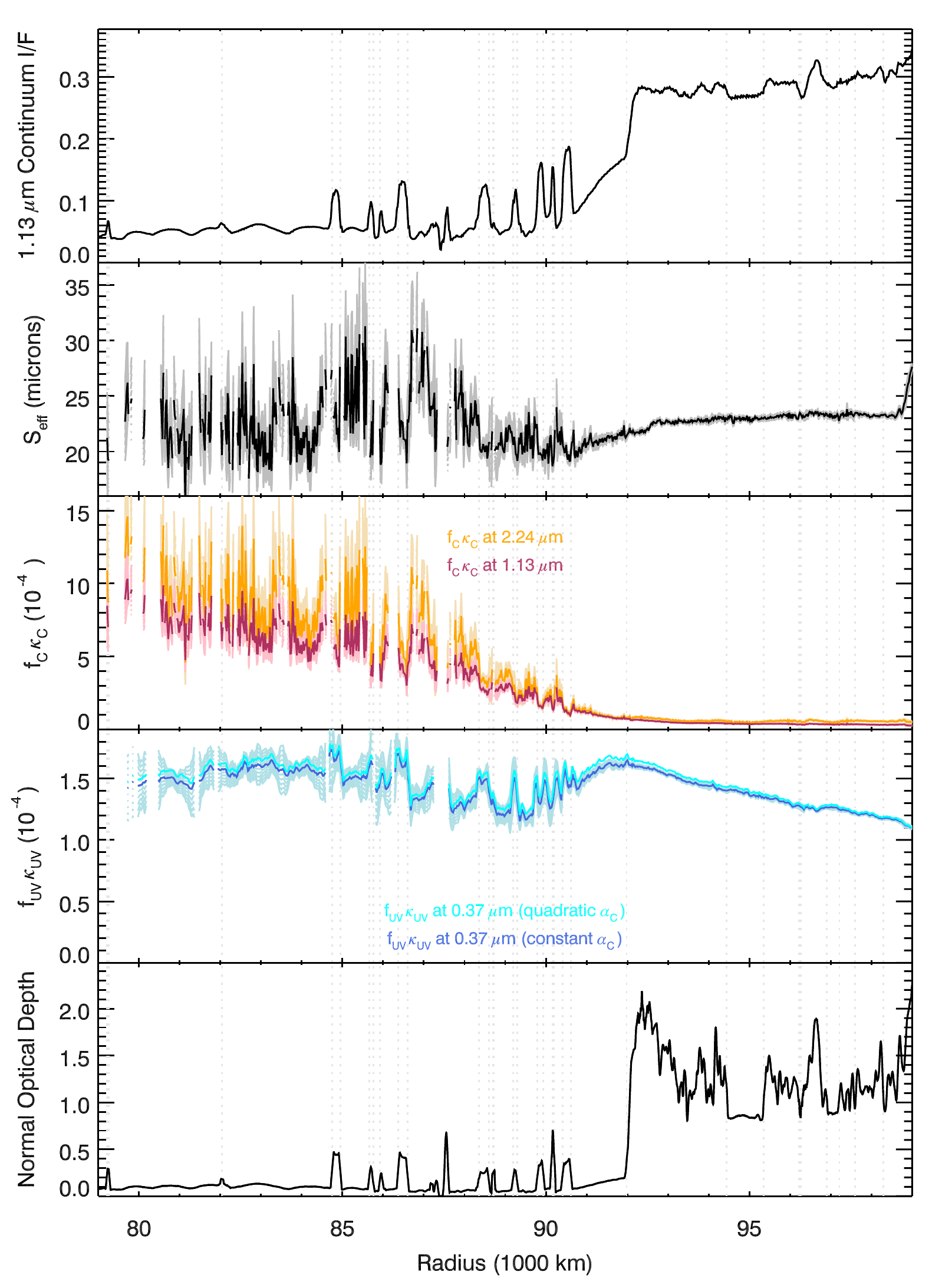}}
\caption{Profiles of typical regolith properties across the C ring and inner B ring derived from the low-phase, lit-face RDHRCOMP observations See Figure~\ref{profov} for information about what each panel shows. Faint vertical dotted lines serve to guide the eye between the different panels. Note that there are no abrupt changes in the regolith parameters at the inner edge of the B ring (92,000 km). }
\label{profovc}
\end{figure}

\begin{figure}
\resizebox{6in}{!}{\includegraphics{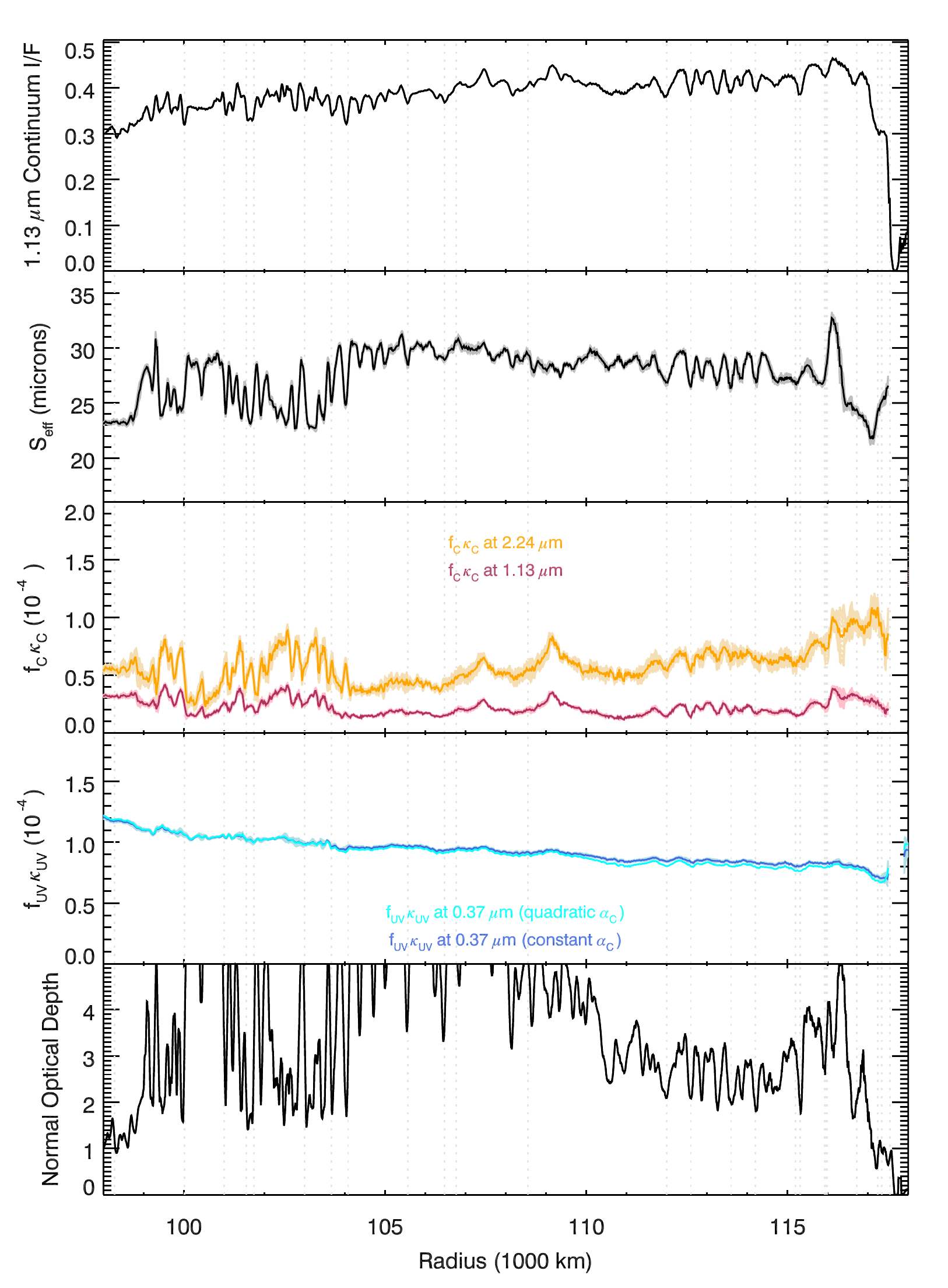}}
\caption{Profiles of typical regolith properties across the B ring  derived from the low-phase,
lit-face RDHRCOMP observations. See Figure~\ref{profov} for information about what each panel shows. Faint vertical dotted lines serve to guide the eye between the different panels.}
\label{profovb}
\end{figure}

\begin{figure}
\resizebox{6in}{!}{\includegraphics{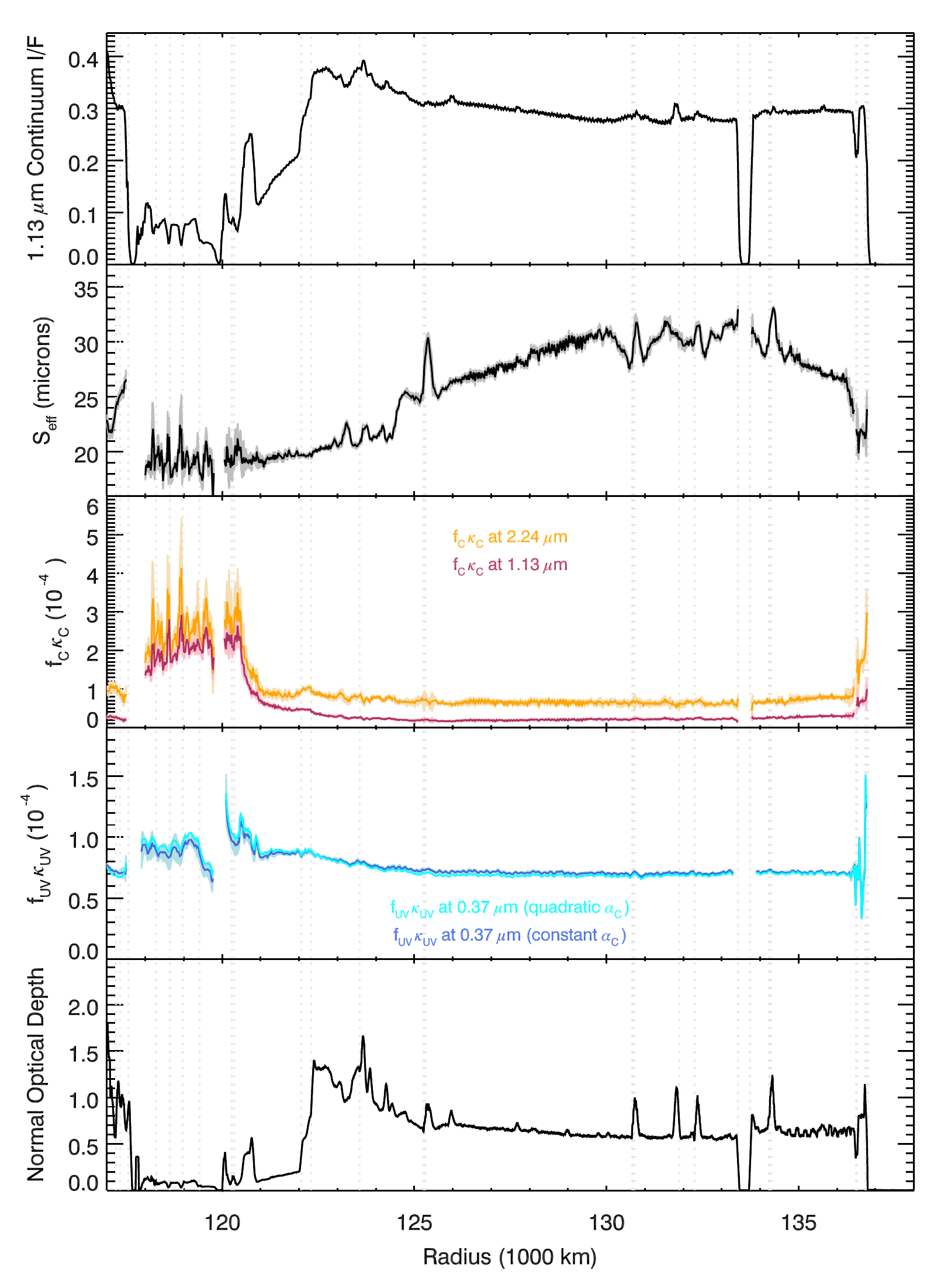}}
\caption{Profiles of typical regolith properties across the A ring and Cassini Division derived from the low-phase,
lit-face RDHRCOMP observations. See Figure~\ref{profov} for information about what each panel shows. Faint vertical dotted lines serve to guide the eye between the different panels. Note that there are no abrupt changes in the regolith parameters at the inner edge of the A ring (122,000 km).}
\label{profova}
\end{figure}

\clearpage
\pagebreak

\section*{Appendix C: Estimates of $A_1$}

\begin{figure}
\centerline{\resizebox{5in}{!}{\includegraphics{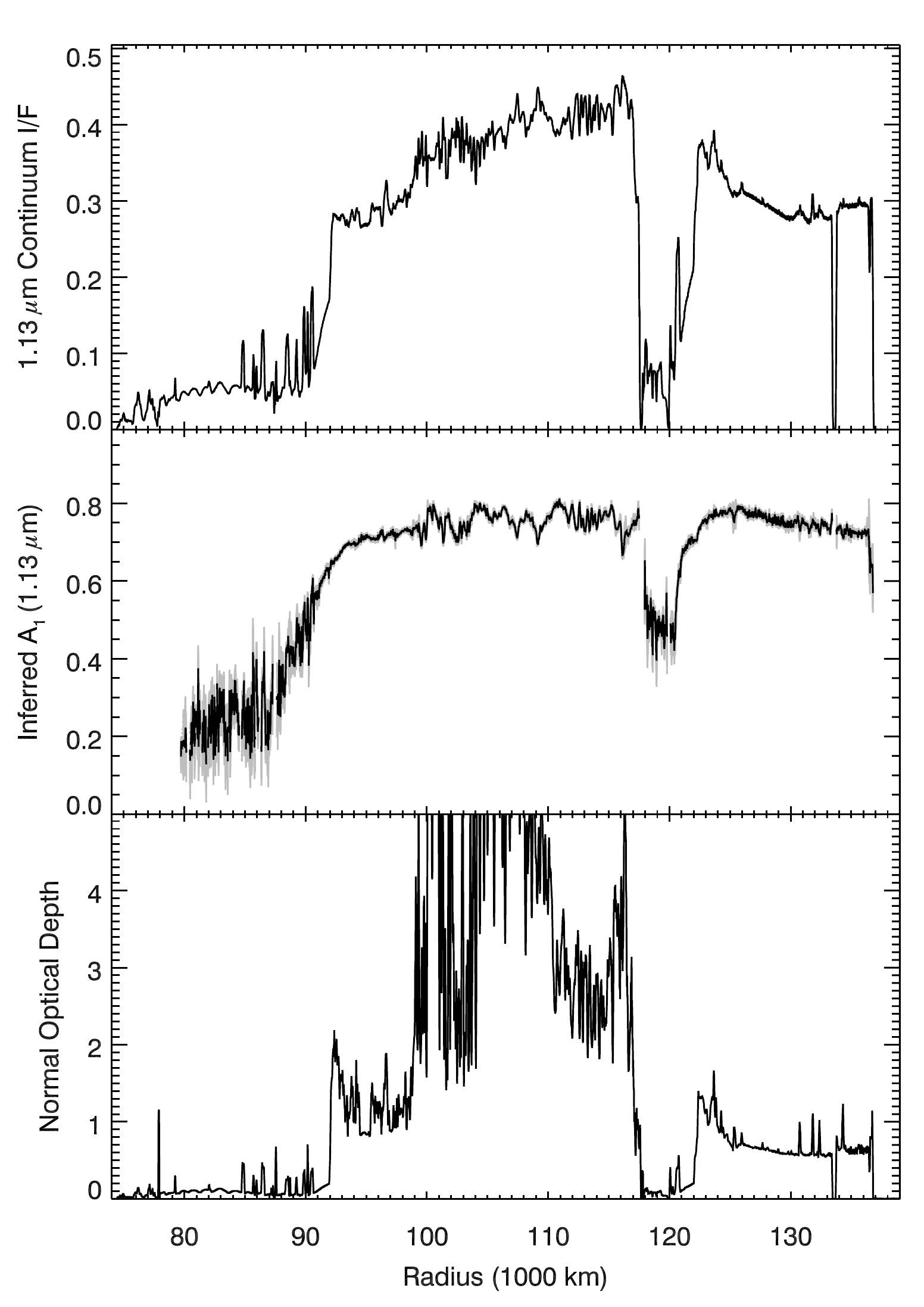}}}
\caption{Plot of the $A_1$ parameter  at 1.13 $\mu$m as a function of position across the ring,  derived from the estimates of $f_C\kappa_C$ and $S_{\rm eff}$ shown in the Figure~\ref{profov}.  Where visible, the statistical uncertainty in this parameter is shown as a gray shaded band. As usual, profiles of the rings' brightness and optical depth have been included for reference (the optical depth data have been smoothed to match the resolution of the spectral data). Note that  the $A_1$ parameter the \citet{Shkuratov99} light-scattering theory is only the albedo of a one-dimensional model for a regolith surface, so this parameter cannot be directly compared with the ring-particle bond albedos derived from various photometric measurements (see text).}
\label{alprof}
\end{figure}

While the above calculations relied entirely on brightness ratios from individual spectra, the resulting estimates of $f_C\kappa_C$ and $S_{\rm eff}$ can be fed back into the appropriate \citet{Shkuratov99} formulae to obtain estimates of the $A_1$ reflectance parameter at a single wavelength. For example, Figure~\ref{alprof} shows the resulting estimates of $A_1$ at 1.13 $\mu$m as a function of position across the ring (this wavelength corresponds to the reference wavelength for our spectral analysis, and thus is the simplest to calculate). Note in the A and B rings, $A_1$ is between 0.6 and 0.8, while in the Cassini Division $A_1$ is around 0.5 and in the C ring $A_1$ falls to as low as 0.2. These numbers are roughly comparable to the ring-particle single-scattering or Bond albedos at long visible and short-infrared wavelengths derived from various photometric analyses of Voyager and Cassini data \citep{Doyle89, Cooke91, Dones93, CE98, Porco05, Deau07, Morishima10}.  However, the detailed shape of this $A_1$ profile cannot be directly compared to those earlier measurements because they involve different photometric quantities. Recall that the $A_1$ parameter is the albedo of a one-dimensional model system that is approximately equal to the reflectance of a regolith surface observed at low phase angles \citep{Shkuratov99}. By contrast, most other published measurements are estimates of the single-scattering or bond albedo of individual ring particles derived from photometric observations of the entire ring (making some assumptions about the spatial distributions of the ring particles). Thus, we would need detailed information about  the ring-particles' scattering properties and spatial distributions before we could rigorously evaluate the consistency of these different measurements.\footnote{Even if we had used the Hapke-based formalism described in \citet{CE98} to compute estimates of the ring-particle's albedos directly, uncertainties in the ring-particles'  spatial distribution (which affects the amount of inter-particle scattering) would still complicate any comparisons with photometric analyses.}  Furthermore, the differences in the shapes of the RDHRCOMP and RDHRSCHP brightness profiles shown in Figures~\ref{prof3band} and~\ref{prof3vis} indicate that the rings' photometric phase function varies with position on a wide range of spatial scales, which would also need to be taken into account before we could make sensible comparisons. Such complex photometric analyses are well beyond the scope of this work, and so we will not attempt to interpret the  trends in the inferred $A_1$ profile here.

\end{document}